\documentclass[twocolumn,iop]{emulateapj}

\usepackage{aas_macros}
\usepackage{natbib}
\usepackage{graphicx}

\usepackage{color}
\usepackage{soul}

\begin{document}
\title{The Q/U Imaging Experiment:  Polarization Measurements of Radio Sources at 43 and 95$\,$GHz}

\author{ QUIET Collaboration --- 
K.~M.~Huffenberger\altaffilmark{1,2,$\star$},
D.~Araujo\altaffilmark{3},
C.~Bischoff\altaffilmark{4,5},
I.~Buder\altaffilmark{4,5},
Y.~Chinone\altaffilmark{6,7},
K.~Cleary\altaffilmark{8},
A.~Kusaka\altaffilmark{9,10},
R.~Monsalve\altaffilmark{11},
S.~K.~N\ae ss\altaffilmark{12,13},
L.~B.~Newburgh\altaffilmark{14},
R.~Reeves\altaffilmark{15},
T.~M.~Ruud\altaffilmark{13},
I.~K.~Wehus\altaffilmark{16},
J.~T.~L.~Zwart\altaffilmark{3,17},
C.~Dickinson\altaffilmark{18},
H.~K.~Eriksen\altaffilmark{13,19},
T.~Gaier\altaffilmark{16},
J.~O.~Gundersen\altaffilmark{2},
M.~Hasegawa\altaffilmark{6},
M.~Hazumi\altaffilmark{6},
A.~D.~Miller\altaffilmark{3},
S.~J.~E.~Radford\altaffilmark{8},
A.~C.~S.~Readhead\altaffilmark{8},
S.~T.~Staggs\altaffilmark{10},
O.~Tajima\altaffilmark{4,6},
K.~L.~Thompson\altaffilmark{20}
}

\altaffiltext{1}{Department of Physics, Florida State University, P.O.~Box 3064350, Tallahassee, FL 32306-4350, USA}
\altaffiltext{2}{Department of Physics, University of Miami, 1320 Campo Sano Drive, Coral Gables, FL 33146, USA}
\altaffiltext{3}{Department of Physics and Columbia Astrophysics Laboratory, Columbia University, New York, NY 10027, USA}
\altaffiltext{4}{Kavli Institute for Cosmological Physics, Department of Physics, Enrico Fermi Institute, The University of Chicago, Chicago, IL 60637, USA}
\altaffiltext{5}{Harvard-Smithsonian Center for Astrophysics, 60 Garden Street MS 42, Cambridge, MA 02138, USA}
\altaffiltext{6}{High Energy Accelerator Research Organization (KEK), 1-1 Oho, Tsukuba, Ibaraki 305-0801, Japan}
\altaffiltext{7}{Department of Physics, University of California, Berkeley, CA 94720, USA}
\altaffiltext{8}{Cahill Center for Astronomy and Astrophysics, California Institute of Technology, 1200 E. California Blvd M/C 249-17, Pasadena, CA 91125, USA}
\altaffiltext{9}{Physics Division, Lawrence Berkeley National Laboratory, 1 Cyclotron Road, Berkeley, CA 94720, USA}
\altaffiltext{10}{Joseph Henry Laboratories of Physics, Jadwin Hall, Princeton University, Princeton, NJ 08544, USA}
\altaffiltext{11}{School of Earth and Space Exploration, Arizona State University, 781 E. Terrace Road, Tempe, AZ 85287, USA}
\altaffiltext{12}{Department of Astrophysics, University of Oxford, Keble Road, Oxford OX1 3RH, UK}
\altaffiltext{13}{Institute of Theoretical Astrophysics, University of Oslo, P.O. Box 1029 Blindern, N-0315 Oslo, Norway}
\altaffiltext{14}{Dunlap Institute, University of Toronto, 50 St. George St., Toronto, ON M5S 3H4}
\altaffiltext{15}{CePIA, Departamento de Astronom\'ia, Universidad de Concepci\'on, Chile}
\altaffiltext{16}{Jet Propulsion Laboratory, California Institute of Technology, 4800 Oak Grove Drive, Pasadena, CA, USA 91109}
\altaffiltext{17}{Physics Department, University of the Western Cape, Private Bag X17, Bellville 7535, South Africa}
\altaffiltext{18}{Jodrell Bank Centre for Astrophysics, Alan Turing Building, School of Physics and Astronomy, The University of Manchester, Oxford Road, Manchester M13 9PL, UK}
\altaffiltext{19}{Centre of Mathematics for Applications, University of Oslo, P.O. Box 1053 Blindern, N-0316 Oslo, Norway}
\altaffiltext{20}{Kavli Institute for Particle Astrophysics and Cosmology and Department of Physics, Stanford University, Varian Physics Building, 382 Via Pueblo Mall, Stanford, CA 94305, USA}
\altaffiltext{$\star$}{Corresponding author: \mbox{\textit{huffenbe@physics.fsu.edu}}}

\slugcomment{
Submitted to ApJ---This paper should be cited as ``QUIET Collaboration (2014)''
}

\begin{abstract}
We present polarization measurements of extragalactic radio sources observed during the Cosmic Microwave Background polarization survey of the Q/U Imaging Experiment (QUIET), operating at 43$\,$GHz (Q-band) and 95$\,$GHz (W-band).  We examine sources selected at 20$\,$GHz from the  public, $>$40$\,$mJy catalog of the Australia Telescope (AT20G) survey.  There are $\sim$480 such sources within  QUIET's four low-foreground survey patches, including the nearby radio galaxies Centaurus A and Pictor A.  The median error on our polarized flux density measurements is 30--40$\,$mJy per Stokes parameter.  At S/N~$> 3$ significance, we detect linear polarization for seven sources in Q-band and six in W-band; only $1.3 \pm 1.1$ detections per frequency band are expected by chance.  For sources without a detection of polarized emission, we find that half of the sources have polarization amplitudes below 90$\,$mJy (Q-band) and 106$\,$mJy (W-band), at 95\% confidence.  Finally, we compare our polarization measurements to intensity and polarization measurements of the same sources from the literature.  For the four sources with \textit{WMAP} and \textit{Planck} intensity measurements $>1$\, Jy, the polarization fraction are above 1\% in both QUIET bands.  At high significance, we compute polarization fractions as much as 10--20\% for some sources, but the effects of source variability may cut that level in half for contemporaneous comparisons.  Our results indicate that simple models---ones that scale a fixed polarization fraction with frequency---are inadequate to model the behavior of these sources and their contributions to polarization maps.
\end{abstract}

\subjectheadings{cosmic microwave background---galaxies: active---galaxies: individual (Cen A, Pict A)---methods: statistical---polarization }

\section{Introduction}
  
Polarized Cosmic Microwave Background (CMB) fluctuations are important for cosmological analysis because they carry information that is complementary to temperature fluctuations, and they can therefore tighten cosmological parameter constraints.  Furthermore, observations of the large-scale odd-parity ($B$-mode) polarization patterns predicted by inflation can constrain inflationary models and, consequently, the underlying GUT-scale physics\footnote{ Upward revisions in the assessment of the Galactic dust contamination \citep{2014arXiv1409.5738P} have tempered initial claims of CMB $B$-mode observations \citep[BICEP2 collaboration,][]{2014PhRvL.112x1101A}.}. On small scales, polarization fluctuations are essential to establish high signal-to-noise ratio measurements of CMB lensing caused by foreground structure \citep{2003PhRvD..67h3002O,2003PhRvD..68h3002H,2012JCAP...06..014S}.  
However, polarized emission---from Active Galactic Nuclei (AGNs) and other extragalactic sources---creates an additional source of fluctuations in the polarized microwave sky, leading to a small-scale systematic effect for CMB polarization experiments \citep[e.g.][]{1999NewA....4..481D,2005A&A...431..893D,2004MNRAS.349.1267T,2005MNRAS.360..935T,2011arXiv1102.2181T,2012AdAst2012E..52T}.  Polarized point sources will limit bispectrum non-Gaussianity parameter $f_{\rm NL}$ studies at lower CMB observation frequencies, $\nu < 100$\,GHz \citep{2013MNRAS.432..728C}.  Measurements of bright polarization sources help us to refine models of the point source emission, and account for the contribution of dimmer, unobserved sources.    Also, if we can identify bright and compact polarized sources, they make valuable calibrators for millimeter wave polarization surveys \citep[e.g.][]{2012A&A...541A.111A}.

In the following we make polarization measurements of extragalactic radio sources at 43 and 95$\,$GHz with data from the Q/U Imaging ExperimenT \citep[][]{2011ApJ...741..111Q,2012ApJ...760..145Q,2013ApJ...768....9B}.  This work is ancillary to QUIET's main aim to measure the CMB polarization, but useful because these measurements provide information about the cores and jets of active galaxies, and they provide a handle on the contamination by such sources to measurements of the CMB \citep{2002A&A...396..463M,2009ApJ...705..868L,2010MNRAS.401.1388J,2011MNRAS.413..132B}.

In unification models, active galaxies that look different are actually similar objects, containing central, supermassive black holes with accretion disks and jets of relativistic plasma \citep{1978Natur.276..768R,1984RvMP...56..255B,1993ARA&A..31..473A,1995PASP..107..803U}.  Their luminosities and spectral lines may differ because they are observed at different angles, have different black hole masses and angular momenta,  accrete at different rates, and possess different interstellar media.   Through synchrotron and inverse Compton emission, these objects radiate across the electromagnetic spectrum, including at millimeter wavelengths (30--300$\,$GHz).   At frequencies $\geq 1\,\textrm{GHz}$, they typically have polarization fractions of a few percent \citep{2014ApJ...787...99S}.  Because they are so luminous, we can watch the evolution of supermassive black holes and their host galaxies over cosmic time.

An advantage of millimeter observations of AGN polarization is that they avoid the undesirable effect of Faraday rotation.  The angle of Faraday rotation ($\beta$) is proportional to the square of the wavelength, $\beta = {\rm RM}\lambda^2$, and the typical rotation measures (${\rm RM}$)  in these objects are $ \sim 10^2$--$10^4$$\,$rad$\,$m$^{-2}$ \citep{2002ApJ...566L...9Z,2003ApJ...589..126Z,2004ApJ...612..749Z}.  Thus only at wavelengths shorter than $\lambda \sim 10^{-3}$$\,$m can we reduce Faraday rotation (and the subsequent depolarization in regions of inhomogeneous magnetic field) to levels that allow order-degree measurements of the intrinsic polarization direction, even for bright sources.  Also, at longer wavelengths, the jet's opacity shrouds its inner parts due to synchrotron self-absorption, while at millimeter wavelengths, measurements like these probe the magnetic field environment in the inner jet regions \citep{2012MNRAS.423..756P}.

However, as of today there are only a few polarization surveys of AGNs $>20$\,GHz, particularly at $<1$$\,$Jy, due to observational challenges.  Compared to radio telescopes, millimeter-wave telescopes and receivers are less sensitive and the sources are often dimmer.
Efforts include those by \citet{2010ApJS..189....1A}, who targeted 145 sources with the Institut de Radioastronomie Millim\'etrique (IRAM) 30$\,$m dish at 90$\,$GHz; \citet{2009ApJ...705..868L}, who identified 22 objects in \textit{Wilkinson Microwave Anisotropy Probe} (\textit{WMAP}) polarization data {\citep{2007ApJS..170..335P}}; and \citet{2011MNRAS.413..132B}, who made VLA measurements  at 8--43$\,$GHz of \textit{WMAP} catalog sources \citep{2009ApJS..180..283W}. 

Thus, much of the information about AGNs at these frequencies must come from CMB surveys themselves.  Although the main science goals of dedicated CMB experiments often target angular scales that are significantly larger than $1'$, they nonetheless make useful contributions to AGN science because they survey large areas of the sky, rather than aiming at known targets.  QUIET's unique dataset, one of the most sensitive to date at these frequencies, will at Q-band continue to be competitive even after the \textit{Planck} polarization data release.  QUIET's beams have Full Widths at Half Maximum (FWHMs) of $27'$ and $13'$, large compared to most AGNs, so for all but a few systems we measure an integrated flux density for the whole system.

Most AGNs, particularly blazars, are known to be variable, so it can be useful to make many short-duration observations with multiple bands simultaneously \citep[][]{2014arXiv1402.0717A}.  By contrast, the long duration campaigns for some CMB observation strategies may provide only season- or year-long average emission\footnote{For example, \textit{WMAP} scanned large areas of the sky rapidly and repeatedly, building up sensitivity gradually.  \textit{Planck}, by contrast, focused on one thin ring on the sky at a time, achieving the full survey sensitivity after a few minutes, but not returning to the same region of sky until the next survey, months later.}.  This includes our QUIET observations, as the Q- and W-band observations took place in subsequent years, 2008--2009 (8 months) and 2009--2010 (17 months). The corresponding maps integrate observations over whole seasons to achieve low noise over broad sky areas, which is important for CMB measurements.
Other on-going low-frequency CMB and CMB-foreground experiments, like C-BASS \citep[5$\,$GHz, $44'$ FWHM;][]{2011arXiv1111.2702H,2014MNRAS.438.2426K} and QUIJOTE \citep[10--40$\,$GHz, $17'$--$55'$ FWHM;][]{2012SPIE.8444E..2YR}, can also provide useful  AGN science.

Our method probes the polarization of objects with known positions.  Typically CMB surveys are less sensitive in polarization than total intensity, and the sources are often just a few percent polarized.  It is therefore useful to detect sources in total intensity, and also use those same measurements to determine the source positions.  Alternatively, one could take source positions from an external catalogs or different frequency bands, where a particular instrument may be more sensitive or the sources are brighter.  For AGN, radio surveys like NVSS and FIRST (and the VLA Sky Survey\footnote{\url{https://science.nrao.edu/science/surveys/vlass}} in the future) can identify sources down to sub-mJy flux levels, and have small positional uncertainty \citep{1998AJ....115.1693C,1995ApJ...450..559B,VLASSpre2014}.  The use of external source catalogs is particularly important for QUIET, because the QUIET detector technology is specifically optimized for dedicated polarization measurements, and deep QUIET temperature maps are not readily available. 

In this work, we adopt the catalog from the Australia Telescope 20$\,$GHz (AT20G) Survey of the Southern sky \citep{2008MNRAS.384..775M,2010MNRAS.402.2403M}, and we measure the polarized flux density at the specified locations in the QUIET sky maps, carefully accounting for the errors caused by instrumental noise and background CMB fluctuations.  We then compare the measured fluxes in the two QUIET bands to intensity and polarization measurements of the same sources in the literature.

The rest of the paper is organized as follows. First, in Section~\ref{sec:data} we describe the QUIET observations and the AT20G source sample.  In Section~\ref{sec:method}, we outline the method for measuring the polarized source flux density from our QUIET maps, before reporting our results in Section~\ref{sec:results}.  Finally, in Section~\ref{sec:conclusions} we present our conclusions.

Throughout we refer to Stokes parameters and angles defined in Galactic coordinates, using the CMB/HEALPix convention\footnote{IAU convention differs by the sign of $U$, so that $S_{U,\rm IAU} = -S_U$ as we have written it.} \citep{2005ApJ...622..759G,1997PhRvD..55.1830Z}.  This facilitates direct comparison to \textit{Planck} and \textit{WMAP} CMB maps at the same frequencies, which are sensitive to sources down to $\sim 1$\,Jy.  We use $Q$ and $U$ to refer to Stokes parameter values in the CMB map in thermodynamic temperature.  Flux density is denoted by $S_Q$ and $S_U$.  For the polarization amplitude we similarly use $P$ and $S_P$, and for the polarization angle we use $\alpha$.

The Stokes parameters and linear polarized flux density are related  by 
\begin{eqnarray} 
S_P &=& (S_Q^2 + S_U^2)^{1/2} \\ \nonumber
\alpha &=& \frac{1}{2}\arctan(S_U/S_Q) \\\nonumber
S_Q &=& S_P \cos(2\alpha) \\ \nonumber
S_U &=& S_P \sin(2\alpha) 
\end{eqnarray}
where $\alpha$ is the polarization angle between the electric field and a fiducial axis, and has range $0 < \alpha < \pi$.  We note that the pairs of quantities $(S_Q,S_U)$ and $(S_P,2\alpha)$ are equivalent to Cartesian and polar coordinates in a plane\footnote{In optics, the mid-plane of the Poincar\'e sphere.}, which aids us in later computations.  Source positions are in the J2000 epoch.

\section{Data}  \label{sec:data}
\subsection{QUIET CMB maps} \label{sec:quiet_data}

The QUIET instrument \citep{2013ApJ...768....9B} consisted of a 1.4$\,$m side-fed Dragonian telescope coupled to an array of correlation polarimeters.  These used High Electron Mobility Transistor (HEMT) amplifiers cooled to 20 K. Situated on the Chajnantor plateau in Chile's Atacama region, QUIET made polarized microwave measurements of six patches of the sky; two in the Galactic plane and four in low-foreground regions.  The latter yielded polarization maps and CMB power spectra at 43$\,$GHz \citep[Q-band,][]{2011ApJ...741..111Q} and 95$\,$GHz \citep[W-band,][]{2012ApJ...760..145Q}.  

The Q-band receiver had 17 polarization-sensitive assemblies with a total sensitivity of 69 $\mu$Ks$^{1/2}$, and was mounted on the telescope during the first observing season beginning in 2008 October.  The W-band receiver had 84 polarization-sensitive assemblies and a sensitivity of 87 $\mu$Ks$^{1/2}$, and was mounted on the telescope during 2009 July.  The second observing season proceeded until 2010 December.  The instrument produced more than 10$\,$000 hours of data between the two bands. For point sources in the low-foreground patches, this yields a median $1\sigma$ sensitivity per Stokes parameter of 32$\,$mJy for Q-band and 39$\,$mJy for W-band.  Over the same area of sky and in the same bands, this compares favorably to the polarization errors for \textit{WMAP}, $> 70$\,mJy, and the upcoming release of Planck polarization data, for which we estimate errors $> 50$\,mJy based on the 2013 catalog (which contained intensity only).

Bandpasses for QUIET were determined with a signal generator and standard gain horn, injecting known signals  into the front window of the cryostat.  Effective frequencies in polarization were found to be 43.0$\,$GHz and 94.4$\,$GHz for Q- and W-bands respectively, with 7.6$\,$GHz and 10.7$\,$GHz bandwidths, for a CMB blackbody spectrum \citep[Tables 6 and 11,][]{2013ApJ...768....9B}.  Below we adopted the CMB values for our computation of flux density.  Note that for AGN-type spectra\footnote{The typical AGN-type spectrum at microwave frequencies is nearly flat, with $S\propto \nu^{\alpha_{\rm SED}}$ and $\alpha_{\rm SED} \sim 0$.
} the effective frequencies are lower by $\sim$0.2 and 0.1$\,$GHz, respectively, but at our relatively low signal-to-noise ratio, this difference is not important.

For both arrays, we obtained beam profiles derived from observations of Jupiter and Tau A, the brightest unpolarized and polarized sources on the sky that are compact compared to the QUIET beams.  Observations with the polarimeter assemblies of the fainter polarization signal from Tau A were found to be consistent with the Jupiter profiles from the temperature assemblies, after accounting for bandpasses, source spectra, and horn positions within the focal plane \citep{2013ApJ...768....9B}.  

The impact of the beam on the map is nearly axisymmetric, a product of the intrinsic roundness of the optical beams and the sky- and active boresight-rotations.  Including the effect of scan-to-scan pointing jitter, the FWHM of the effective beam for Q-band is $27.3'$; for W-band it is $12.8'$. We compute the beam window function with a Legendre transform of a one-dimensional Hermite expansion of the symmetrized beam \citep{2010SPIE.7741E..75M}.  A pixel window function captures the integration over the finite-sized pixels.  This is folded into our expected profile for a point-like source.  Any error in pointing reduces the measured source flux density.  The uncertainty on the absolute pointing calibration is $3.5'$ for Q-band; in simulations, such an error in pointing decreases the measured flux from a source by about 2.1\%, much smaller than our other errors.  For W-band, the uncertainty is $2'$, but the beam is smaller, and this position error decreases the measured flux by 11\% in simulations.  For our bright sources ($\sim 100$\,mJy), this is less than a third of the typical statistical error, but for most sources it is much less.



We characterized the instrumental polarization with Jupiter, decomposing the polarization leakage maps into Gauss-Hermite moments \citep{2010SPIE.7741E..75M}.  The largest moment was the Q-band temperature to polarization monopole, for which the module-median leakage was $\sim$1\%. The W-band median leakage was 0.25\%, and typical values for the higher moments ranged between 0.2 and 0.4\%.  Sky rotation and instrumental rotation around the boresight help to mitigate this leakage.  For instance, in the W-band power spectrum analysis \citep{2012ApJ...760..145Q}, these effects yielded a factor of $\sim$40 improvement in the systematic $BB$ contamination (assuming zero input $B$-modes signal) and a factor of $\sim$10 improvement for $EE$\footnote{A smaller improvement is seen for $EE$ than $BB$, because intensity-to-polarization leakage tends to primarily induce $E$-mode residuals.}.  For Q-band, with less variety in rotation about the telescope boresight, the mitigation degrades by roughly a factor 1.3 in the power spectrum.

To control systematic errors, the QUIET analysis used a pair of map-making and power spectrum estimation pipelines. This paper uses products from the maximum likelihood pipeline (pipeline ``B'' in earlier papers), which projects the time-ordered data into the map domain while accounting for its noise covariance, the telescope pointing model, and a bandpass filter that suppressed both low frequency noise and  scan-synchronous contamination.  In addition to the actual map, the maximum-likelihood method also produces a full pixel--pixel noise covariance, which forms the noise model adopted below. From these maps and covariance matrices, we excise miniature maps containing only those pixels within a radius of $60'$ (Q-band) or $30'$ (W-band) of the source catalog positions, depending on the band's beam size, and we retain only the pixel noise covariance for these small submaps. Finally, because of the high-pass filters applied in the original QUIET analysis, the very largest scales in these maps are associated with significant uncertainties, and we therefore marginalize over a baseline offset for each source, adopting independent offset planes in the Stokes $Q$ and $U$ parameters.  

Table \ref{tab:summary} summarizes our observations for point sources, which we detail below.
\begin{table}
\caption{Summary of QUIET source observations}
\begin{center}
\begin{tabular}{ccccc}
\hline
Band & $\nu$ & FWHM  & Pol.~sensitivity & Count \\ 
 & (GHz) &  (arcmin) & (mJy) & (S/N $>3$) \\ 
\hline
Q & 43.0 & 27.3 & 32 mJy & 7 \\
W & 94.4 & 12.8 & 39 mJy & 6 \\
\hline
\end{tabular}
\end{center}
\tablecomments{For each band, the effective frequency, beam full-width at half-maximum, median polarization sensitivity to point sources with known position (per Stokes parameter), and the number of compact sources in which we detect polarized emission at S/N~$ > 3$.}
\label{tab:summary}
\end{table}

\subsection{AT20G source sample} \label{sec:at20g_data}

The Australia Telescope 20$\,$GHz Survey \citep[AT20G,][]{2008MNRAS.384..775M,2010MNRAS.402.2403M,2011MNRAS.412..318M} covered sources across the entire Southern Hemisphere. The source catalog includes 5890 sources brighter than 40mJy at 20$\,$GHz.  All sources in the catalog have S/N~$\geq 8.0$, and the median S/N~$= 19.3$. Each source was measured simultaneously in intensity and polarization.  Many sources have also near-simultaneous measurements at 5 and 8$\,$GHz, and polarization was detected for 1559 sources in at least one band. Some of the brightest sources were re-observed with 1\,mJy polarization sensitivity at 20 GHz \citep{2013MNRAS.436.2915M}.  Sky areas within $12^\circ$ of the nominal centers of the low-foreground QUIET patches contain 531 sources, 86 of which have a 20$\,$GHz polarization measurement (Table \ref{tab:patch_sources}).  
\begin{table}
\caption{AT20G counts near QUIET patches}\label{tab:patch_sources}
\begin{center}
\begin{tabular}{ccccc}
\hline
Patch& RA & dec. & $N$ sources & $N$ (20$\,$GHz pol.)\\
\hline
CMB-1 
&12:04:00&-39:00:00&108&12\\
CMB-2 
&05:12:00&-39:00:00&143&26\\
CMB-3 
&00:48:00&-48:00:00&130&18\\
CMB-4 
&22:44:00&-36:00:00&95&11\\
\hline 
& & \textbf{Totals}   & 476 & 67 \\
\hline
\end{tabular}
\end{center}
\tablecomments{QUIET low foreground patch centers, the number of sources in the AT20G catalog within 12$^\circ$ of the center (after cuts for QUIET Q-band coverage), and the number of sources with measured 20$\,$GHz polarization. 
}
\end{table}
Because our cut-out maps may intersect the edge of our survey, we additionally require that each source map must contain a minimum number of pixels ($7'$ size), namely 120 for Q-band and 30 for W-band.  This costs us minimal statistical power, because the survey sensitivity declines gradually at the patch edges, and those sources are only poorly measured.  This cut reduces the number of candidate sources to 476 for Q-band and 480 for W-band. For sources that have AT20G-measured polarization at 20$\,$GHz, 67 fall in our Q-band map and 71 fall in our W-band map, due to slight differences in sky coverage.

About 23\% of the sources have a nearby source located within $30'$, so to note possible contamination, these sources are flagged in our summary statistics.  Since AT20G is 91\% complete above 100$\,$mJy in total intensity \citep{2010MNRAS.402.2403M}, we expect that they have captured nearly everything brighter than a few mJy in polarization.  QUIET's $1\sigma$ sensitivity is typically 30--40$\,$mJy per Stokes parameter, so we have not considered source confusion noise beyond this flagging; especially after averaging over polarization directions, it will be small compared to our uncertainty.

\section{Method} \label{sec:method}


\subsection{Estimation of Stokes parameters} \label{sec:estStokes}
For a point source of radiation, we model our data as
\begin{equation}
  \mathbf{d} = \mathbf{F} \mathbf{s} + \mathbf{n},
\end{equation}
where the vector $\mathbf{d}$ contains the measured polarization values for pixels near the source, the matrix $\mathbf{F}$ is the two-dimensional  template for a point source in those pixels, and the vector $\mathbf{s}$ contains the polarized flux density and parameters for the template.  The vector $\mathbf{n}$ denotes instrumental and background noise. The minimum-variance unbiased source flux density estimate is
\begin{equation}
  \mathbf{\tilde s} = (\mathbf{F}^T\mathbf{N}^{-1}\mathbf{F})^{-1}  \mathbf{F}^T   \mathbf{N}^{-1}\mathbf{d},
\end{equation}
where $\mathbf{N}$ is the instrumental noise covariance.

The template for each source is built from an axisymmetric source profile, sampled at the distances from the source catalog position to the map pixel centers \citep{2012MNRAS.424.3028S}.  The source profile combines the effects of the beam and pixel window function, which are combined in harmonic space, then converted to real space with a Legendre transform.  We convert from flux density to temperature units in the source template, using the effective band frequency.  As a rule of thumb, a 100$\,$mJy source creates a $24\,\mu$K peak signal in Q-band, and a 29 $\mu$K peak signal in W-band.  As mentioned above, to reduce our sensitivity to large scale modes, we additionally fit and marginalize over a constant planar offset, independent for each Stokes parameter, for every source.  Thus our model for the vector $\mathbf{s}$ includes $(S_Q, S_U, A_Q, A_U)^T$, where $(S_Q,S_U)$ are the linear polarized flux densities and $(A_Q,A_U)$ are the constant offsets.

For point sources, the effective noise matrix $\mathbf{N}$ includes contributions from two terms, instrumental noise and CMB fluctuations.  The instrumental contribution dominates, and has a standard deviation of $\sim$20$\,\mu$K for a 7' Q-band pixel near the center of the map; for a W-band pixel it is typically lower by a factor of {$\sim 0.6$}. Due to the QUIET scanning strategy, the covariance is nearly diagonal, although slightly anticorrelated for adjacent pixels. The second, and smaller, contribution to the covariance is from the CMB, an unavoidable background for these sources.  This contribution is evaluated from the CMB power spectrum \citep{1997PhRvD..55.7368K} based on the \textit{WMAP}7 best-fit spectrum \citep{2011ApJS..192...16L}, accounting for beam window and pixel window functions. The resulting covariance is dense, and has an RMS of $\sim$1$\mu\textrm{K}$ for Q-band and  $\sim$2$\mu\textrm{K}$ for W-band; the different amplitudes is due to the different beams. Because the CMB is highly correlated, the Pearson's correlation coefficient is $r \sim 0.8$ or higher for adjacent pixels. Including the CMB fluctuation in the covariance matrix accounts for CMB polarization modes which could otherwise masquerade as source flux, and potentially bias the source amplitude.

\begin{figure}
\includegraphics[width=\columnwidth]{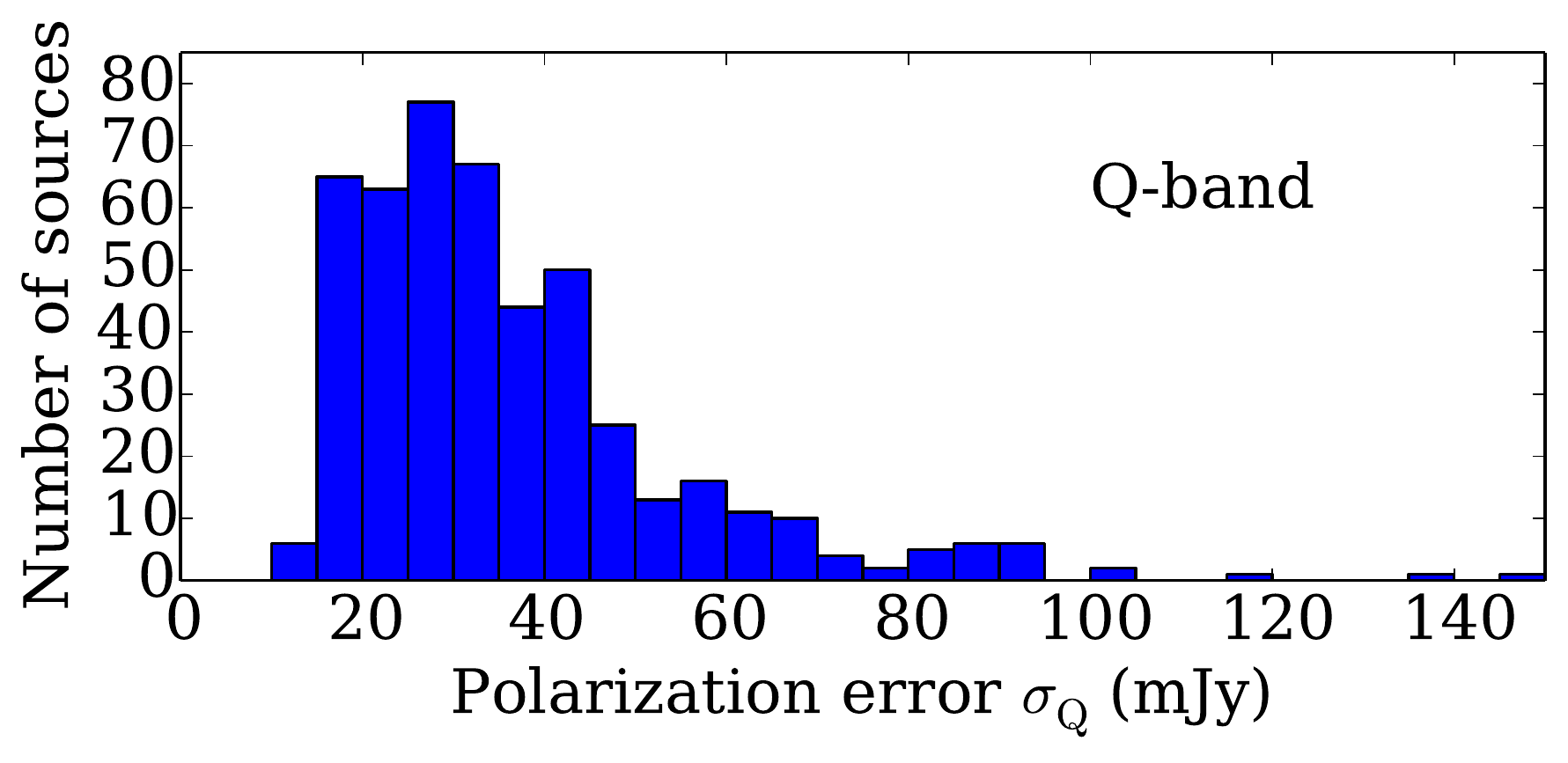}
\includegraphics[width=\columnwidth]{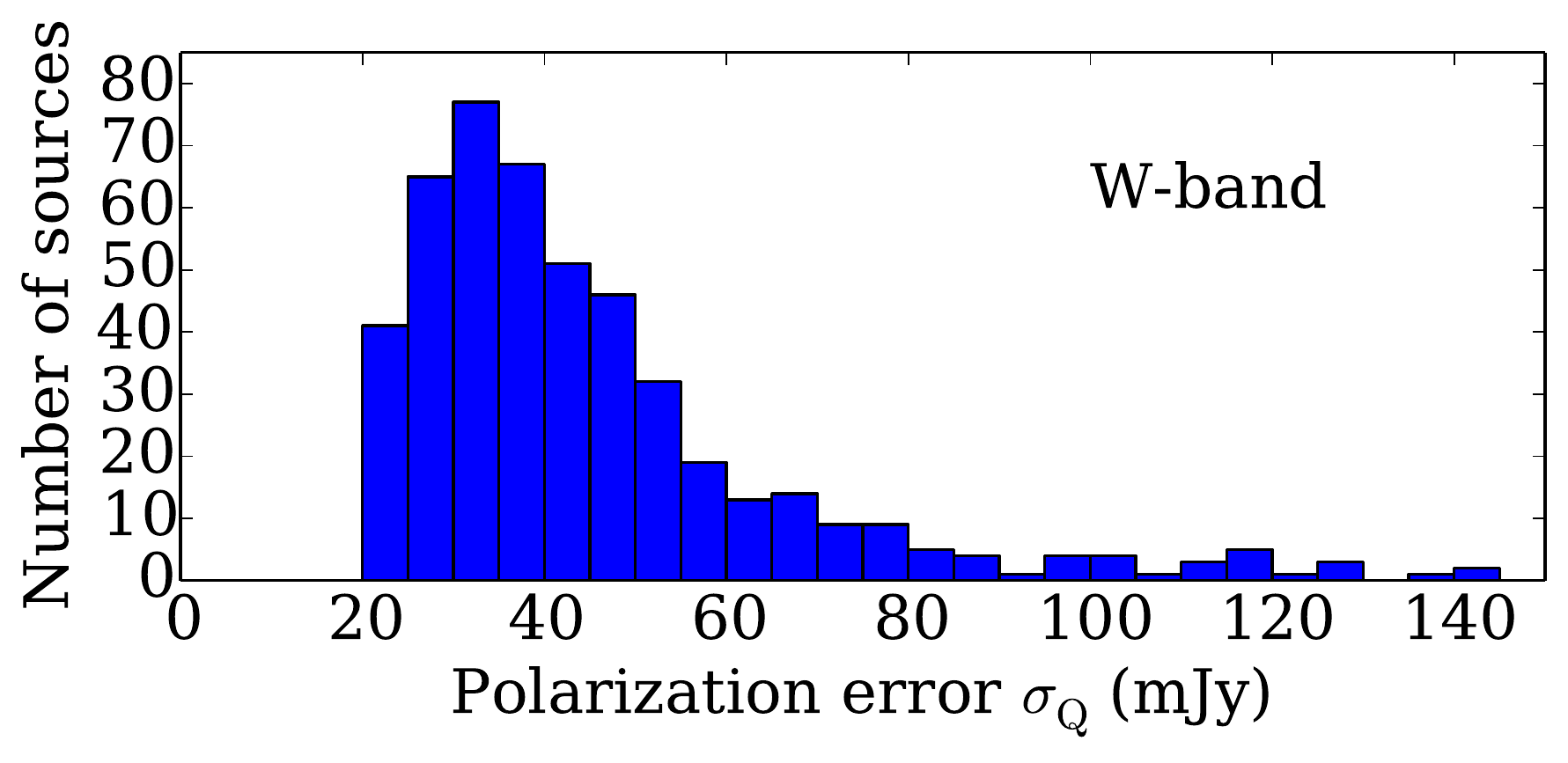}
\caption{Uncertainty distribution for the Stokes parameter $S_Q$, as defined by the square root of the $QQ$-component of the covariance. Distributions for $\sigma_U$ are similar. 
}\label{fig:errors}
\end{figure}

The polarization estimate $\mathbf{\tilde s}$ has covariance matrix 
\begin{equation}
  \mathbf{C} = (\mathbf{F}^T\mathbf{N}^{-1}\mathbf{F})^{-1}.
\end{equation}
For each source this is a symmetric $4\times4$ matrix, listing every combination of Stokes parameters and offsets.  Marginalizing over the offsets amounts to retaining the four entries for the Stokes parameter combinations $QQ$, $QU$, $UQ$, and $UU$ to build the $2\times2$ marginal covariance $\mathbf{C}_m$.  This matrix is nearly diagonal, and the marginal uncertainty for most sources in our sample is less than 100$\,$mJy (see Figure \ref{fig:errors}), with a median uncertainty per Stokes parameter of $32$$\,$mJy for Q-band and $ 39$$\,$mJy for W-band\footnote{{For W-band the noise in temperature units is typically lower than in Q-band by a factor $\sim 0.6$, while the smaller beam concentrates the source signal by a factor of the solid angle ratio $\sim 4$.  However, the number of pixels that meaningfully contribute to the estimate is also smaller by the same factor, which increases the error by a factor $\sim \sqrt{4} = 2$.  The blackbody conversion factor $dB/dT$ to flux density units from temperature is roughly a factor of 5 larger for W-band.  So the crude expectation is that W-band errors will be  $0.6 / 4 \times 2 \times 5 = 1.5$ times larger than Q-band errors, while we observe a factor $\sim 1.4$.}}.  We may expect such a wide range in errors because the map sensitivity drops near the edge of each patch.
By comparison, in the \textit{WMAP} 9-year point source catalog, the median error on total intensity is 50$\,$mJy in Q-band and 170$\,$mJy in W-band \citep{2009ApJS..180..283W}, and polarization errors would be bigger by a factor $\sqrt{2}$.  For the \textit{Planck} catalog of compact sources \citep{2013arXiv1303.5088P}, the median intensity error is 110$\,$mJy at 44$\,$GHz and 60$\,$mJy at 100$\,$GHz after two sky surveys.   Errors  will improve from the inclusion of three more surveys in the upcoming releases, but for polarization they will also be degraded by a factor $\sqrt{2}$.  Thus QUIET's uncertainty per source is competitive with both \textit{WMAP} and \textit{Planck}; however, they have the advantage that they cover 40 times more sky area.

For every source we compute 
\begin{equation}
\chi^2 =  (S_Q \  S_U) \mathbf{C}^{-1}_m (S_Q \  S_U)^T,
\end{equation}
 based on the marginalized covariance and the null hypothesis that there is no polarized flux.  We then compute the probability-to-exceed (PTE) for the resulting $\chi^2$ due to chance alone, $  P(>\chi^2)$,  where the two Stokes parameters are the two degrees of freedom.  Low probability indicates significant flux density.  We also compute the signal-to-noise ratio in terms of the equivalent significance in standard deviations for a Gaussian distribution with the same probability,
\begin{equation}
    P(>\chi^2) = 1 - {\rm erf} \left( \frac{\mbox{S/N}}{\sqrt{2}} \right).
\label{eq:significance}
\end{equation}
where $\rm erf$ is the error function.
Whenever we discuss the significance of polarized emission, we compute it from the measured Stokes parameters using Gaussian/$\chi^2$ statistics defined in this way.

\subsection{Estimation of polarized amplitude}

The amplitude of polarization is positive-definite, requiring non-Gaussian statistics especially when signal-to-noise ratio is low.
Our estimates of the Stokes parameters $S_Q$ and $S_U$ are unbiased, but computing the polarized flux naively as $S_P = (S_Q^2 + S_U^2)^{1/2}$ is biased by the noise.  This bias is treated elsewhere in the literature: \citet{1985A&A...142..100S} account for the biases when Stokes parameters are uncorrelated and have the same errors, using properties of the Rice distribution; \citet{2014MNRAS.tmp..429P} develop an analytic, approximate distribution for the general case of correlated errors; \citet{2014arXiv1410.4436V} treat a case with an already-known polarization angle; and \citet{2014arXiv1407.0178M} compare several estimation methods.  Our Stokes parameters are approximately uncorrelated, but we develop a simple alternative approach which retains the full correlated information via a Monte Carlo method.  

For each source, we seek the posterior probability (${\cal P}$) for the true Stokes parameters ($S_Q, S_U$), the polarized flux ($S_p$), and polarization angle ($\alpha$), all conditioned on our observed Stokes parameters.  The covariance we computed gives the likelihood (${\cal L}$) of an observation based on the true value.  Using Bayes theorem and a uniform prior, we write the posterior as
\begin{equation}
{\cal P}(S_Q,S_U|S_Q^{\rm obs},S_U^{\rm obs}) \propto {\cal L}(S_Q^{\rm obs},S_U^{\rm obs}|S_Q,S_U),
\end{equation}
which is a 2-d Gaussian distribution with mean and covariance given by the measurement.
To compute the distribution of $S_P$ (and $\alpha$), we generate samples of the Stokes parameters from this Gaussian distribution.  
We then multiply with the Jacobian to transform the sampled distribution while conserving probability,
\begin{equation}{\cal P}(S_Q, S_U)= {\cal P}(S_P, 2\alpha) \left| \frac{\partial (S_Q, S_U)}{\partial (S_P, 2\alpha)} \right| =  {\cal P}(S_P, 2\alpha) S_p.
\end{equation}
Finally, we marginalize the right-hand distribution to produce the 1-d posterior for the polarized flux ${\cal P}(S_P|S_Q^{\rm obs},S_U^{\rm obs})$.  This simply amounts to binning the samples, dividing the bin value by $S_P$, then normalizing.  We construct the distribution for the angle in a similar way.

\begin{figure}
\includegraphics[width=\columnwidth]{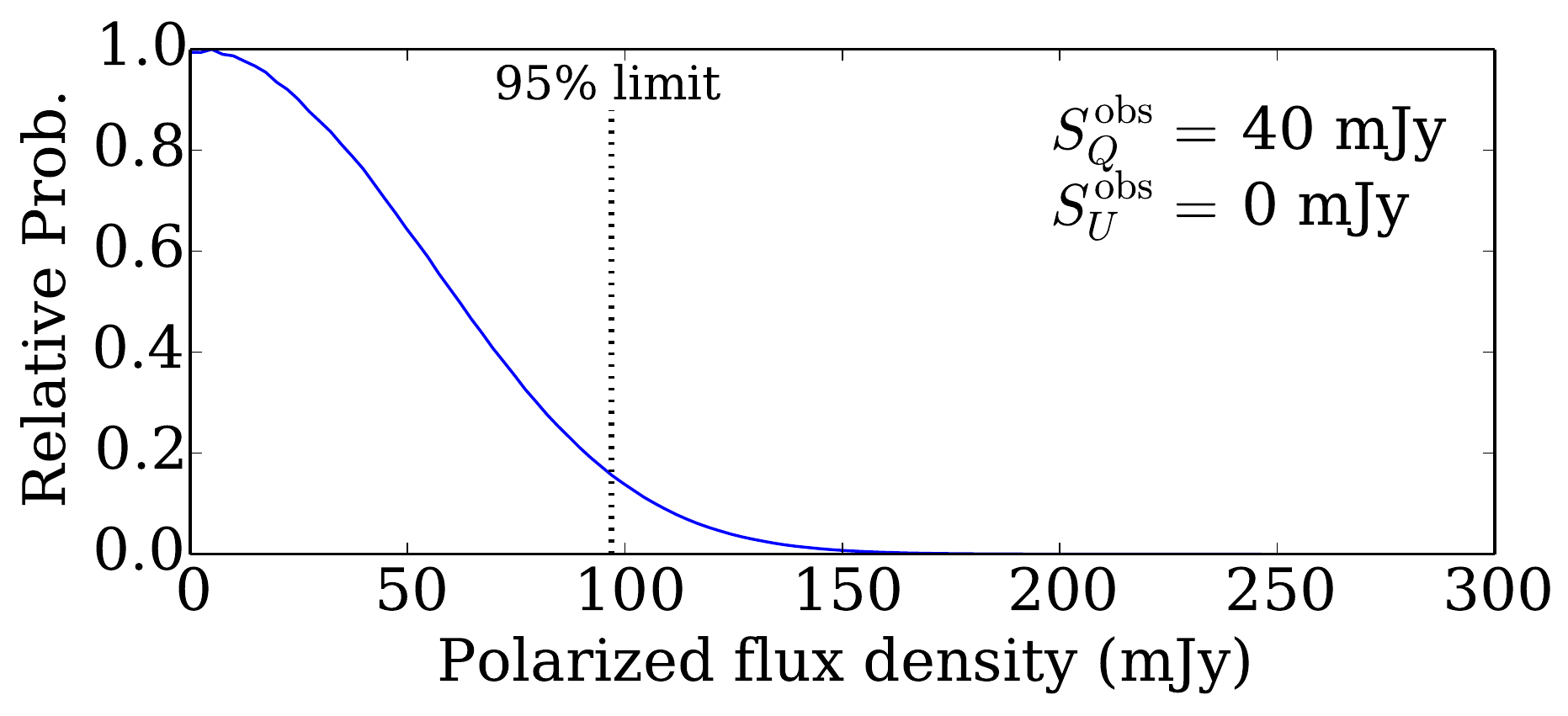}
\includegraphics[width=\columnwidth]{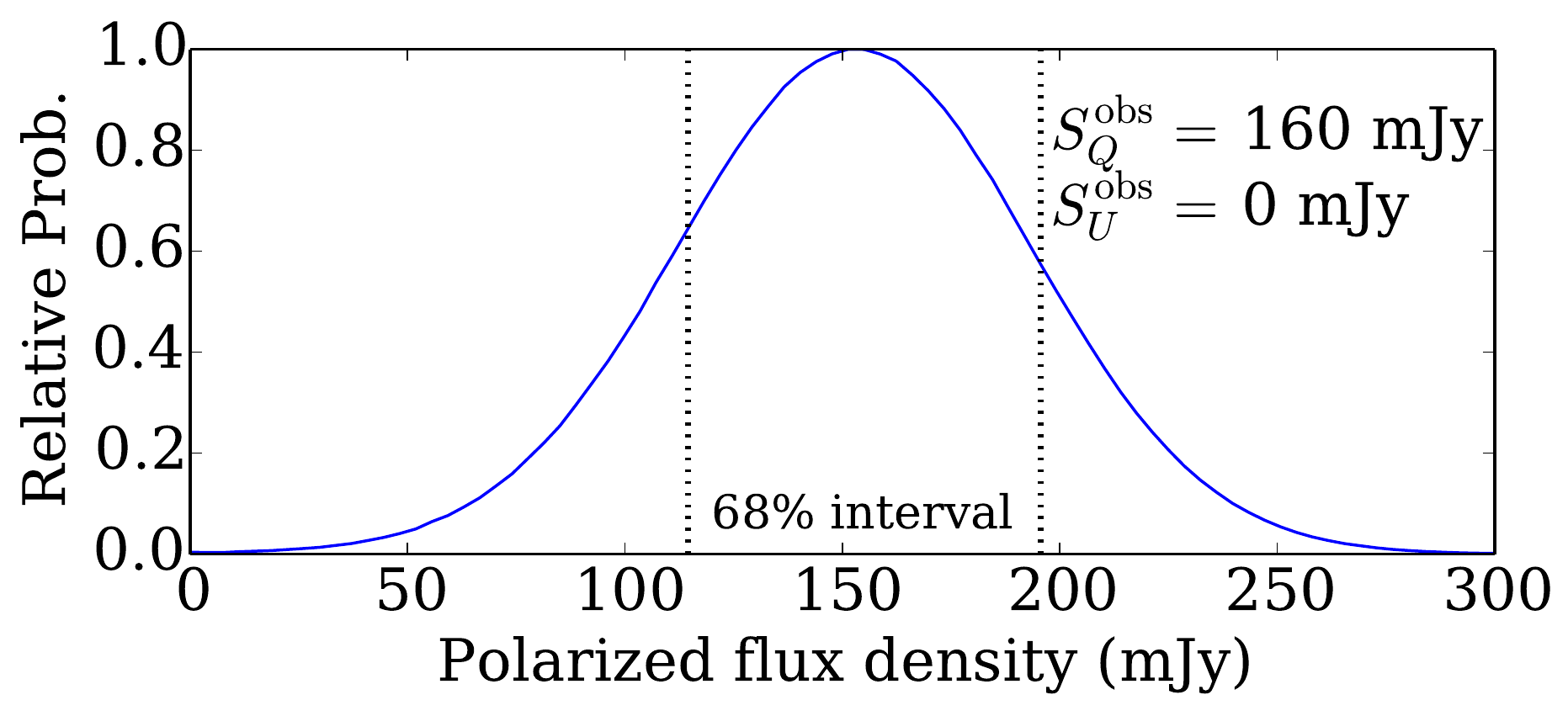}
\caption{Example posteriors for polarized flux density measurements.  Both have diagonal covariances with $\sigma_Q = \sigma_U = 40\,$mJy.  Top: For a low S/N ratio, in this case a 40$\,$mJy observation, the maximum probability point for the true polarization is near zero, and we accordingly quote only a 95\% upper limit.  Bottom: For a higher S/N ratio, such as a 160$\,$mJy observation, the maximum probability for the true polarization occurs near the observed value.  In cases like these we quote the 68\% interval.}\label{fig:Pml+errors}
\end{figure}

We use $10^7$ samples for each source, which probes the shape of the likelihood  sufficiently in only a few seconds per source.  In Figure~\ref{fig:Pml+errors} we show that this method reproduces the Rice distribution behavior from \citet[Figure~2]{1985A&A...142..100S}.  As there, for sources with measured polarized emission less than $\sim$1.41 times the error per Stokes parameter, the maximum likelihood value for the true polarization is zero, and the measured polarization is likely the result of noise alone.  For sources with larger measured polarization, the maximum likelihood value exceeds zero, and at high signal-to-noise ratio, the maximum likelihood estimate approaches the true value.  

When the signal-to-noise ratio for a given source is high, the likelihood peaks well away from zero polarized flux.  Like \citet{2006PASP..118.1340V}, we base confidence intervals for these cases on the integrated probability, and report the median 68\% interval (from the 16th to 84th percentile) of the posterior, which contains the maximum posterior point. When the signal-to-noise ratio for a source is low, the polarization likelihood peaks near zero polarized flux, and the maximum posterior point is typically closer to zero than the median 68\% interval.  In this paper, we report a 95\% upper limit for any source with S/N$<2$.

\section{Results} \label{sec:results}

The two brightest objects in our fields, Centaurus A and Pictor A, have structures that are extended compared to our beams.  We report results for these first, followed by the larger population of fainter objects, which we treat as point sources.

\subsection{Extended sources}

\def\imagetop#1{\vtop{\null\hbox{#1}}}
\newlength{\CenA}
\setlength{\CenA}{0.41\columnwidth}

\begin{figure*}
\begin{center}
\begin{tabular}{ccccc}
& \quad QUIET Stokes $Q$ & QUIET Stokes $U$ & \qquad QUIET Polarization & \textit{WMAP} $T$\\
\imagetop{\rotatebox{90}{Q-band\hspace{0.15\columnwidth}}}&
\imagetop{\includegraphics[height=\CenA]{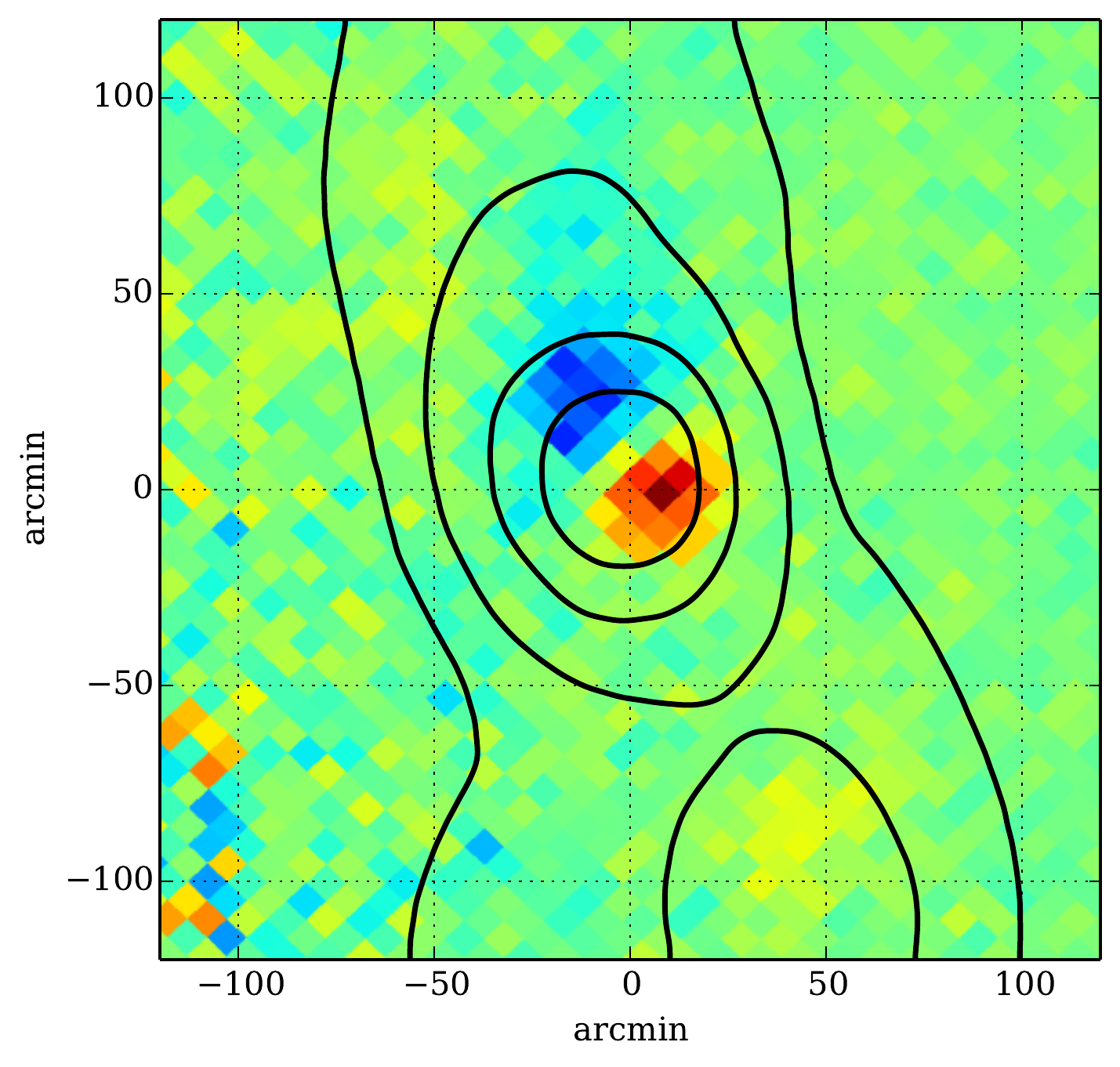}}&
\imagetop{\includegraphics[height=\CenA]{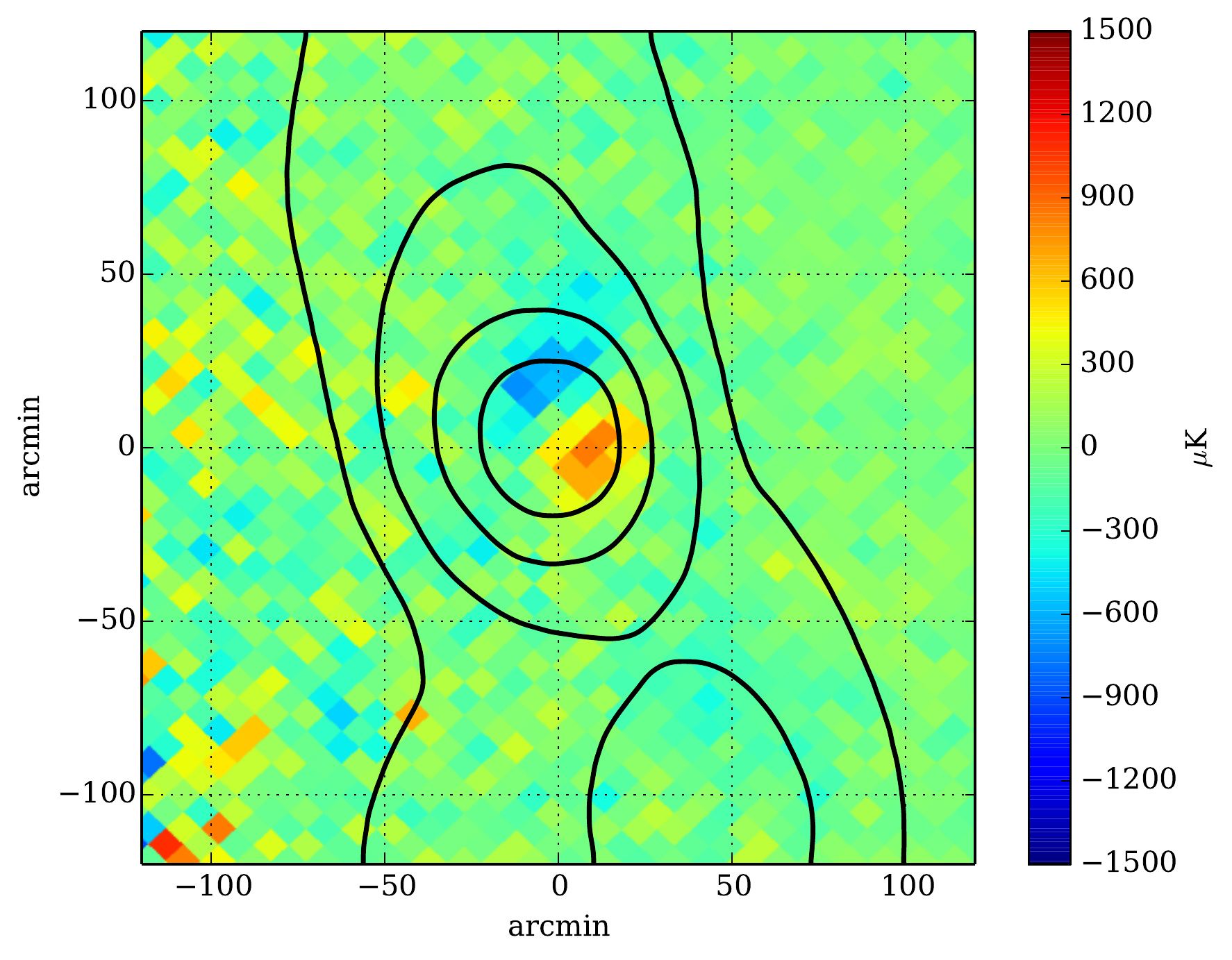}}&
\imagetop{\includegraphics[height=\CenA]{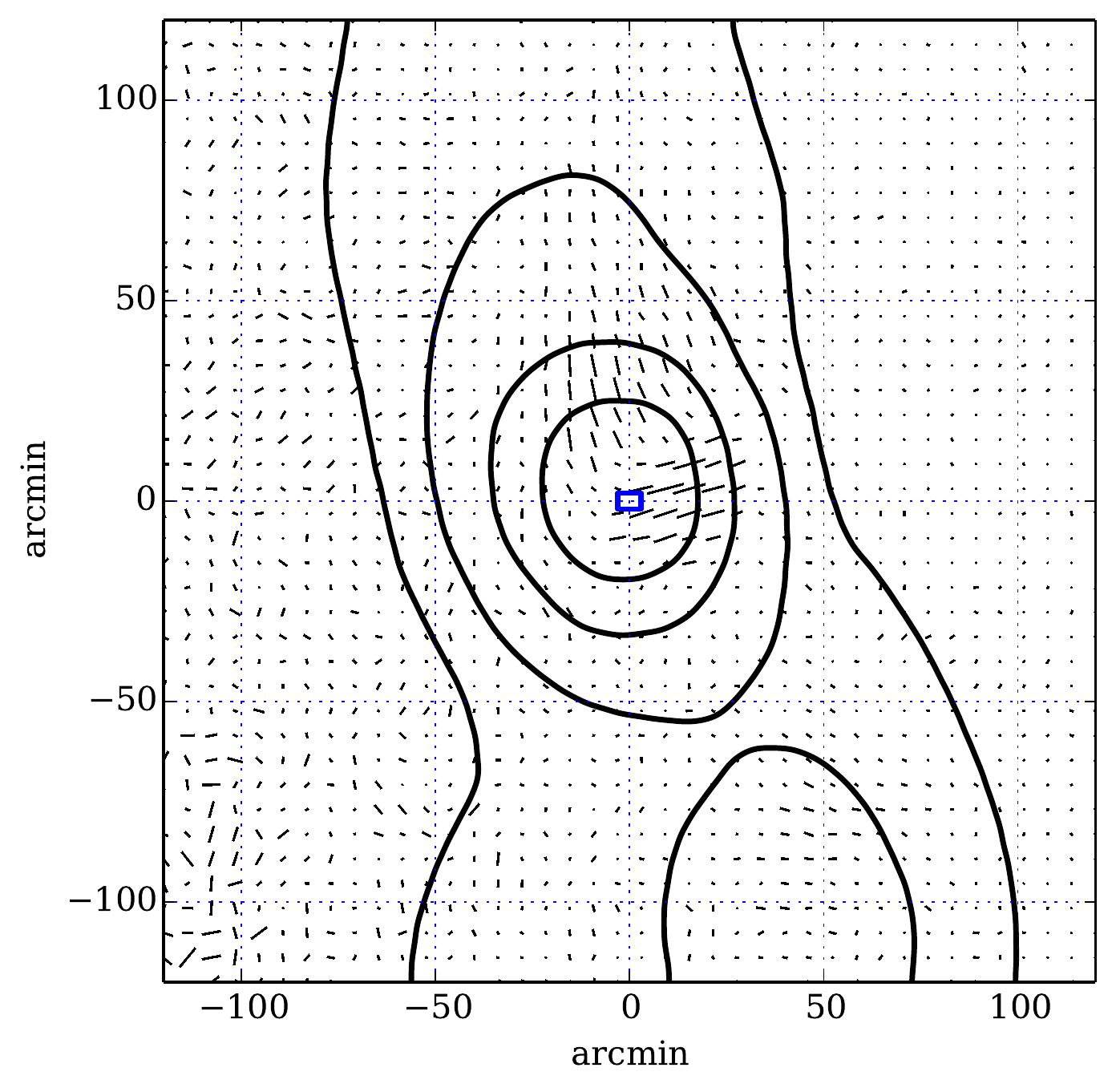}}&
\imagetop{\includegraphics[height=\CenA]{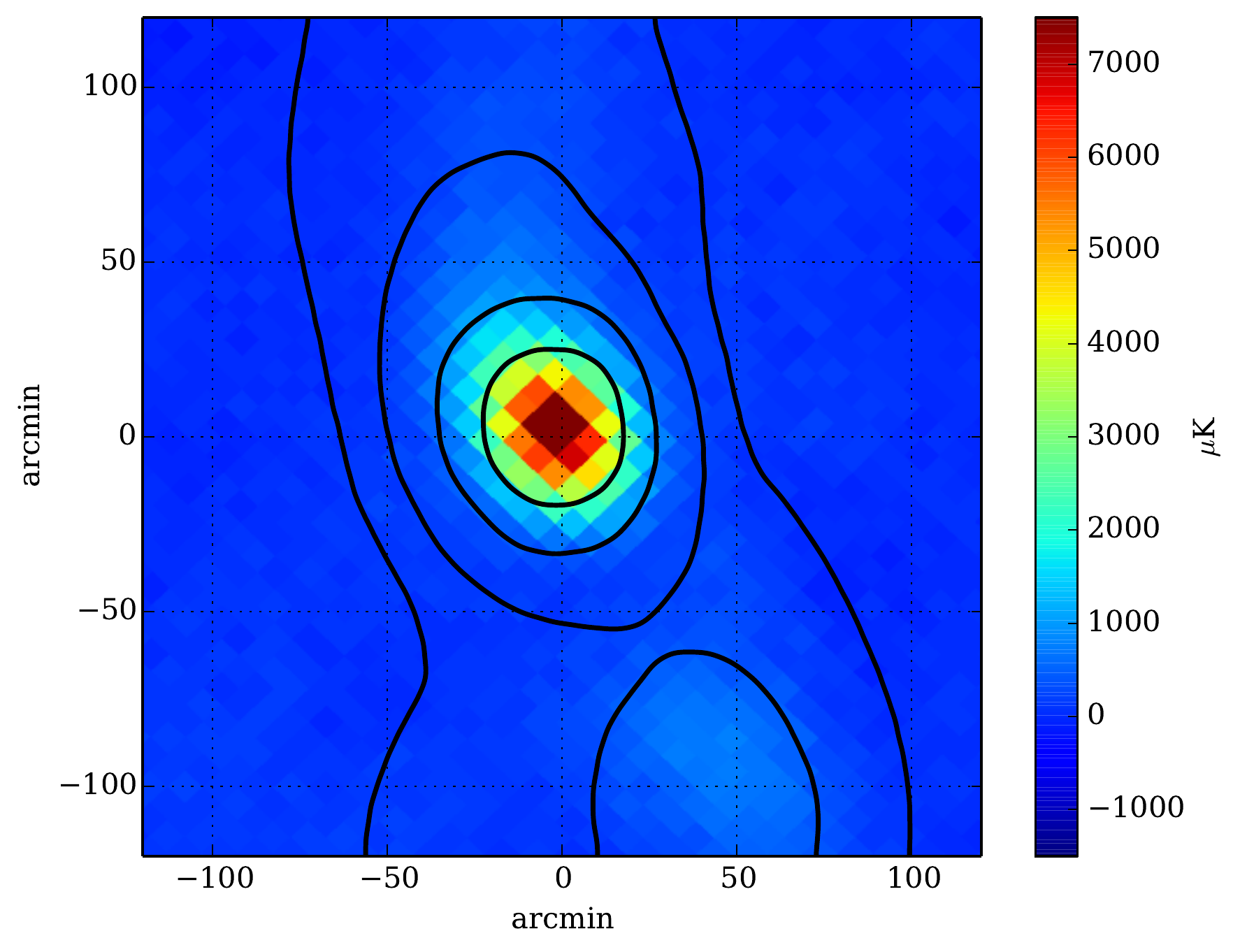}}\\
\imagetop{\rotatebox{90}{W-band\hspace{0.15\columnwidth}}}&
\imagetop{\includegraphics[height=\CenA]{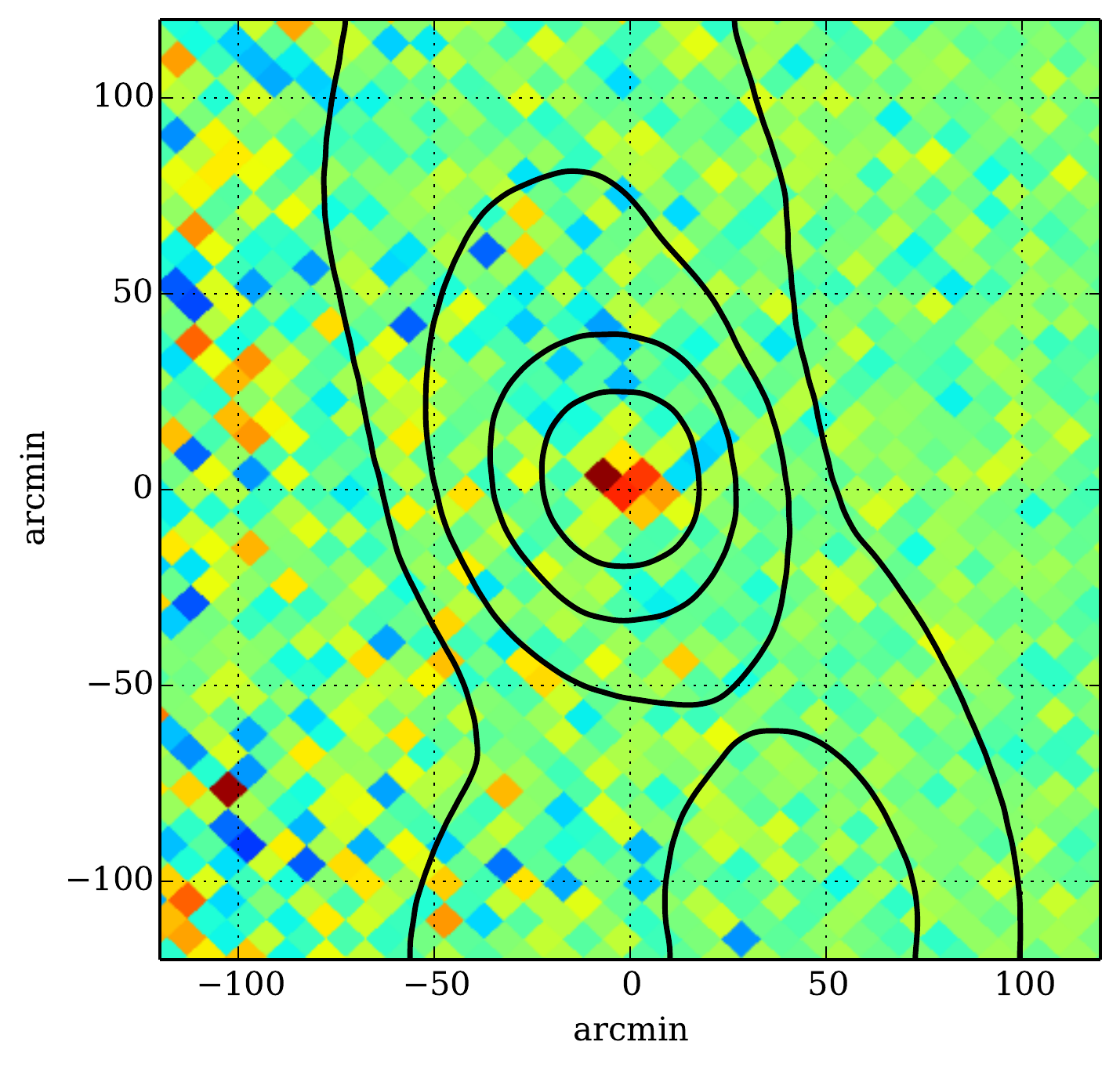}}&
\imagetop{\includegraphics[height=\CenA]{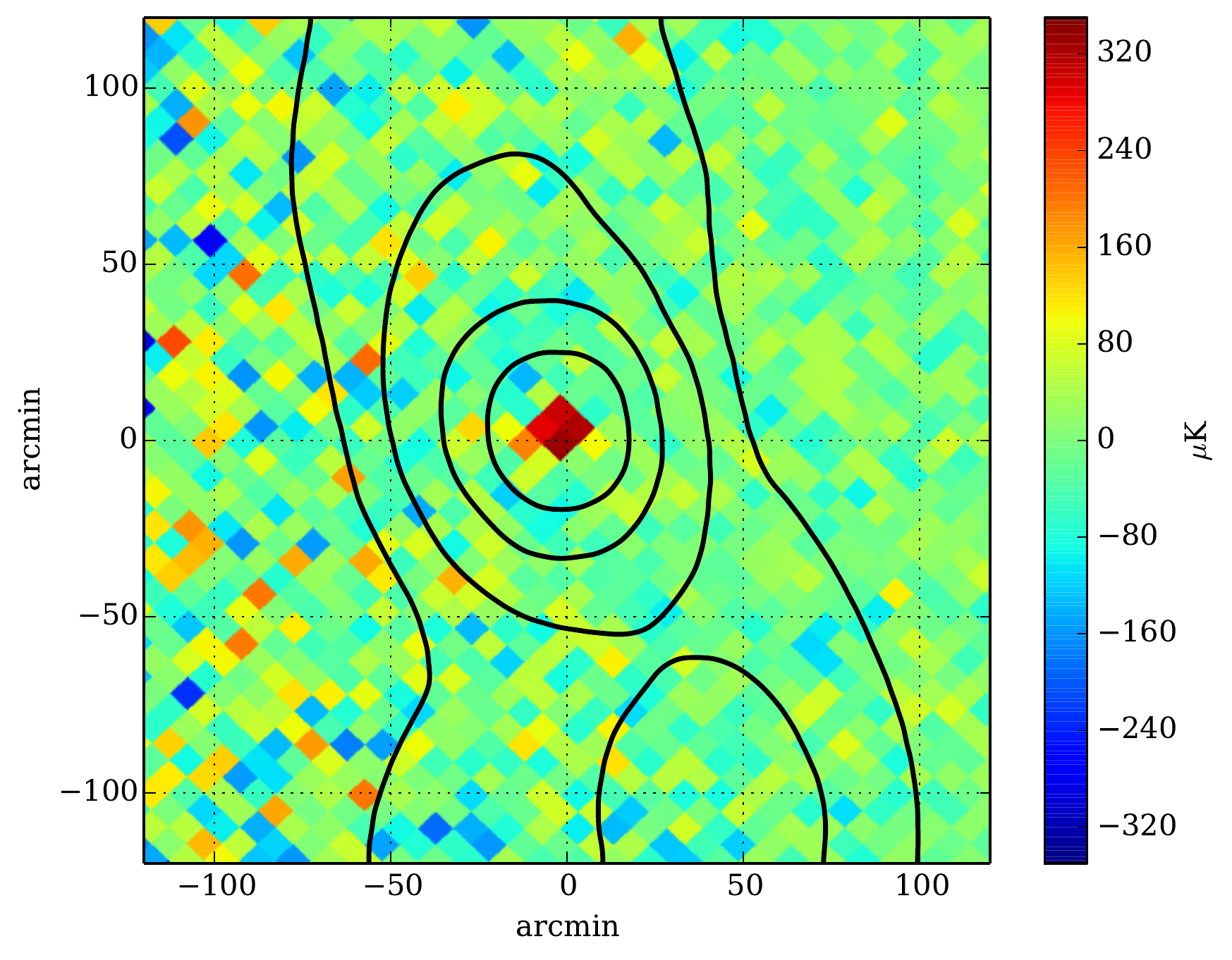}}&
\imagetop{\includegraphics[height=\CenA]{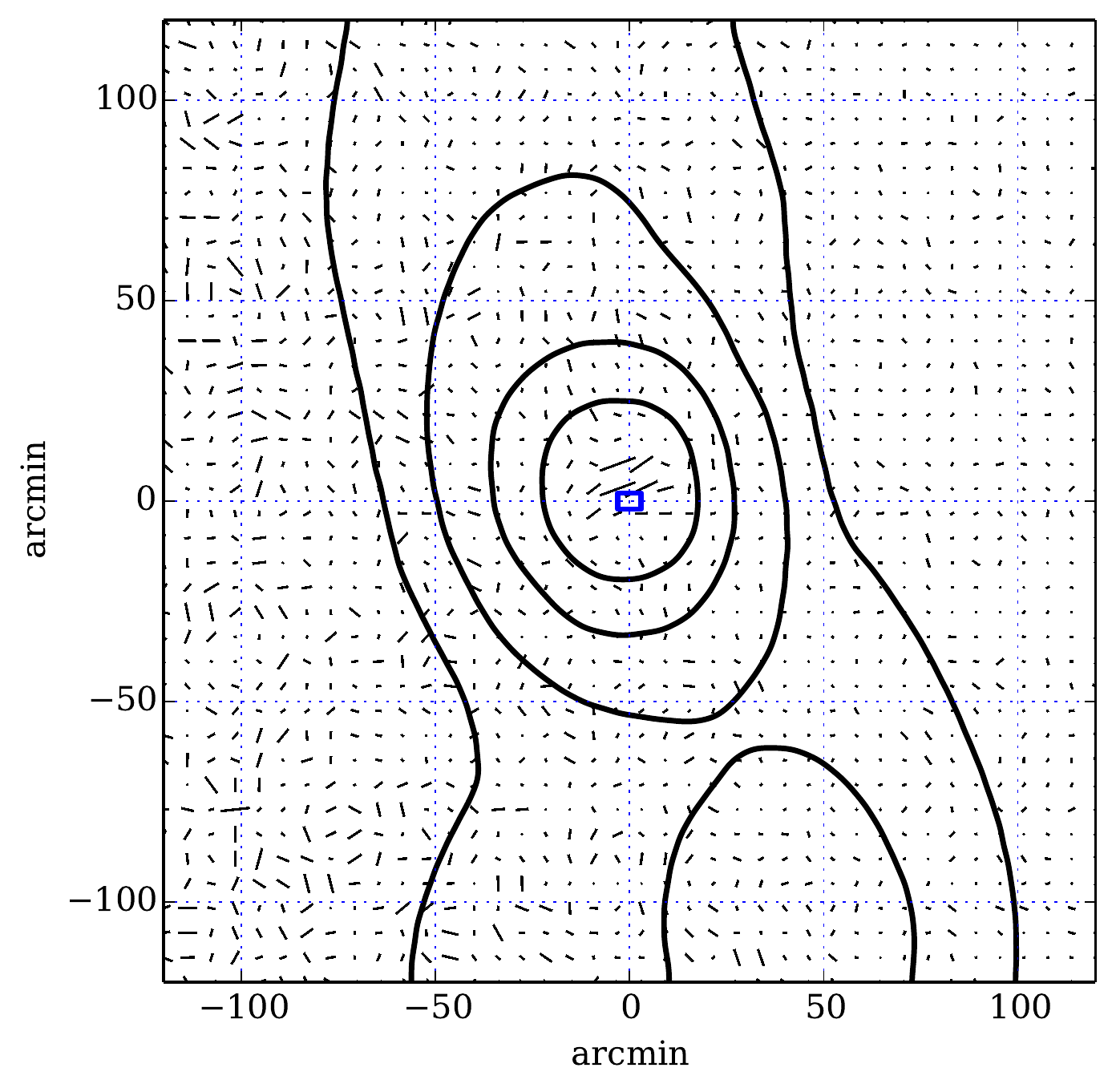}}&
\imagetop{\includegraphics[height=\CenA]{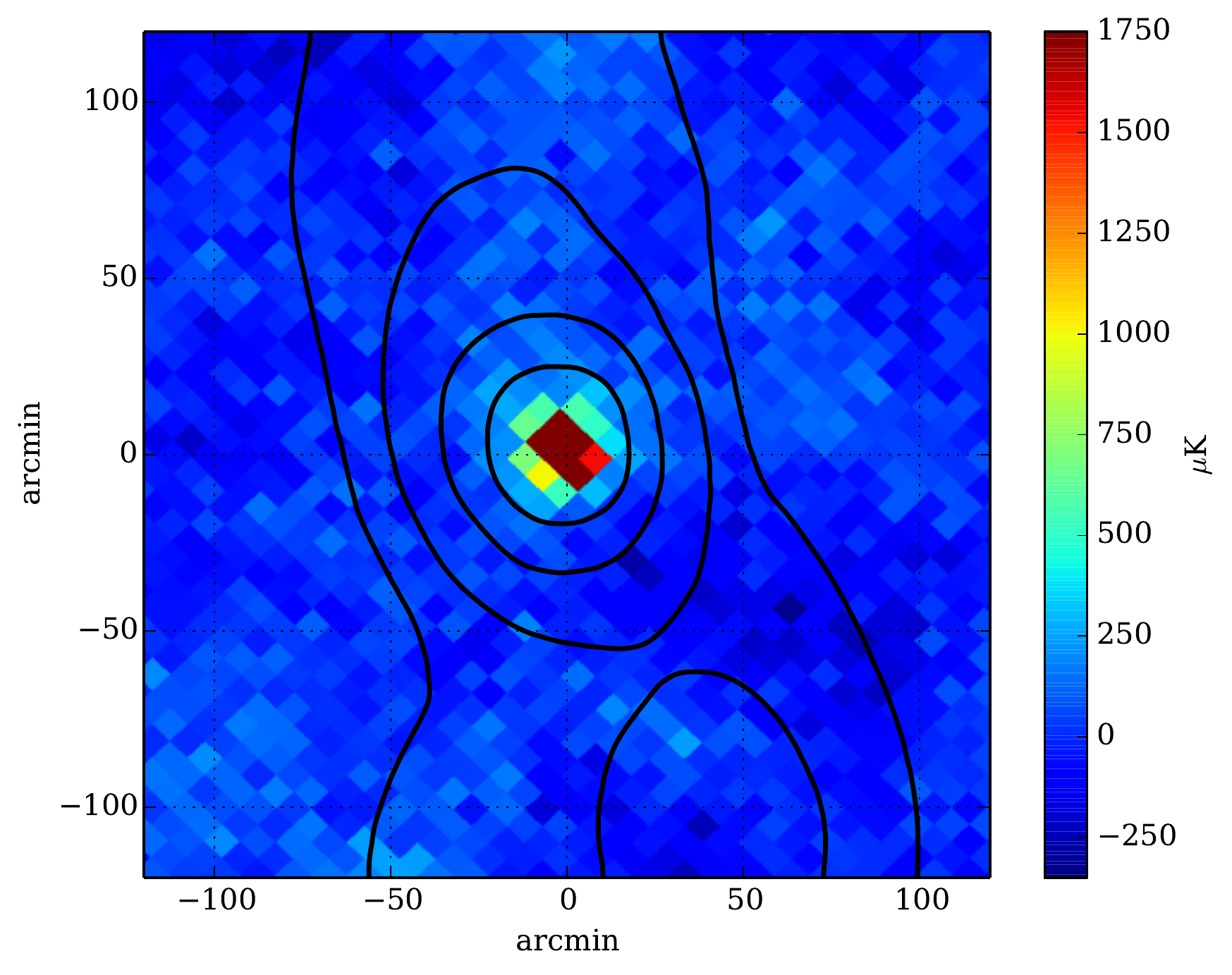}}\\
\end{tabular}
\end{center}
\caption{Radio galaxy Cen A ($l=309.516,b=19.417$), showing Stokes $Q$ and $U$, polarization vectors, and total intensity from \textit{WMAP} \citep{2013ApJS..208...20B}, for Q-band (43$\,$GHz, above) and W-band (95$\,$GHz, below).  Images are $4^{\circ}$ on a side, and show naively binned maps in gnomonic projection; Galactic longitude varies in the horizontal direction.  For comparison, we plot contours from the  \citet{1982A&AS...47....1H} radio map at 408 MHz, indicating radio brightness temperatures $T_B = 70,140,280,420$\,K (from the outside in).
The small blue box in the vector image, the central $6 \times 4$$\,$sq.~arcmin, spans the extent of the galaxy in 2MASS imaging.}
\label{fig:CenA}
\end{figure*}

The brightest object in our fields is the radio galaxy Centaurus A.  In radio images this object is nearly 10 degrees across, and has several major components \citep{1993A&A...269...29J}, including the  Northern and Southern Giant Outer Lobes, the Northern Middle Lobe ($30'$ North of the core), the inner lobes, and the nuclear region.  With our angular resolution, we can only resolve the outer and middle lobes.  Also, Cen A lies in the outskirts of our patch CMB-1, in a region with fewer observations, less cross-linking, and higher noise than the bulk of our survey.  It lies outside the normal processing mask for our maximum likelihood pipeline. For that reason we limit ourselves here to qualitative discussion in the following.  Nonetheless, Cen A is so bright that  a naive (binned) map of our time-ordered data is sufficient to obtain useful results. For QUIET in polarization, this naive map differs from the full solution only on large scales.  The binned maps act as highpass-filtered versions of the solved maps.

This map is shown in Figure~\ref{fig:CenA}. The images are centered on the peak of the galaxy's light from a 2MASS K$_{\rm S}$-band image \citep{2006AJ....131.1163S}.  Polarized emission is brighter in Q-band than in W-band.   The noise is higher to the left, a gradient expected from the position of the source at the outskirt of the QUIET field.  In Q-band, polarized emission is apparent in the outer lobes even $2^{\circ}$ away from the galaxy.  The lobes have the same slight ``S'' shape as seen in total intensity (in \textit{WMAP}, \textit{Planck}, and multiple radio observations) and in radio polarization.  The peak emission is offset from the galaxy. In Q-band, the peak emission is centered on two spots on either side of the galaxy, at $(\Delta l \sin b,\Delta b) \sim (-5',30')$ (near the Northern Middle Lobe) and another spot at $\sim (0',15')$.    Comparing \textit{WMAP} and \textit{Planck} images of the lobe, the peaks of polarized emission lie nearer to the galaxy than the peaks of total intensity. In W-band, the brightest emission is nearer to the galaxy than in Q-band, and the brightest emission is in fact concentrated in one spot, at $(\Delta l \sin b,\Delta b) \sim (0',5')$.  

Note that the peaks of the polarized emission in Q-band and W-band differ in location and intensity, and the polarization direction varies strongly as a function of position. This is instructive as we later consider the population of unresolved sources.  For the same reason, our measurements of point-like sources below cannot easily  probe rotation measure and magnetic fields; for unresolved sources, we cannot be sure that we are always probing regions with the same intrinsic polarization direction.  This may be why, for our larger set of point-like sources, the polarization intensities and the polarization directions show little correlation between the two frequencies.

\begin{figure}
\includegraphics[width=\columnwidth]{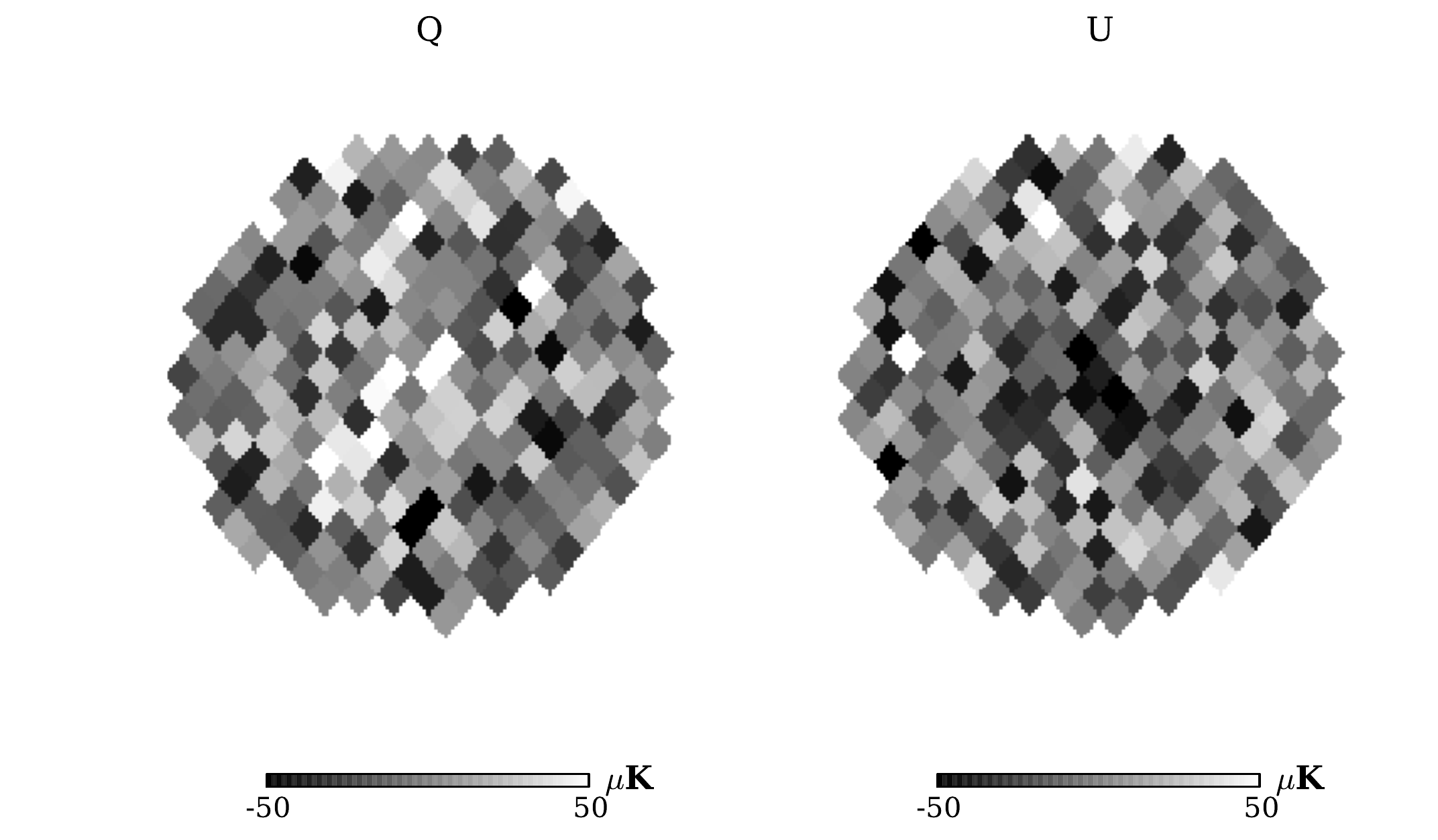}
\caption{
The data are noisy, but Pict A, at the center of this Q-band image, is nonetheless detected at S/N~$= 7.75$, with a positive Stokes $Q$ and negative Stokes $U$. 
The unmasked region has a $120'$ diameter, and the beam FHWM is $27.3'$.
}
\label{fig:PictA}
\end{figure}


Another bright source, radio galaxy Pictor A, lies in our CMB-2 field. This object also has multiple components; the central nucleus and both radio lobes appear as separate entries in the AT20G catalog\footnote{Western lobe: AT20G J051926-454554; nucleus: AT20G J051949-454643; eastern lobe: AT20G J052006-454745.}, with a total lobe-to-lobe separation of about $8'$.  This separation is smaller than the FWHM of our beams, so we cannot separately resolve the components.  Figure \ref{fig:PictA} shows that our data are noisy, but Pict A is nonetheless detected in polarization at S/N~$=7.75$.  At 20$\,$GHz, \citet{2010MNRAS.402.2403M} find that the nucleus ($6320 \pm 110$$\,$mJy) is much brighter than the western lobe ($1464 \pm 55$$\,$mJy) in total intensity. However, they detect no significant polarization from the nucleus, while the western lobe has a large polarized flux density ($423\pm 5$$\,$mJy).  The western lobe also has stronger polarization at lower frequencies $<5$$\,$GHz \citep{1997A&A...328...12P}.  Using our standard photometry method, as described in Section \ref{sec:estStokes}, our observations also yield the largest polarization toward the western lobe, with a maximum likelihood polarization of $S_P=205^{+31}_{-21}\,$mJy at Q-band and $S_P=89^{+33}_{-33}\,$mJy at W-band.   

However, because of our beam sizes and the close proximity of the components, our standard photometry method is sub-optimal, as it does not account for overlapping emission.  Therefore, for Pict A we also compute the flux density by integrating over top-hat apertures of increasing sizes, which provides some notion of the total polarization, even though we lack the resolution to examine the three components of Pict A individually.  For Q-band, the polarization signal increases as the aperture expands up to $\sim 30'$ in radius, and is then constant within the errors.  For that aperture, we find $(S_Q,S_U) = (105 \pm 40, -146 \pm 39)$$\,$mJy, consistent to our standard fit, but with larger error bars\footnote{Aperture photometry is less restrictive than the fitted template from section \ref{sec:method}.}.   In the same $30'$ aperture for W-band, we find $(S_Q,S_U) = (104 \pm 77, -249 \pm 77)\,$mJy, but the signal is not constant as the aperture is increased further, so we are unable to fit a constant background offset.  This is due to large scale features in the map, and so this W-band aperture photometry may be unreliable.
To avoid multiple counting in our summary statistics below, we either include only the western lobe of Pict A, or exclude it altogether.

\subsection{Point sources}
For the remainder of the AT20G sources, which are selected at 20$\,$GHz, we measure the values of the Stokes parameters at 43 and 95$\,$GHz.  Depending on detection significance, we provide a measurement or an upper limit on the polarized flux density, as described above. Further, for some of the AT20G sources that exceed S/N~$>2.0$ in our measurements, an additional AT20G source lies within  $30'$.  However, except for the components of Pict A, all of those neighboring sources have S/N~$<2.0$ significance in our measurements, limiting the potential for contamination.

\begin{table*}
\caption{ S/N~$>2.7$ detections, Q-band.}
\begin{center}
\begin{tabular}{rrrrrrrrrrrrrrrrrrr}
\hline
ID && R.A. & Dec. & $S_Q$ (mJy) & $S_U$ (mJy)& $S_{P,{\rm ML}}$ (mJy)  & $\alpha$ ($^\circ$) & ${\rm Pr}(>\chi^2)$ & S/N\\
\hline
AT20GJ010613-504421 & $^{}$ & 1:06:13.26 & -50:44:21.7 & $-55 \pm 19$ & $-29 \pm 19$ & ${61}_{- 23}^{+ 18}$ & ${-77}_{- 8}^{+ 12}$ & $0.0063$ & $2.73$ \\ 
AT20GJ010645-403419 & $^{}$ & 1:06:45.11 & -40:34:19.5 & $16 \pm 42$ & $160 \pm 41$ & ${156}_{- 41}^{+ 44}$ & ${42}_{- 8}^{+ 8}$ & $0.00052$ & $3.47$ \\ 
AT20GJ015358-540653 & $^{}$ & 1:53:58.37 & -54:06:53.5 & $-133 \pm 63$ & $175 \pm 63$ & ${208}_{- 65}^{+ 71}$ & ${63}_{- 6}^{+ 10}$ & $0.0025$ & $3.02$ \\ 
AT20GJ042840-375619 & $^{}$ & 4:28:40.37 & -37:56:19.2 & $102 \pm 28$ & $-8 \pm 28$ & ${100}_{- 30}^{+ 27}$ & ${-3}_{- 6}^{+ 10}$ & $0.0014$ & $3.19$ \\ 
AT20GJ050838-330853 & $^{}$ & 5:08:38.05 & -33:08:53.5 & $-86 \pm 23$ & $4 \pm 23$ & ${85}_{- 24}^{+ 22}$ & ${88}_{- 6}^{+ 10}$ & $0.0008$ & $3.35$ \\ 
AT20GJ051926-454554 & $^{ab}$ & 5:19:26.34 & -45:45:54.6 & $153 \pm 26$ & $-143 \pm 26$ & ${205}_{- 21}^{+ 31}$ & ${-22}_{- 3}^{+ 4}$ & $9.5 \times 10^{-15}$ & $7.75$ \\ 
AT20GJ051949-454643 & $^{ab}$ & 5:19:49.70 & -45:46:43.7 & $141 \pm 26$ & $-126 \pm 26$ & ${185}_{- 21}^{+ 31}$ & ${-21}_{- 3}^{+ 5}$ & $3.3 \times 10^{-12}$ & $6.96$ \\ 
AT20GJ052006-454745 & $^{ab}$ & 5:20:06.47 & -45:47:45.5 & $125 \pm 26$ & $-113 \pm 26$ & ${164}_{- 23}^{+ 31}$ & ${-21}_{- 4}^{+ 5}$ & $1.1 \times 10^{-9}$ & $6.09$ \\ 
AT20GJ053757-461430 & $^{}$ & 5:37:57.60 & -46:14:30.3 & $29 \pm 30$ & $-90 \pm 30$ & ${92}_{- 34}^{+ 29}$ & ${-36}_{- 8}^{+ 10}$ & $0.0063$ & $2.73$ \\ 
AT20GJ111301-354947 & $^{}$ & 11:13:01.51 & -35:49:47.5 & $43 \pm 36$ & $-108 \pm 36$ & ${110}_{- 39}^{+ 36}$ & ${-34}_{- 8}^{+ 10}$ & $0.0065$ & $2.72$ \\ 
AT20GJ113855-465342 & $^{a}$ & 11:38:55.60 & -46:53:42.6 & $91 \pm 36$ & $-98 \pm 36$ & ${128}_{- 35}^{+ 38}$ & ${-23}_{- 8}^{+ 8}$ & $0.0011$ & $3.25$ \\ 
AT20GJ123045-312123 & $^{}$ & 12:30:45.02 & -31:21:23.1 & $109 \pm 30$ & $-26 \pm 31$ & ${109}_{- 30}^{+ 33}$ & ${-6}_{- 8}^{+ 8}$ & $0.001$ & $3.29$ \\ 
AT20GJ224326-393352 & $^{}$ & 22:43:26.04 & -39:33:52.6 & $-100 \pm 33$ & $46 \pm 31$ & ${106}_{- 34}^{+ 34}$ & ${77}_{- 8}^{+ 10}$ & $0.0034$ & $2.93$ \\ 

\hline
\end{tabular}
\end{center}
\tablecomments{AT20G catalog locations show 43$\,$GHz polarized flux density at a statistical significance equivalent to S/N~$>2.7$ in QUIET Q-band data.  Stokes parameters and angle are given in Galactic coordinates, adopting the CMB convention (see text).\\
$^a$ Another AT20G source lies within $30'$. \\
$^b$ Component of Pict A.
}
\label{tab:qboomers}
\end{table*}

\begin{table*}
\caption{ S/N~$>2.7$ detections, W-band. }
\begin{center}
\begin{tabular}{rrrrrrrrrrrrrrrrrrr}
\hline
ID && R.A. & Dec. & $S_Q$ (mJy) & $S_U$ (mJy) & $S_{P,{\rm ML}}$ (mJy) & $\alpha$ ($^\circ$) & ${\rm Pr}(>\chi^2)$ & S/N\\
\hline
AT20GJ000601-423439 & $^{a}$ & 0:06:01.95 & -42:34:39.8 & $135 \pm 45$ & $44 \pm 45$ & ${133}_{- 46}^{+ 49}$ & ${9}_{- 8}^{+ 10}$ & $0.0067$ & $2.71$ \\ 
AT20GJ005645-445102 & $^{}$ & 0:56:45.80 & -44:51:02.4 & $83 \pm 26$ & $-37 \pm 26$ & ${85}_{- 25}^{+ 30}$ & ${-12}_{- 8}^{+ 8}$ & $0.0026$ & $3.01$ \\ 
AT20GJ042840-375619 & $^{}$ & 4:28:40.37 & -37:56:19.2 & $248 \pm 37$ & $5 \pm 37$ & ${247}_{- 38}^{+ 38}$ & ${1}_{- 4}^{+ 4}$ & $2.8 \times 10^{-10}$ & $6.31$ \\ 
AT20GJ052257-362730 & $^{a}$ & 5:22:57.94 & -36:27:30.4 & $90 \pm 25$ & $-10 \pm 25$ & ${85}_{- 22}^{+ 29}$ & ${-3}_{- 8}^{+ 8}$ & $0.0015$ & $3.17$ \\ 
AT20GJ053850-440508 & $^{}$ & 5:38:50.35 & -44:05:08.7 & $103 \pm 30$ & $73 \pm 30$ & ${124}_{- 31}^{+ 31}$ & ${18}_{- 6}^{+ 7}$ & $0.00014$ & $3.81$ \\ 
AT20GJ054922-405107 & $^{}$ & 5:49:22.79 & -40:51:06.9 & $86 \pm 33$ & $-76 \pm 33$ & ${109}_{- 34}^{+ 34}$ & ${-22}_{- 6}^{+ 10}$ & $0.0027$ & $3.00$ \\ 
AT20GJ234038-344249 & $^{a}$ & 23:40:38.63 & -34:42:49.4 & $-241 \pm 116$ & $-323 \pm 116$ & ${388}_{- 121}^{+ 121}$ & ${-63}_{- 8}^{+ 8}$ & $0.0025$ & $3.03$ \\ 

\hline
\end{tabular}
\end{center}
\tablecomments{AT20G catalog locations show 95$\,$GHz polarized flux density at a statistical significance equivalent to S/N~$>2.7$ in QUIET W-band data.  Stokes parameters and angle are given in Galactic coordinates, adopting the CMB convention (see text).\\
$^a$ Another AT20G source lies within $30'$. \\
}
\label{tab:wboomers}
\end{table*}


\begin{figure}

\includegraphics[width=\columnwidth]{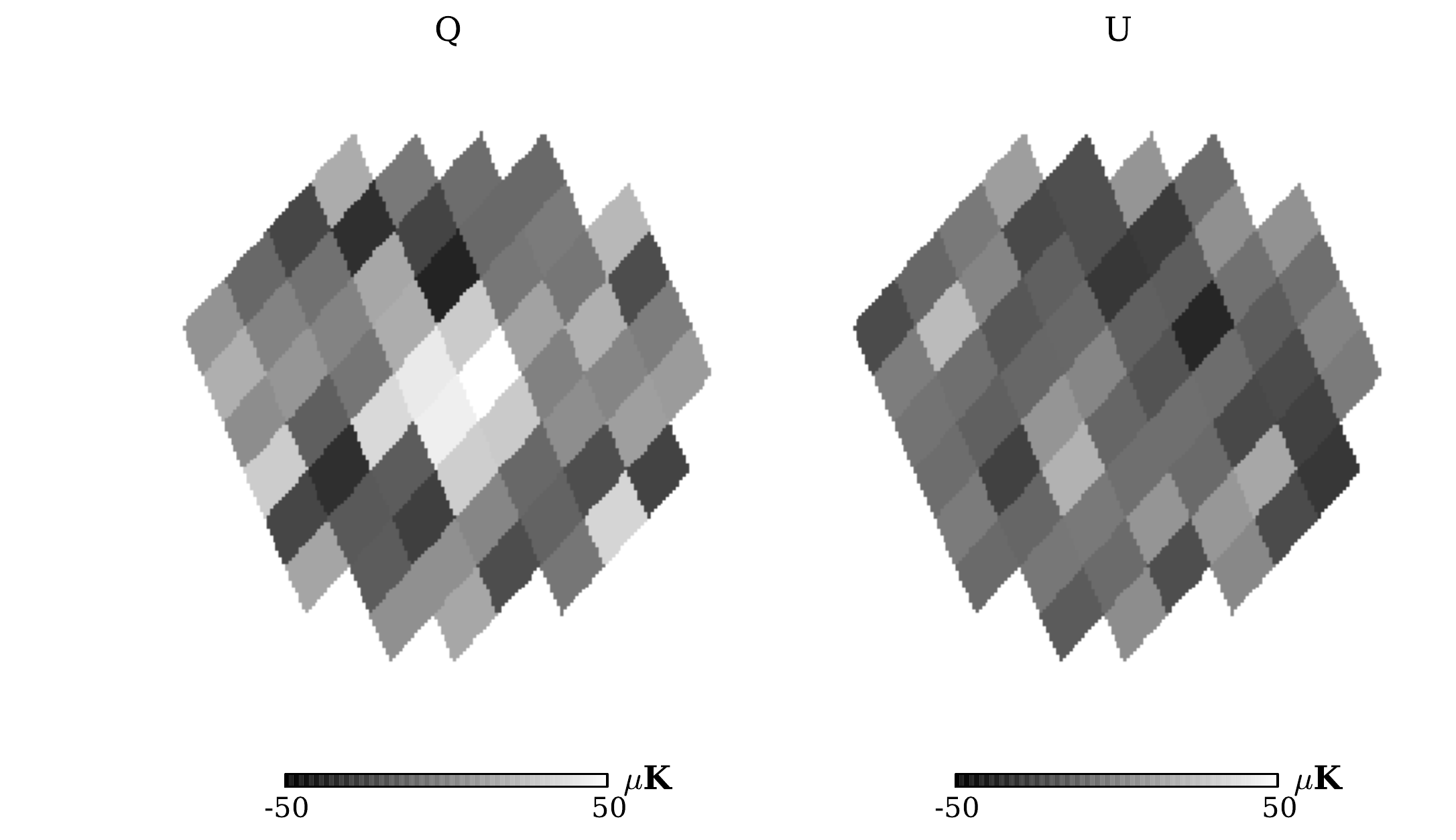}
\includegraphics[width=\columnwidth]{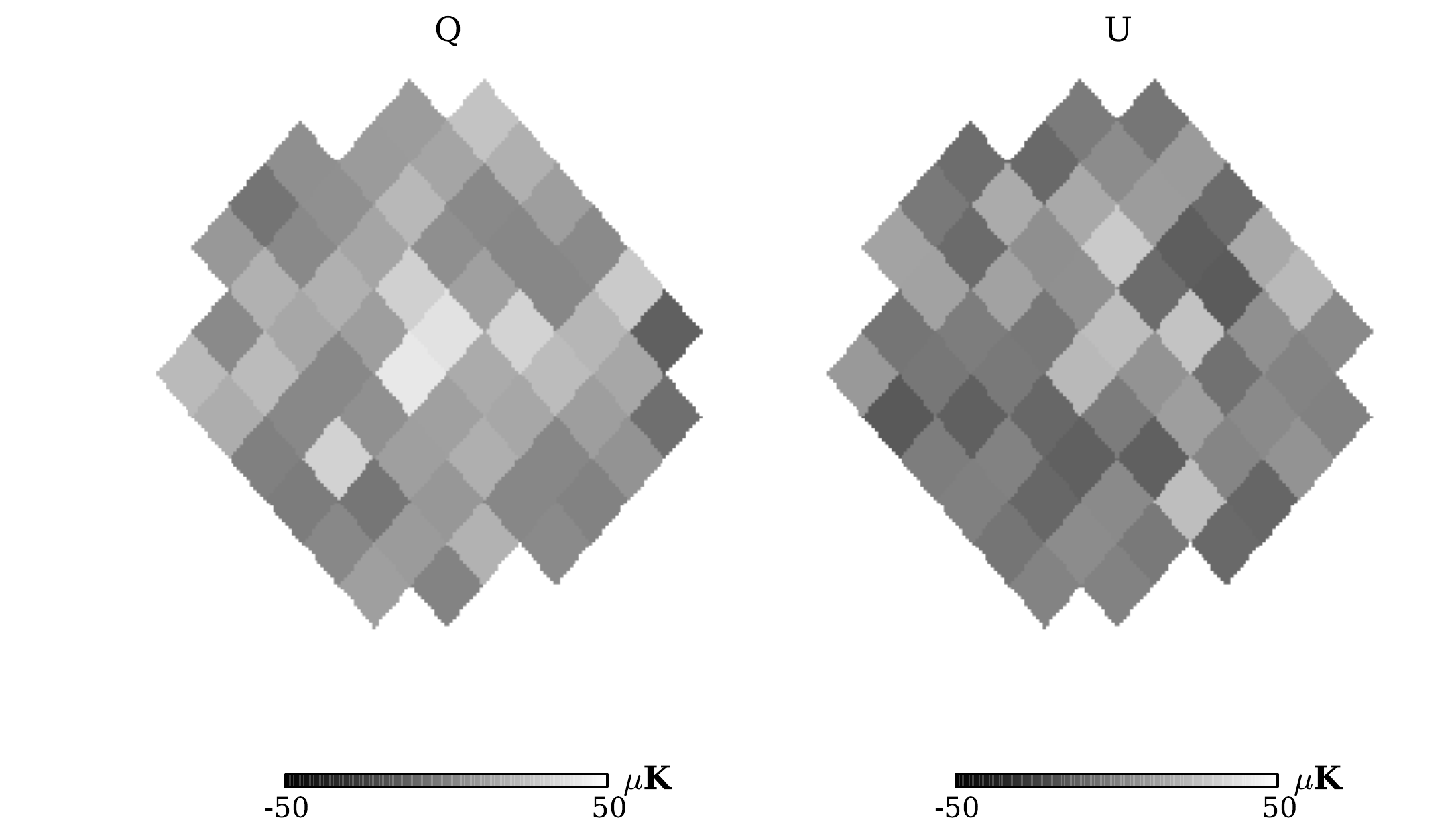}
\includegraphics[width=\columnwidth]{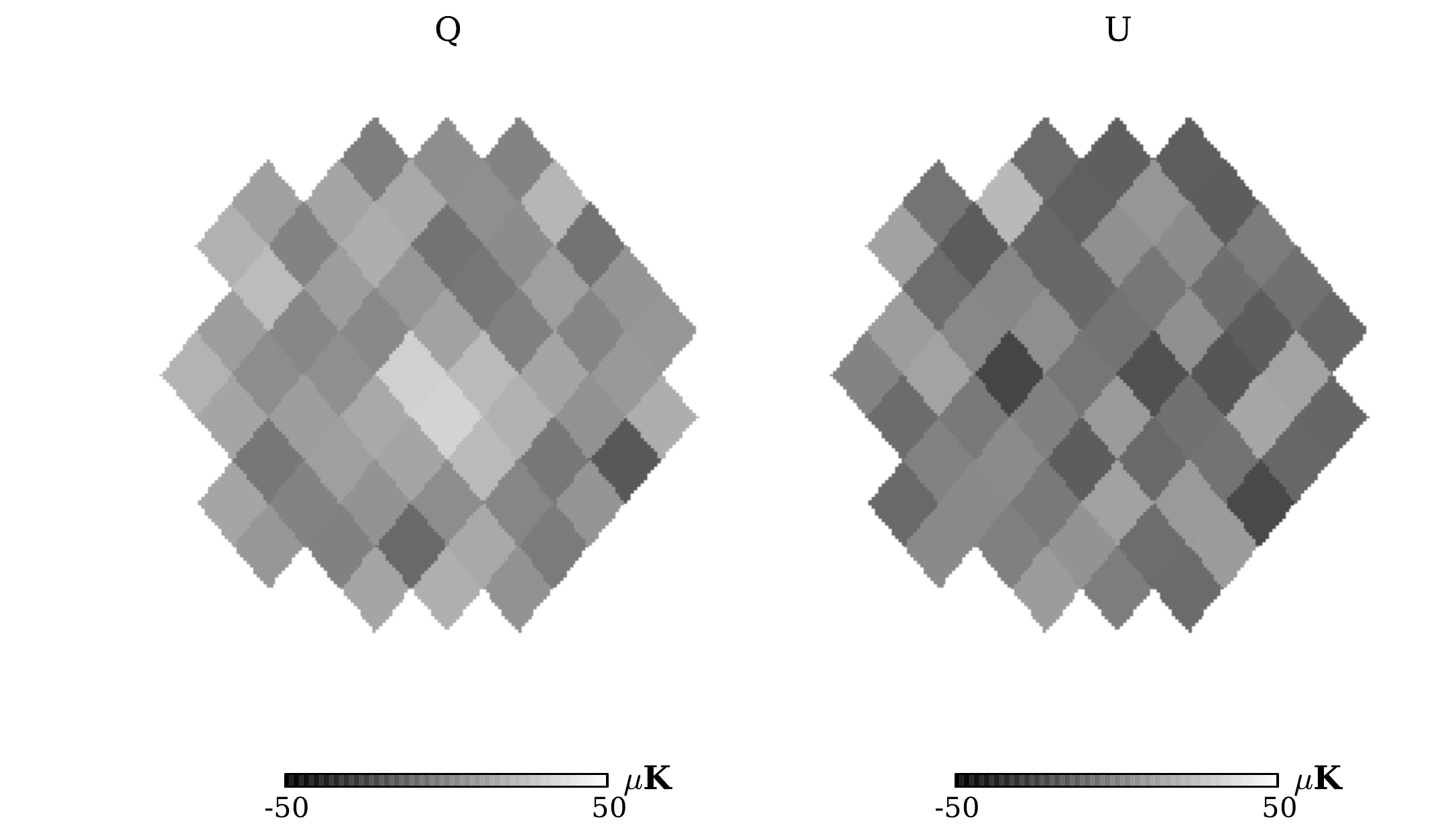}
\caption{
The three sources detected in W-band at greatest significance: AT20GJ042840-375619, AT20GJ053850-440508, and AT20GJ052257-362730.  The unmasked regions have a $60'$ diameters, and the beam FWHM is $12.8'$.
}
\label{fig:wbandboomers}
\end{figure}

We find  a handful of point-like sources producing polarized flux density at high signal-to-noise ratio, as listed in Tables \ref{tab:qboomers} and \ref{tab:wboomers}.  We adopt a threshold of S/N~$>2.7$ for these tables, which is a compromise that keeps the table short and limits the number of spurious detections.  At that level of significance or greater,  we find 11 independent sources in Q-band and 7 sources in W-band.  Only AT20GJ042840-375619 (the well-known quasar PKS 0426-380) exceeds the significance cut in both Q- and W-bands.
Accounting for the total number of AT20G sources, noise alone should account for  $3.3 \pm 1.8$ detections of polarized emission in each table, based on Poisson statistics.  At S/N~$>3.0$ or greater, only $1.3 \pm 1.1$ detections should be spurious, while we find seven in Q-band and six in W-band.  We therefore conclude that the greater part of the listed sources record genuine polarized emission.  Figure \ref{fig:wbandboomers} shows the brightest sources for W-band.

For a sensitivity cut of S/N~$>3$, it is interesting to note that we find similar numbers of sources in the more sensitive Q-band at 43$\,$GHz and the slightly less sensitive W-band at 95$\,$GHz (see Figure~\ref{fig:errors}).    If these sources were equally bright in both bands we would tend to see fewer in W-band.  Of course, we cannot draw strong conclusions based on this small number of sources, but this observation may suggest that source polarization is fairly flat over this frequency range.  \textit{WMAP} found that the mean spectral index in intensity is also nearly flat (and slightly negative): $S \propto \nu^{\alpha_{\rm SED}}$ with  mean $\alpha_{\rm SED} \pm \Delta \alpha_{\rm SED} = -0.09 \pm 0.28$, where the range indicates source-to-source scatter \citep{2009ApJS..180..283W}.

\begin{figure*}
\includegraphics[width=\textwidth]{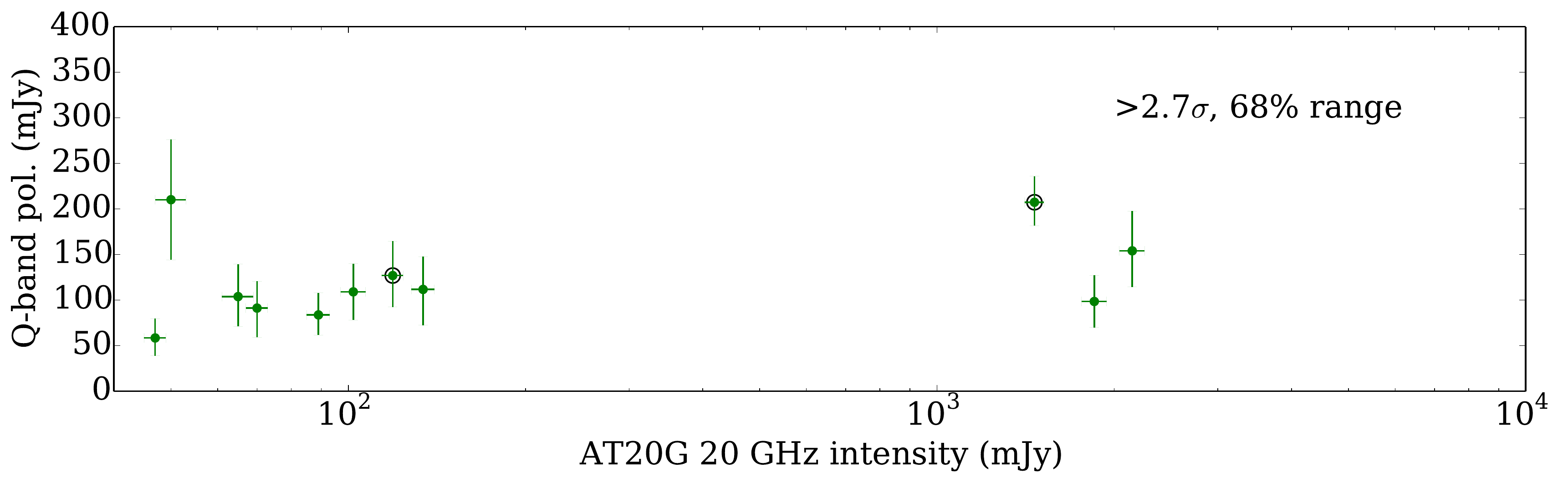}
\includegraphics[width=\textwidth]{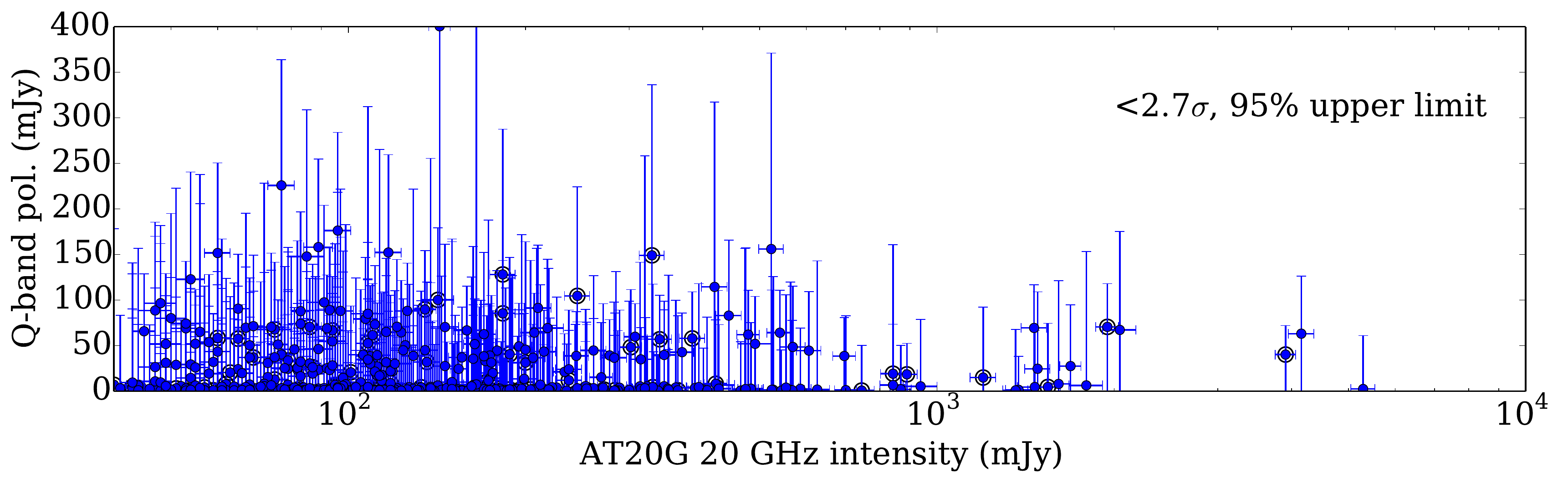}
\includegraphics[width=\textwidth]{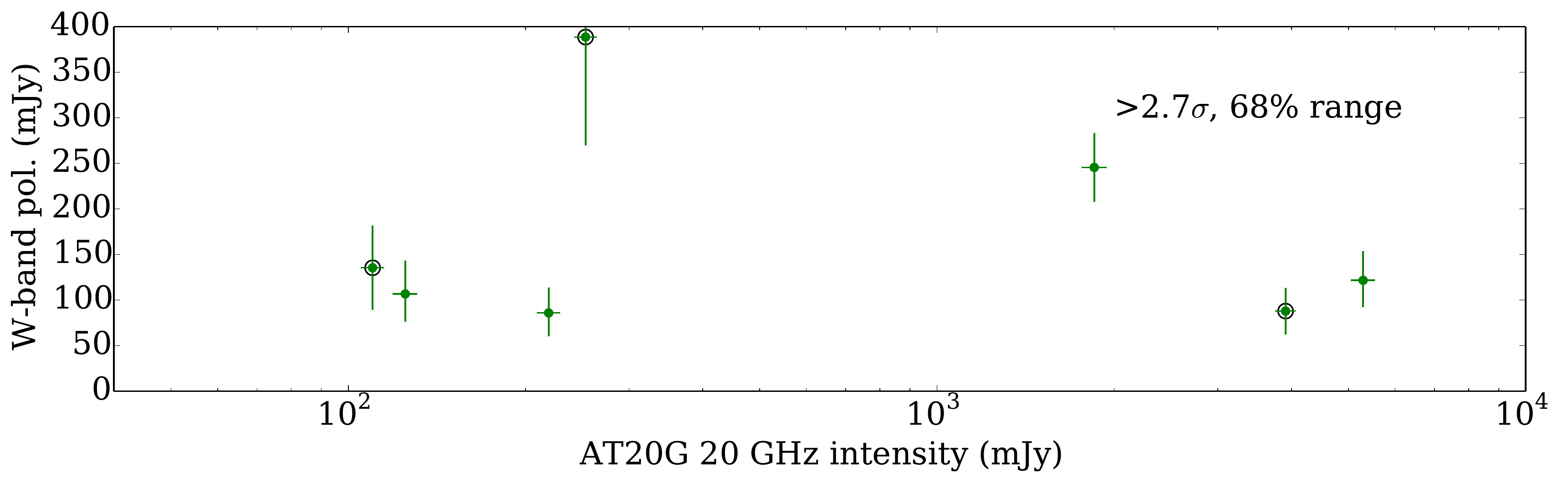}
\includegraphics[width=\textwidth]{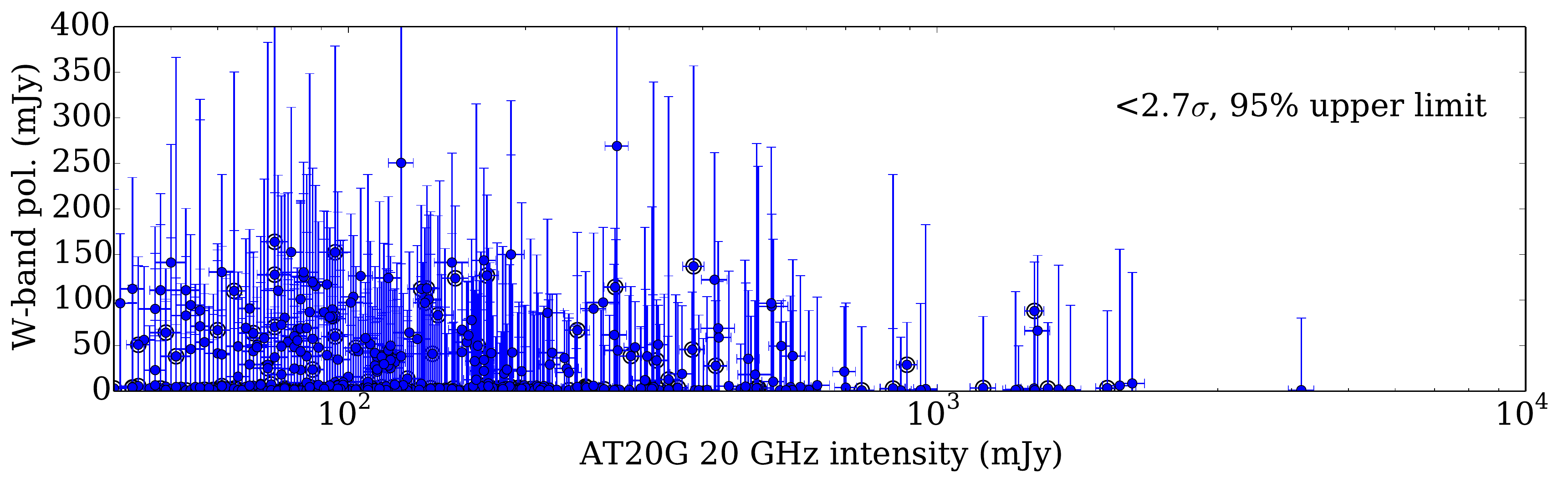}
\caption{Polarized flux density measured with QUIET versus the total intensity at 20$\,$GHz, where the sources are selected from the AT20G catalog.  The top two plots are Q-band (43$\,$GHz, see Table~\ref{tab:qboomers}) and the bottom two plots are W-band (95$\,$GHz, see Table~\ref{tab:wboomers}).   The higher-significance subsets, with S/N~$>2.7$, show 68\% confidence intervals around the maximum likelihood point, and are plotted separately from the upper limits.  In the upper limit plots, each vertical error bars shows the 95\% confidence interval, and connects to zero through the points that mark the maximum likelihood value.  Sources marked with a ring have another AT20G source within $30'$.}\label{fig:P90_vs_S20}
\end{figure*}

In Figure~\ref{fig:P90_vs_S20}, we show our polarization measurements of sources compared to the 20$\,$GHz total intensity on which they were selected.  For each band, we separate sources observed with S/N~$>2.7$ from the lower significance sources to make the plots clearer.   For higher signal-to-noise ratio sources, we show the maximum likelihood point and the 68\% interval, as illustrated in Figure~\ref{fig:Pml+errors}.  Conversely, for lower signal-to-noise ratio sources, we only plot the maximum likelihood point and a 95\% upper limit.  The sensitivity of our upper limit depends on the data and the errors at that position on the map.  Half of the sources have polarizations below 90$\,$mJy in Q-band and below 106$\,$mJy in W-band, at 95\% confidence.

No trend is immediately clear.  Our strongest detections tend to come from sources with 20$\,$GHz $S > 1$$\,$Jy, but among those 19 sources, several yield only upper limits in our data.  This may indicate diversity in the higher-frequency polarization properties of the sources selected from the 20$\,$GHz catalog.  Simple models, which fix both the frequency scaling ($\alpha_{\rm SED}$) and the polarization fraction (at a few percent), will not reproduce this result.


\subsection{Validation and robustness}

We now consider the robustness of these results by means of null-maps and simple simulations. The null-map used in the following was produced in the course of the CMB power spectrum analysis for the Q-band, in which the data were split into the first and second halves of the first observing season, comprising 232 days from 2008 October--2009 June.  Subtracting (rather than adding) these half-datasets during mapmaking produces the null map.  Ideally, any time-independent signal should cancel in this map, leaving only residual noise and, possibly, variable sources.  


\begin{figure}
\includegraphics[width=\columnwidth]{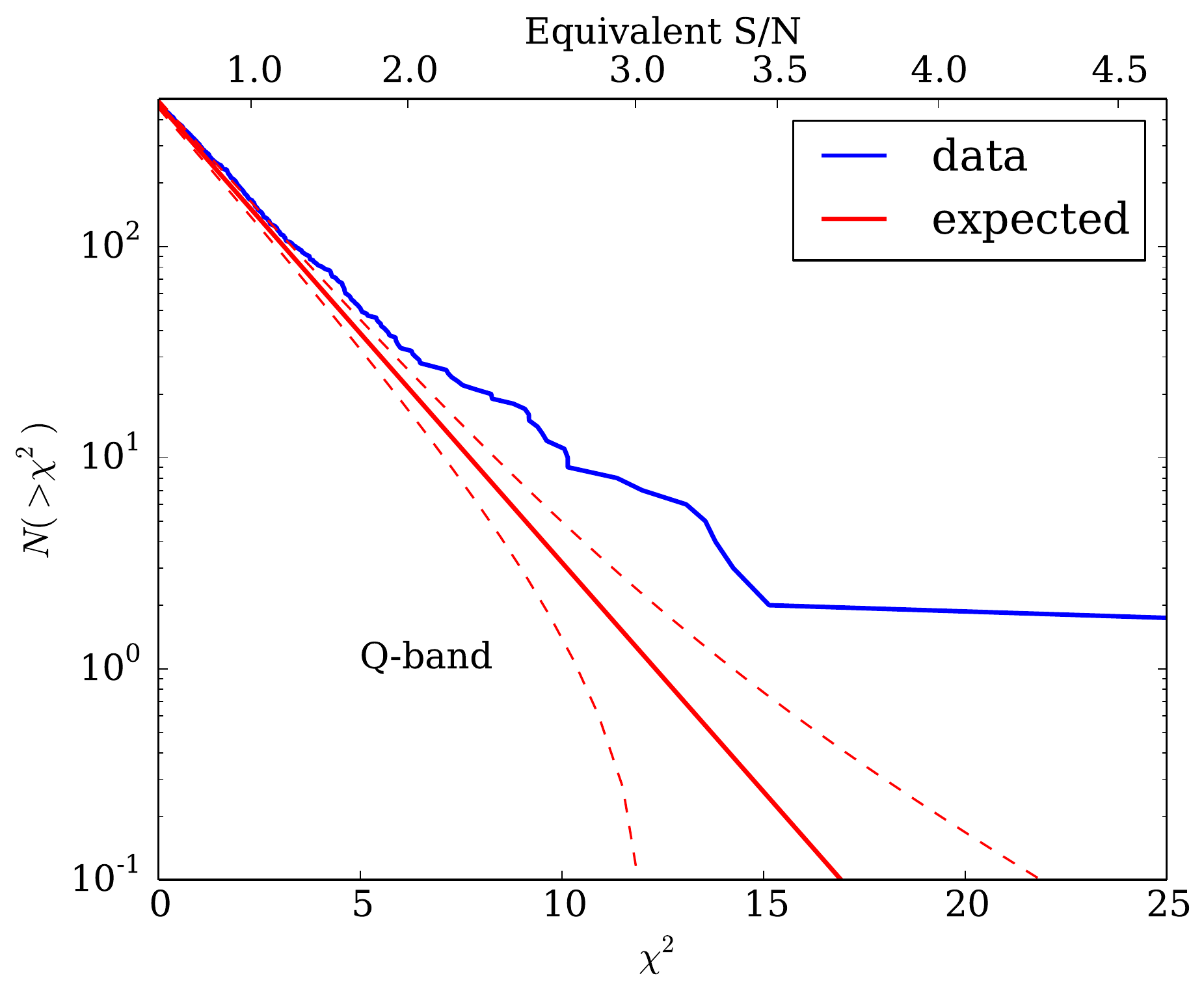}
\includegraphics[width=\columnwidth]{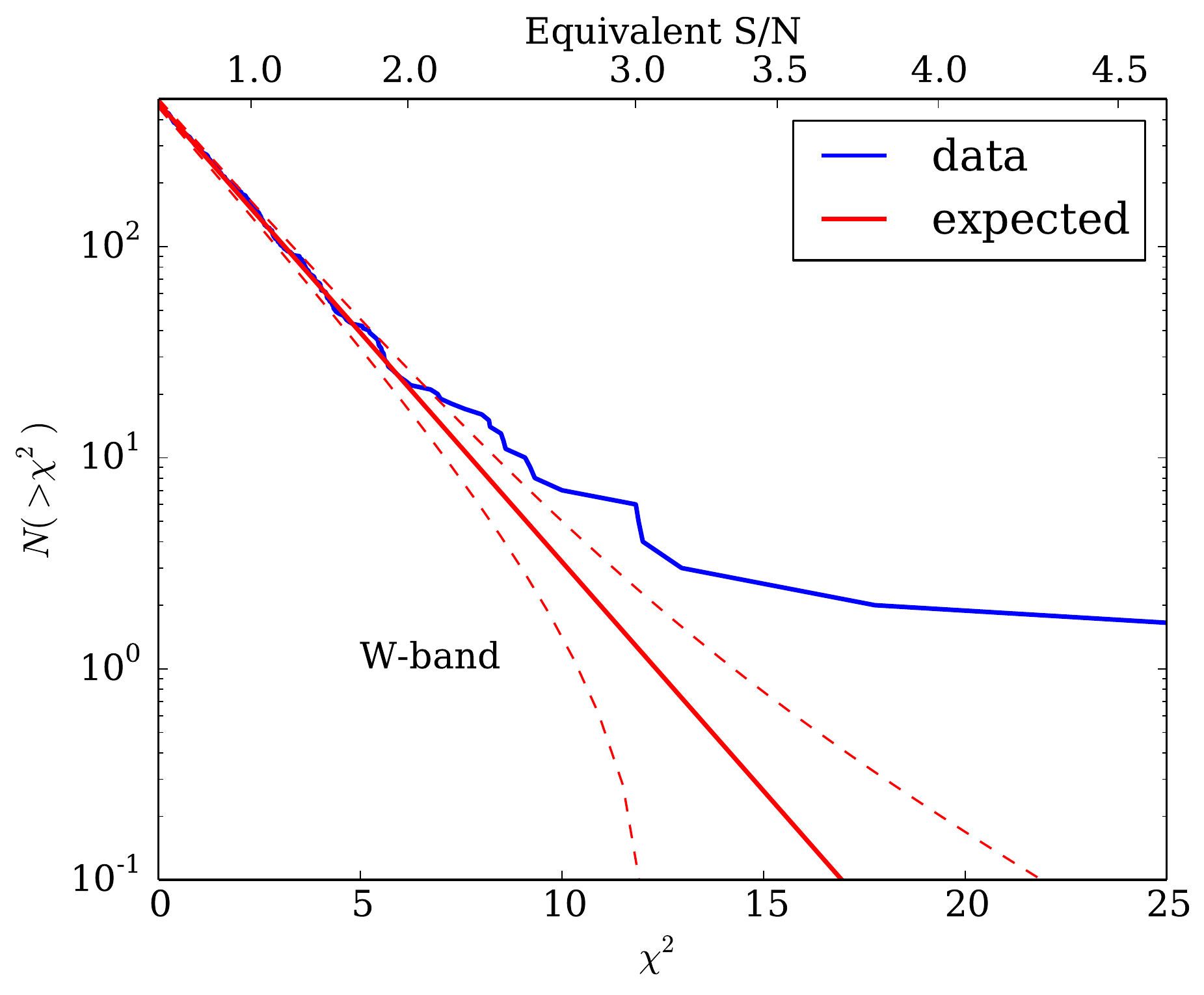}
\includegraphics[width=\columnwidth]{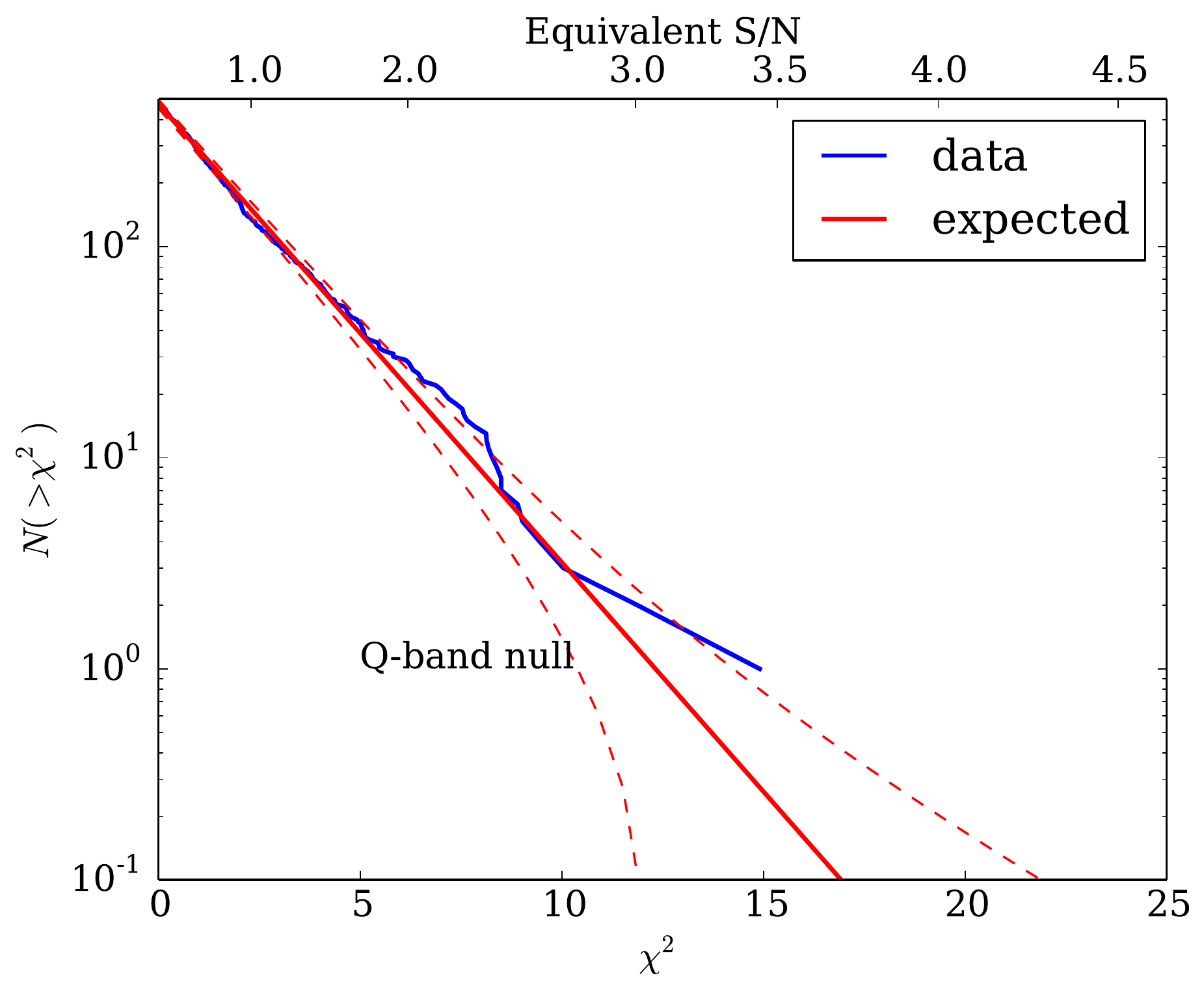}
\caption{Cumulative number of sources versus $\chi^2$, also showing the equivalent significance, compared to the expected distribution. The line is the expected distribution from a cumulative $\chi^2$-distribution under the null hypothesis that source positions show no excess polarized flux density.  The dashed error bars are given by Poisson statistics.  The null hypothesis is a good fit for the null map, but a poor fit for the standard maps; each shows excess high-significance objects, evidence of polarized emission.}
\label{fig:num_vs_chisq}
\end{figure}

We run the null map cutouts through the same pipeline as the main analysis, and show in Figure~\ref{fig:num_vs_chisq} the resulting distributions of source significances. Here we see that the Q-band and W-band maps show an excess of high-significance polarization measurements compared to the expectation from noise alone.  By contrast, for the source positions in the Q-band null map, we find fewer indications of additional polarized flux beyond noise:  At S/N~$>3$ we expect $1.3 \pm 1.1$ spurious sources, and find two. These consistent values for the null map strengthen our confidence in the covariance matrices.
 
Both the S/N~$\geq 3$ sources seen in the null map could be spurious and yet still consistent with noise.
However, one source, AT20GJ111301-354947, which has S/N~$=3.0$ in the null map,  has S/N~$=2.7$ in the standard summed map and also appears in Table~\ref{tab:qboomers}.  The Stokes parameters are nearly the same in the null map and the standard map.  This may be explained by source variability: if a source produces emission during just one half of the observing season, it does not matter if the quiescent half of the data is added (standard map) or subtracted (null map). 
The other significant source in the null map is AT20GJ053850-440508, which has S/N~$=3.4$ in the null map and S/N~$=0.55$ in the standard map.  In intensity, this source is very bright and highly variable, varying during 2009 September--2010 March from $6.141\pm{0.089}$\,Jy to $14.814\pm{0.194}$\,Jy at 39.8\,GHz according to \citep[][]{2011MNRAS.415.1597M}, although that work does not include a measurement of polarization.  Although it is suspicious that this is one of the brightest sources in our region, it is hard to reconcile our large polarized detection in the null map with the lack of detection in the standard map, unless it is a noise fluctuation.  Given the number of source positions in our fields, noise alone could mimic a S/N~$>3.4$ source with a chance of one in four. 

Our next tests of robustness use synthetic maps.  Specifically, we simulate maps in which every source is assigned zero flux density, but has the same noise and covariance properties given by the maximum likelihood pipeline.  The measured fluxes from this simulation are statistically consistent with noise.  Finally, we ran two simulations including simulated sources with either 100$\,$mJy or 1$\,$Jy amplitudes and randomized polarization directions, and find that our recovered Stokes parameters are unbiased.



\subsection{Inter-frequency comparisons}

We now compare the polarization measurements between our bands and also to other measurements of the same sources in the literature.

\paragraph{43$\,$GHz polarization versus 95$\,$GHz polarization}


We expect bright sources will tend to be bright in both bands as a result of both intrinsic luminosity and distance effects.  However, the low signal-to-noise ratio of our measurements complicates our ability to measure this correlation in all but the brightest sources.

Overall, there are 460 sources that overlap the sky area covered by both frequency bands.
Imposing a S/N~$>1.0$ cut in both bands, we expect $47.3 \pm 6.9$ sources but find 60.  At S/N~$>1.5$ in both bands, we expect $8.3 \pm 2.9$ sources but find 17.  At S/N~$>2.0$ in both bands, we begin to run out of sources, expecting $1.2 \pm 1.1$ sources but finding 3.

Because of the statistics of the polarization amplitude, we must be very cautious interpreting our measurements.  For example, for S/N~$>1$, the polarized flux density shows a positive correlation between Q- and W-band data, with Pearson's $r=0.46$.  However, our tests with a synthetic catalog indicate that this effect is mostly statistical.  It results because noise rms have a similar trend as a function of position for Q-band and W-band due to scanning depth, so the ability for a particular location to have large upward fluctuations in polarization amplitude is correlated between the bands.  With a  higher signal-to-noise ratio cut, S/N~$>1.5$, the correlation coefficient drops to $r=0.11$. 
Similarly, among the full set of 460 sources, 252 have a smaller maximum likelihood polarization amplitude in Q-band than in W-band, and among the 60 sources with S/N~$>1$, 36 have a smaller amplitude in Q-band.  This too seems to be a statistical effect, and is not due to the source spectral energy distribution (SED), as our tests based on noise-only synthetic catalogs find similar or greater numbers.  This bias toward W-band can be caused by (1) the positive definite probability distribution for polarization amplitude and (2) the larger error bars in W-band which cause larger excursions (which skew positive) from the true flux density.


Another approach we advocate is to  compare the Stokes parameters individually between the bands. Because the errors in the Stokes parameters are Gaussian distributed, they are simpler statistics, and each set is an independent probe.  This should be useful for future experiments with larger number of high signal-to-noise ratio sources.  For a large number of well-measured sources, any trend should be the same for each Stokes parameter, due to the random orientation of objects on the sky.  Since we only measure a few sources well, we are limited by our significant scatter.
%
Here the correlation among S/N~$> 1$ sources (in both bands) is weak, with $r=0.020$ for Stokes $Q$ and $r=-0.12$ for Stokes $U$. 
%
%
So we do not find any evidence that the individual Stokes parameters correlate between the bands.  

This we might expect if---as in Cen A---different physical parts of the AGN are bright in polarization at different frequencies.  If true generally, it means that predicting accurately the polarization angle of an unresolved source, based on measurements at other frequencies, will not be straightforward without a detailed physical model.

\paragraph{QUIET versus 20$\,$GHz polarization}

\begin{figure*}
\includegraphics[width=.33\textwidth]{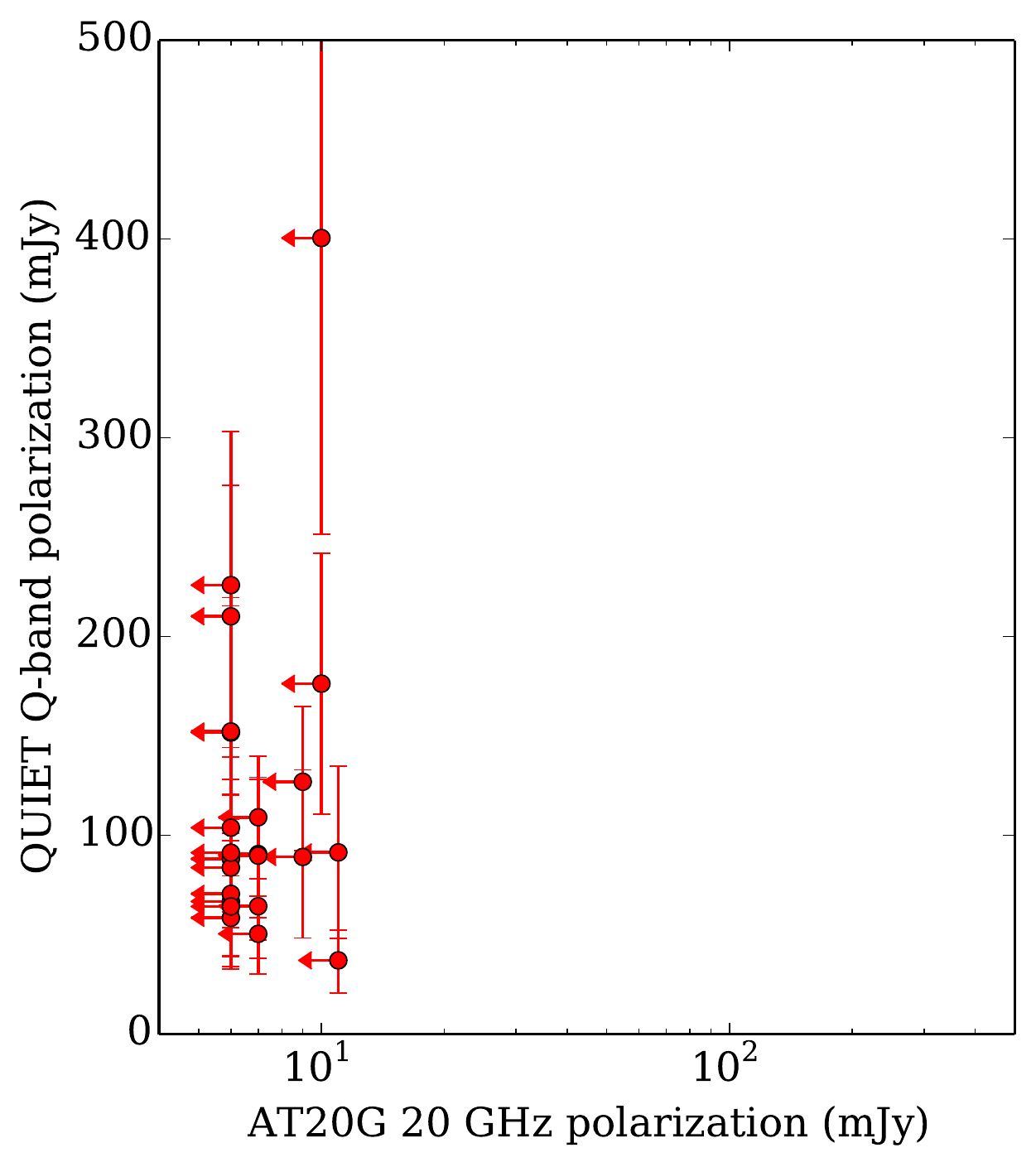}
\includegraphics[width=.33\textwidth]{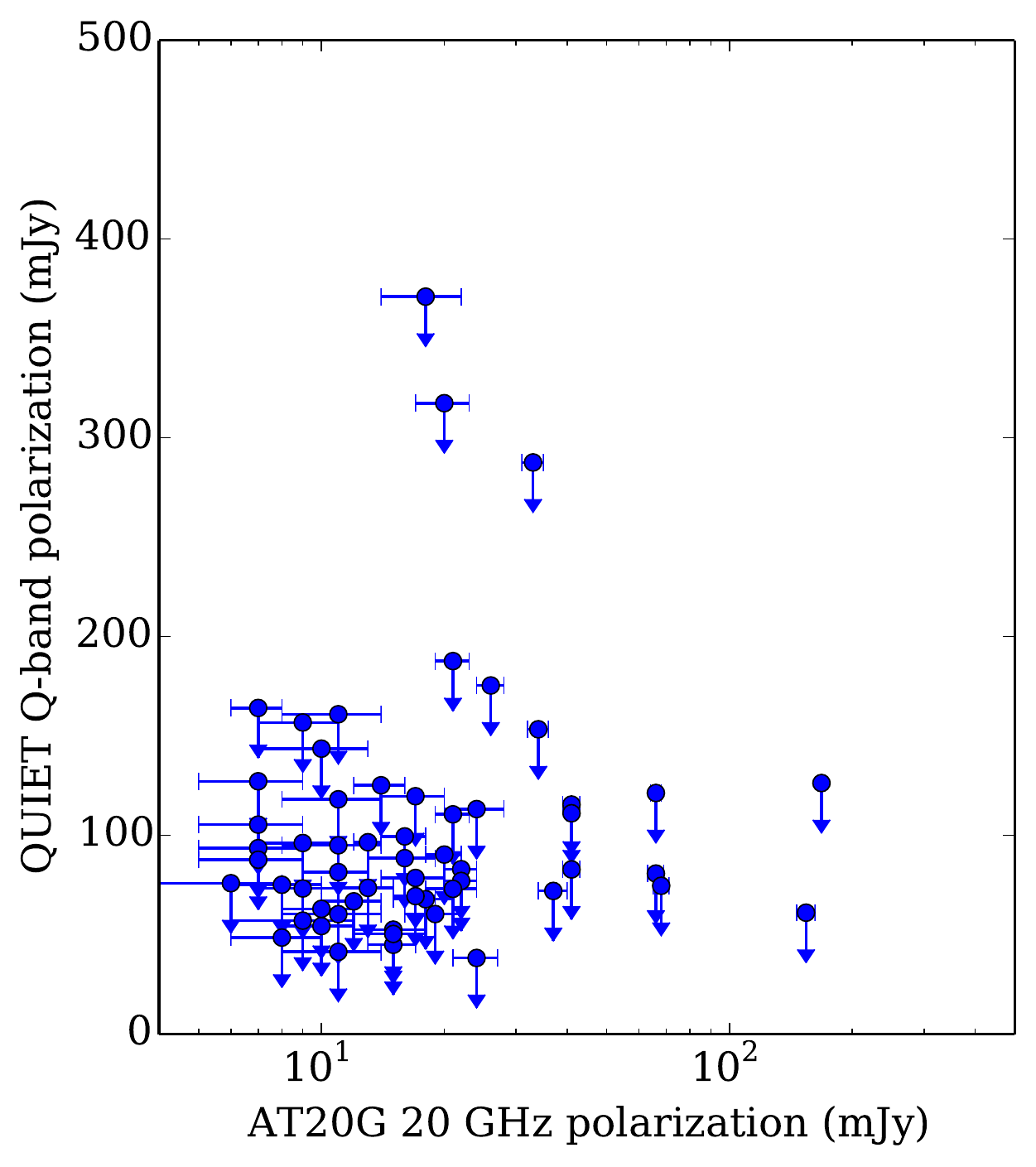}
\includegraphics[width=.33\textwidth]{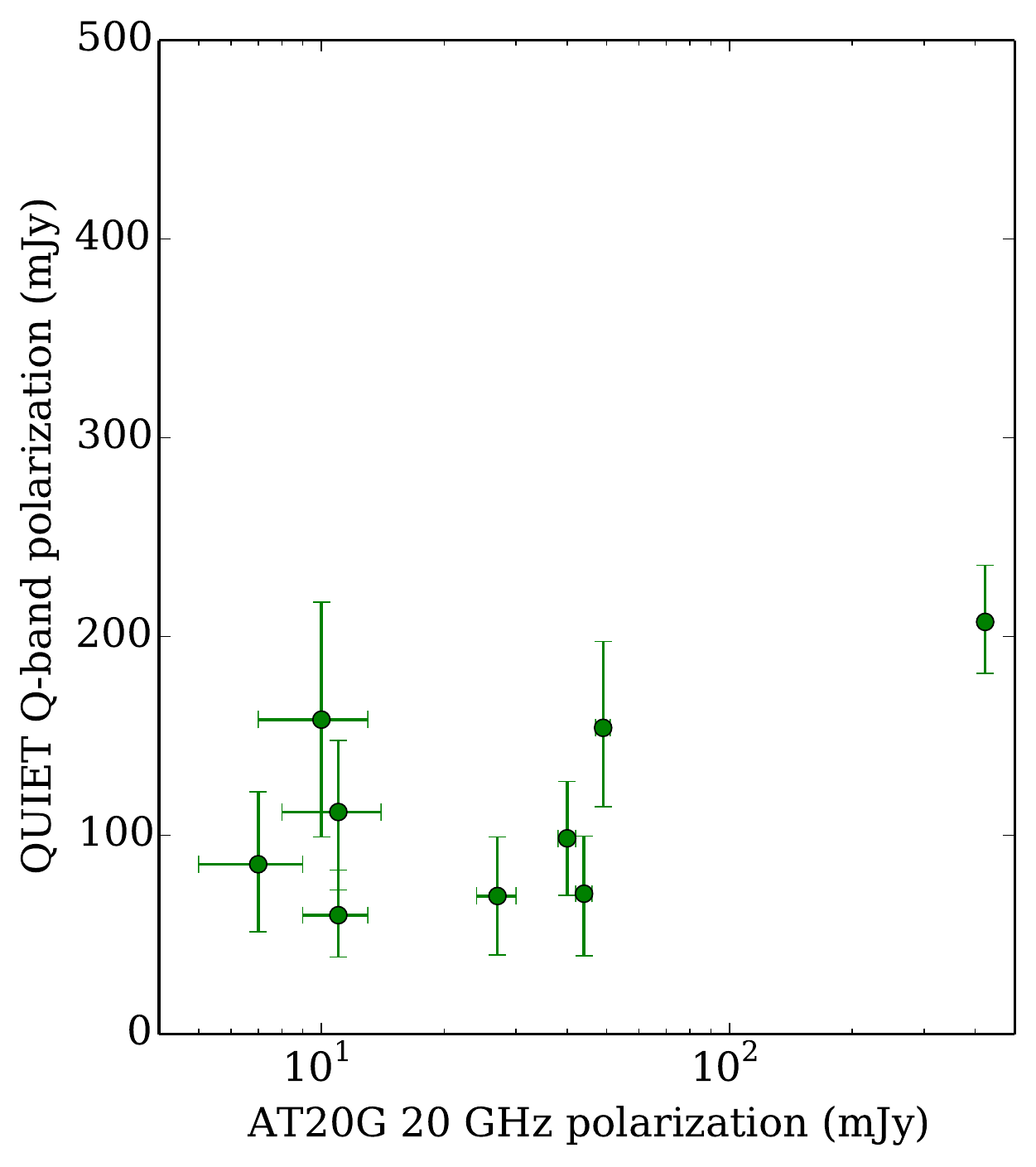}\\
\includegraphics[width=.33\textwidth]{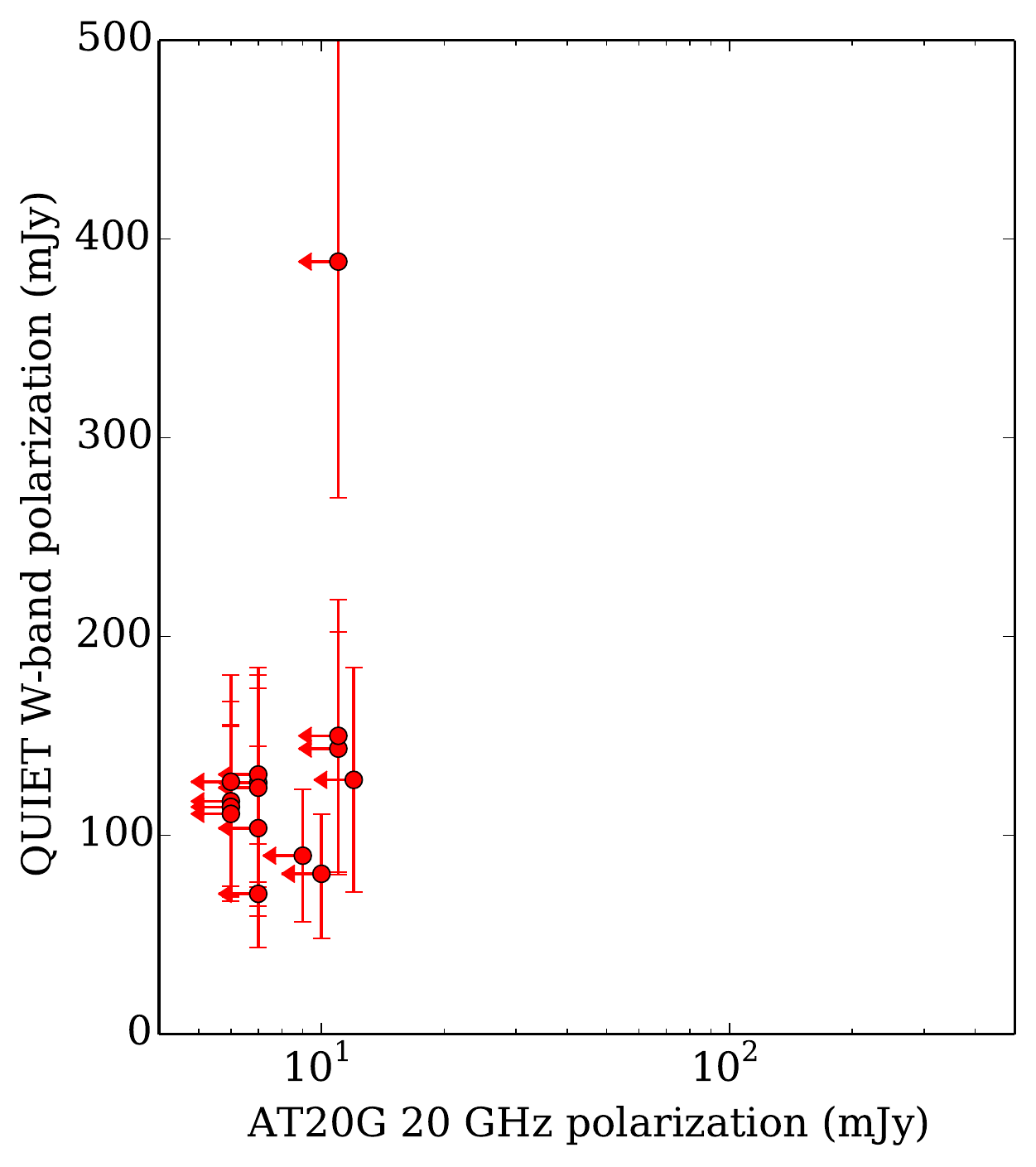}
\includegraphics[width=.33\textwidth]{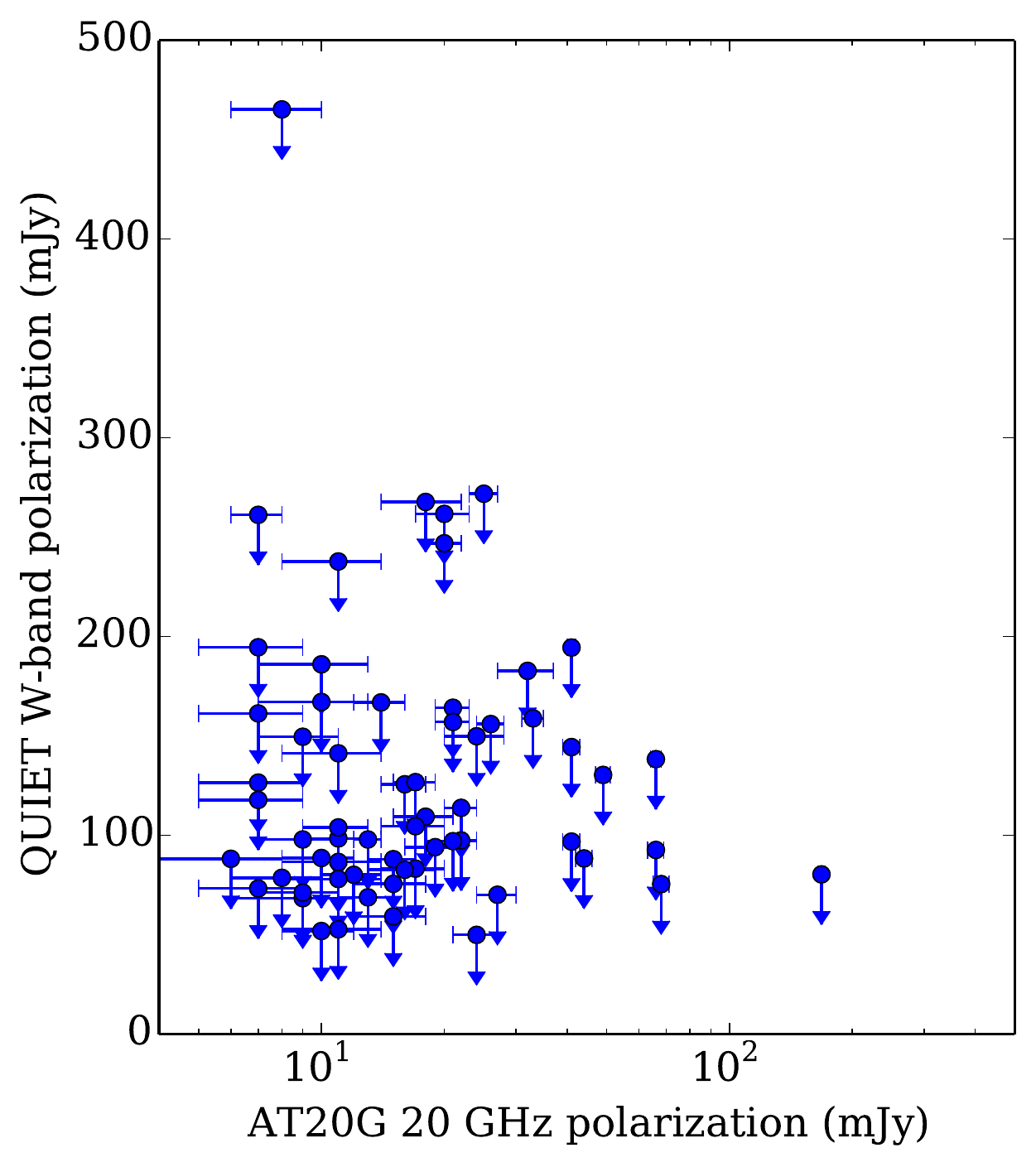}
\includegraphics[width=.33\textwidth]{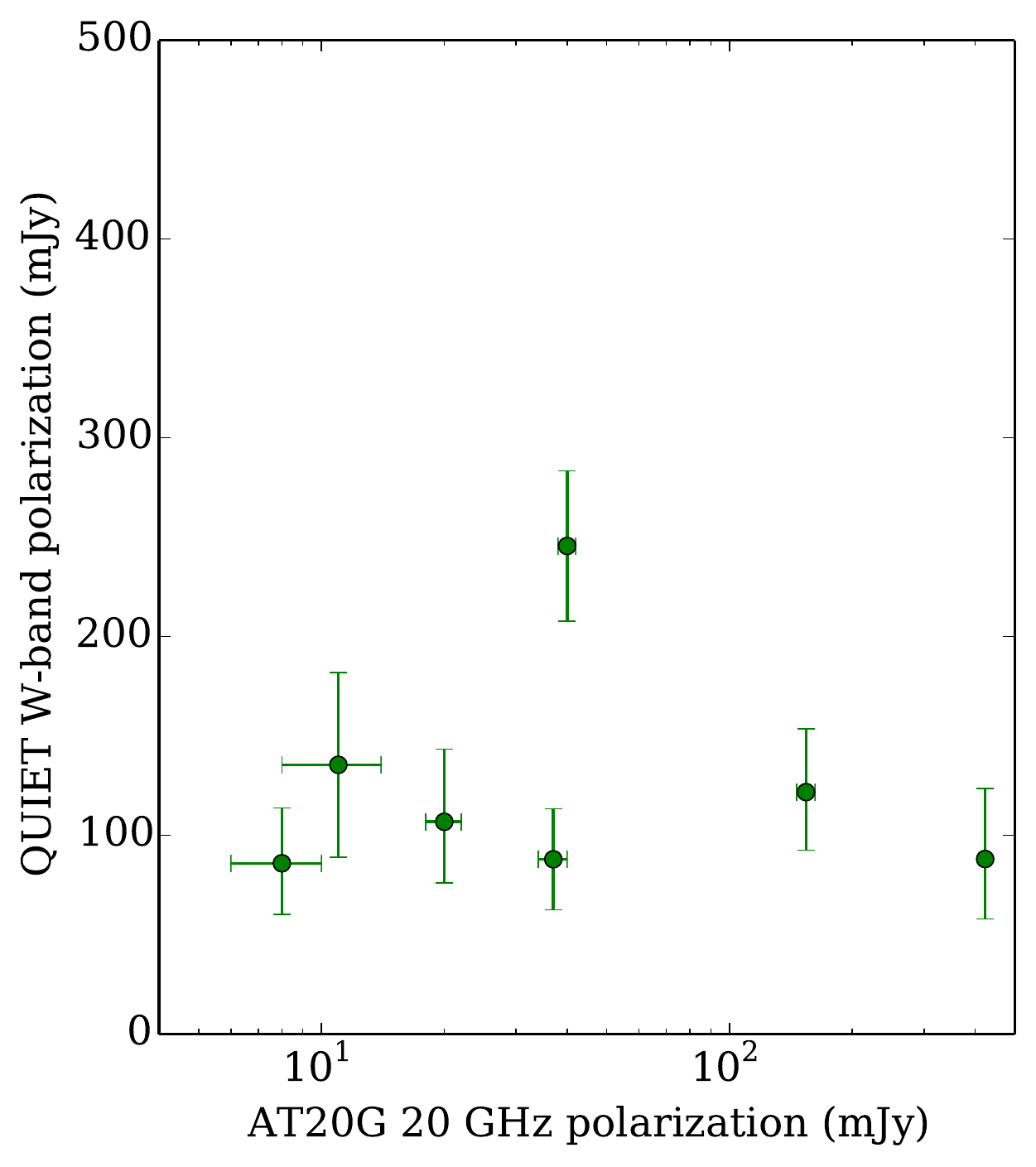}
\caption{QUIET data versus 20$\,$GHz polarization.  Top row: QUIET Q-band (43$\,$GHz) measurements.  Left, limits for AT20G and errors for QUIET; center, errors for AT20G and limits for QUIET; right, errors for both AT20G and QUIET.  Bottom row: the same but for QUIET W-band (95$\,$GHz).}
\label{fig:P20}
\end{figure*}

Of the 476 AT20G sources included in our Q-band analysis, 67 have a 20$\,$GHz polarization listed in the catalog.  For W-band there is a largely overlapping group of 69 sources.  We separate these sources into four groups according to the following criteria: (1) sources for which both AT20G and QUIET set an upper limit on polarization, (2) sources for which AT20G sets an upper limit and QUIET detects, (3) sources AT20G detects and for which QUIET sets an upper limit, and (4) sources which both AT20G and QUIET detect.  The last three groups---which include at least one detection---are plotted in Figure~\ref{fig:P20}.  As in intensity, we see no clear correlation between 20 GHz polarization and our polarization measurements.

\paragraph{Spectral energy distribution for bright sources}

For the ten sources with S/N~$\geq 3$ in either of the QUIET bands, we plot spectral energy distributions using our measurements and those from the literature.  All sources have 20$\,$GHz intensity from the AT20G survey, and many have 5 and 8$\,$GHz total intensity.  Many of these bright sources also have some polarization information from AT20G.  We plot \textit{WMAP} and \textit{Planck} total intensity measurements if a matching catalog source can be found within $4'$, a generous approximation of the positional uncertainty.  

Variability can be significant for these sources, and complicates the interpretation of the SED.  Flux variations are often fairly uniform across millimeter bands \citep{2013A&A...553A.107C,2009MNRAS.400..995F,2006MNRAS.370.1556B}.  The change at higher frequency almost always precedes the change at lower frequency and the polarization fraction could be driven higher or lower as a source flares, due to the physical mechanism of the flux variation.  These changes are almost always associated with a new component propagating through the core or along a jet \citep[e.g.][]{1979ApJ...232...34B}.  Sometimes the polarization is dominated by the core and sometimes it is dominated by the jet, and a new component can change the balance of the emitted flux.

\begin{figure}
\includegraphics[width=\columnwidth]{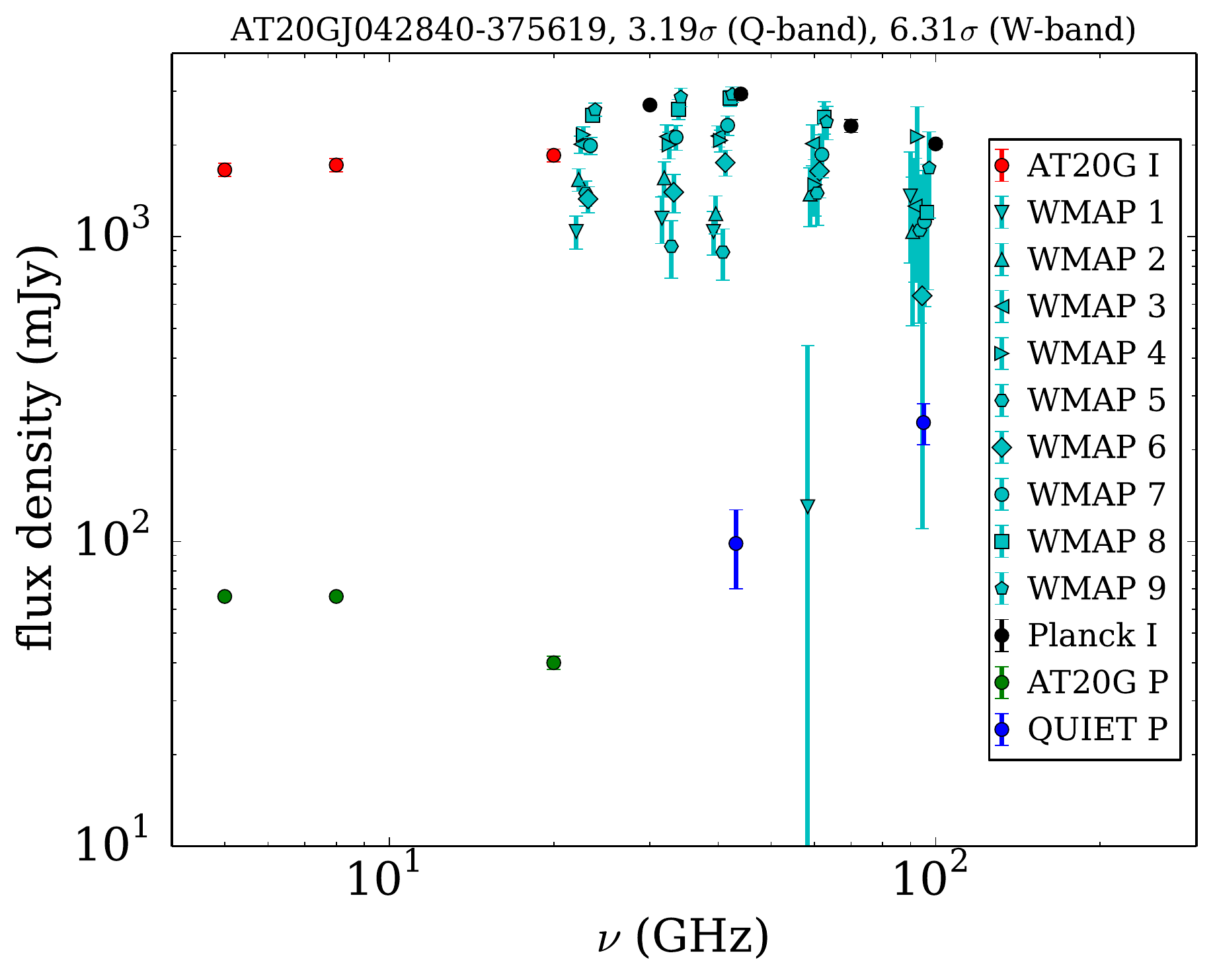}
\caption{Intensity and polarization measurements for AT20GJ042840-375619.  \textit{WMAP} measurements are catalog values for total intensity, by year of \textit{WMAP} data (see text).}
\label{fig:variability}
\end{figure}
To show variability, in Figure~\ref{fig:variability} we depict measurement of intensity and polarization for AT20GJ042840-375619 (PKS 0426-380; \textit{WMAP} J0428-3757; also called PLANCK044 G240.73-43.59 and PMN J0428-3756), highlighting the change in the intensity in \textit{WMAP} measurements over the nine years of the mission, beginning in the second half of 2001.  QUIET took data in 2008--2010, with the Q-band first and the W-band second.  The AT20G catalog data were taken from 2004 to 2008.  \textit{Planck} data from the 2013 release covers two sky surveys from 2009 and 2010.  However, unlike \textit{WMAP} and QUIET, which make long time averages, \textit{Planck}'s scan strategy yields two snapshots of the sources, taken $\sim 6$ months apart, which are then averaged in the catalog.

\begin{figure*}
\begin{center}
\includegraphics[width=0.33\textwidth]{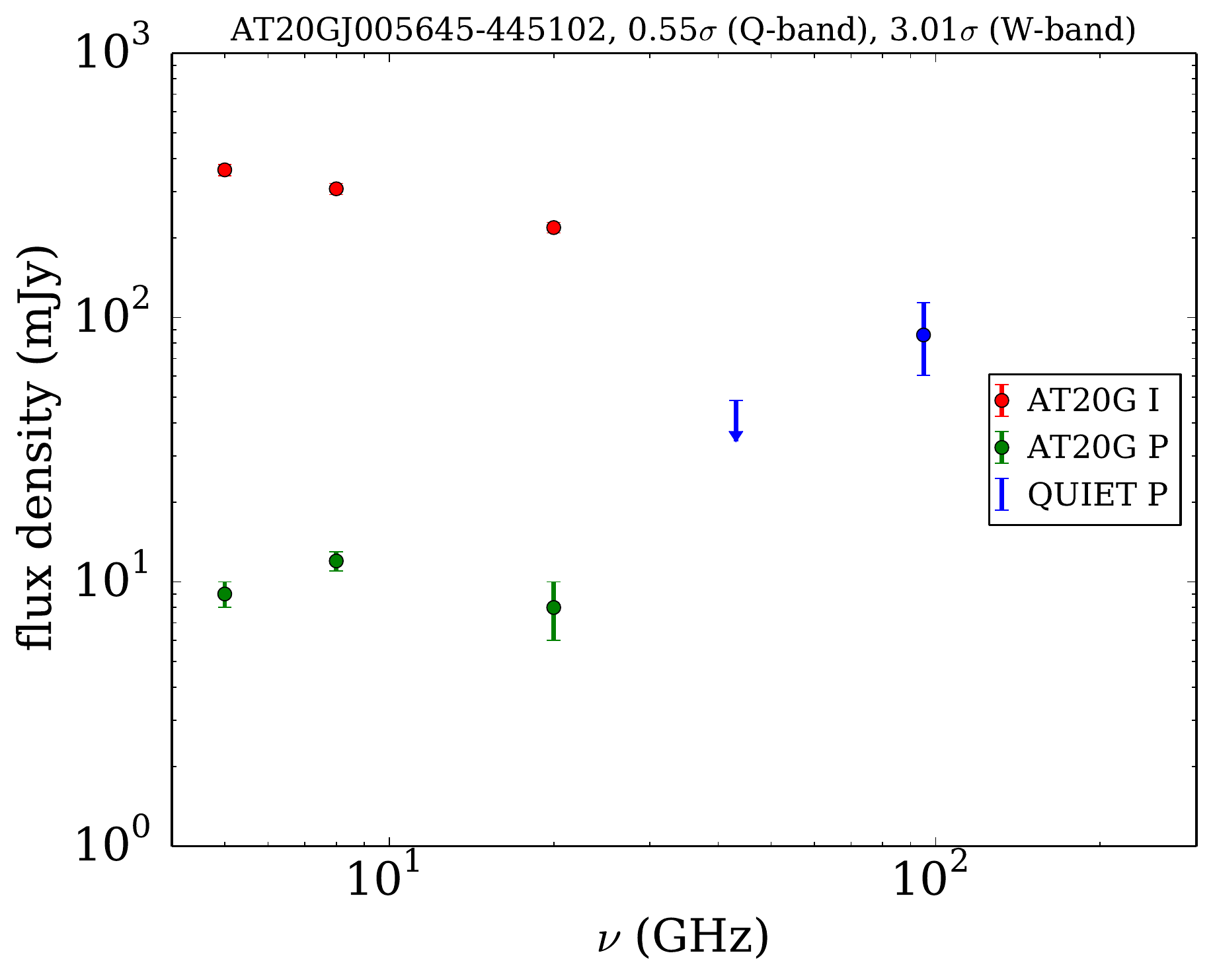}
\includegraphics[width=0.33\textwidth]{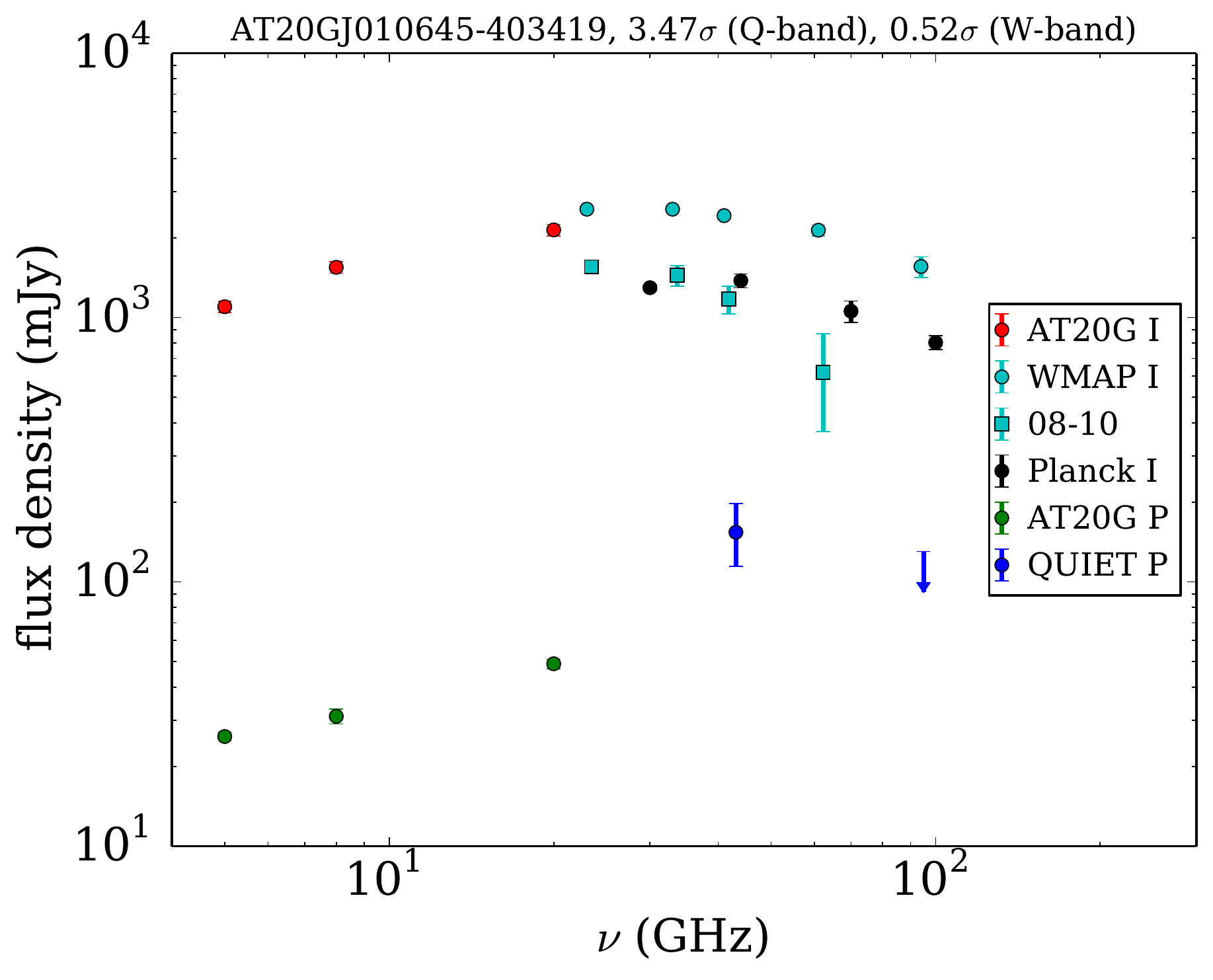} \\
\includegraphics[width=0.33\textwidth]{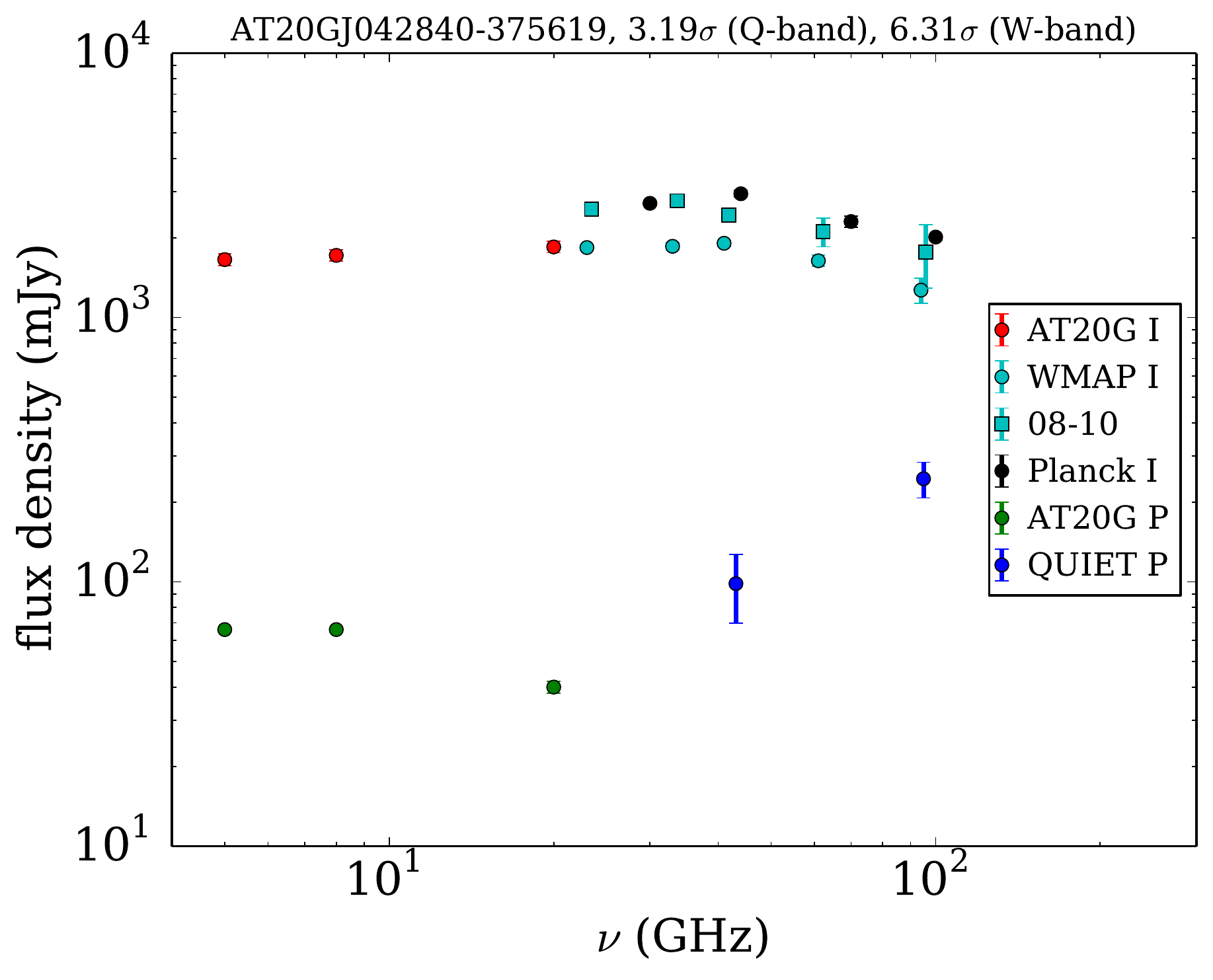}
\includegraphics[width=0.33\textwidth]{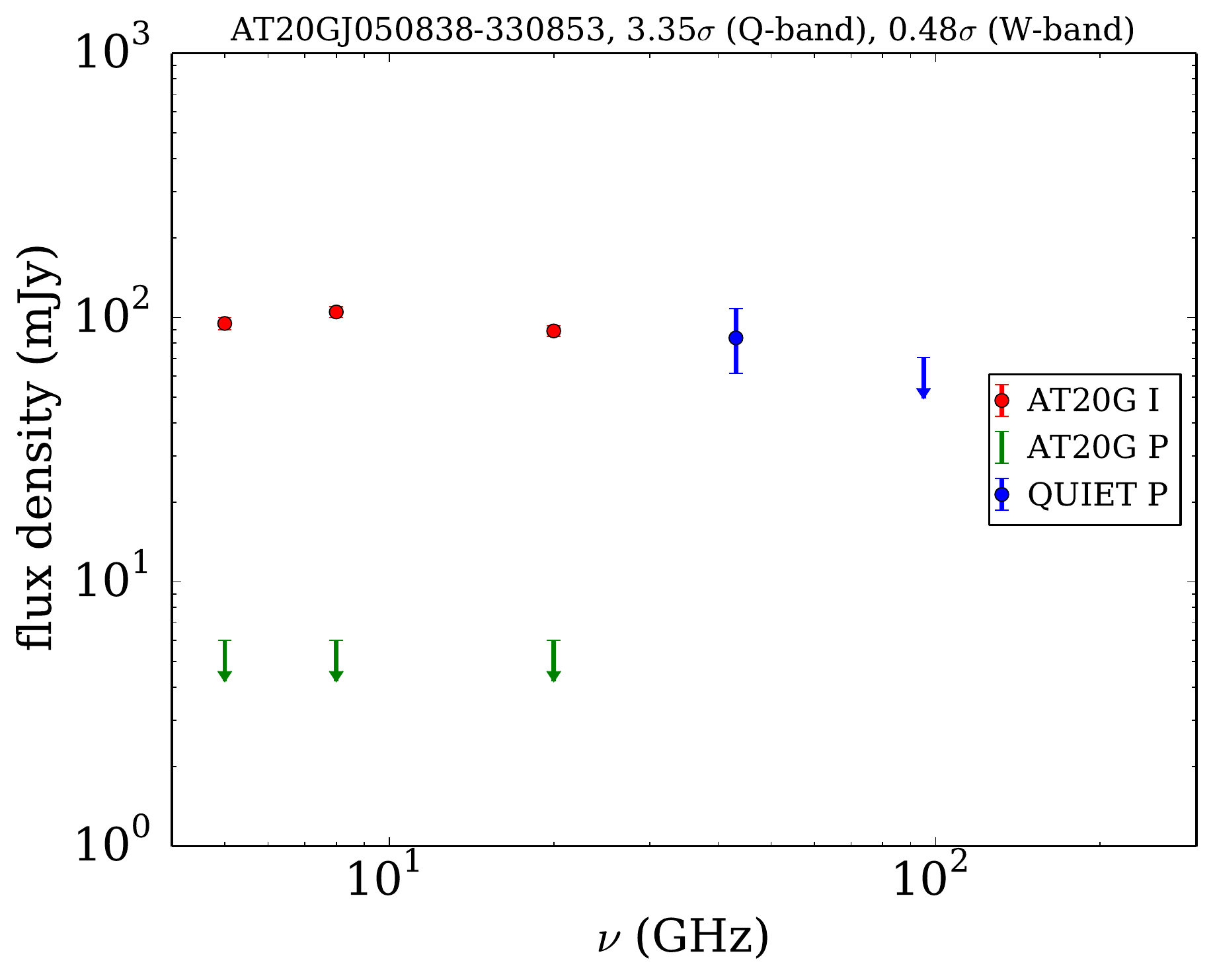}
\includegraphics[width=0.33\textwidth]{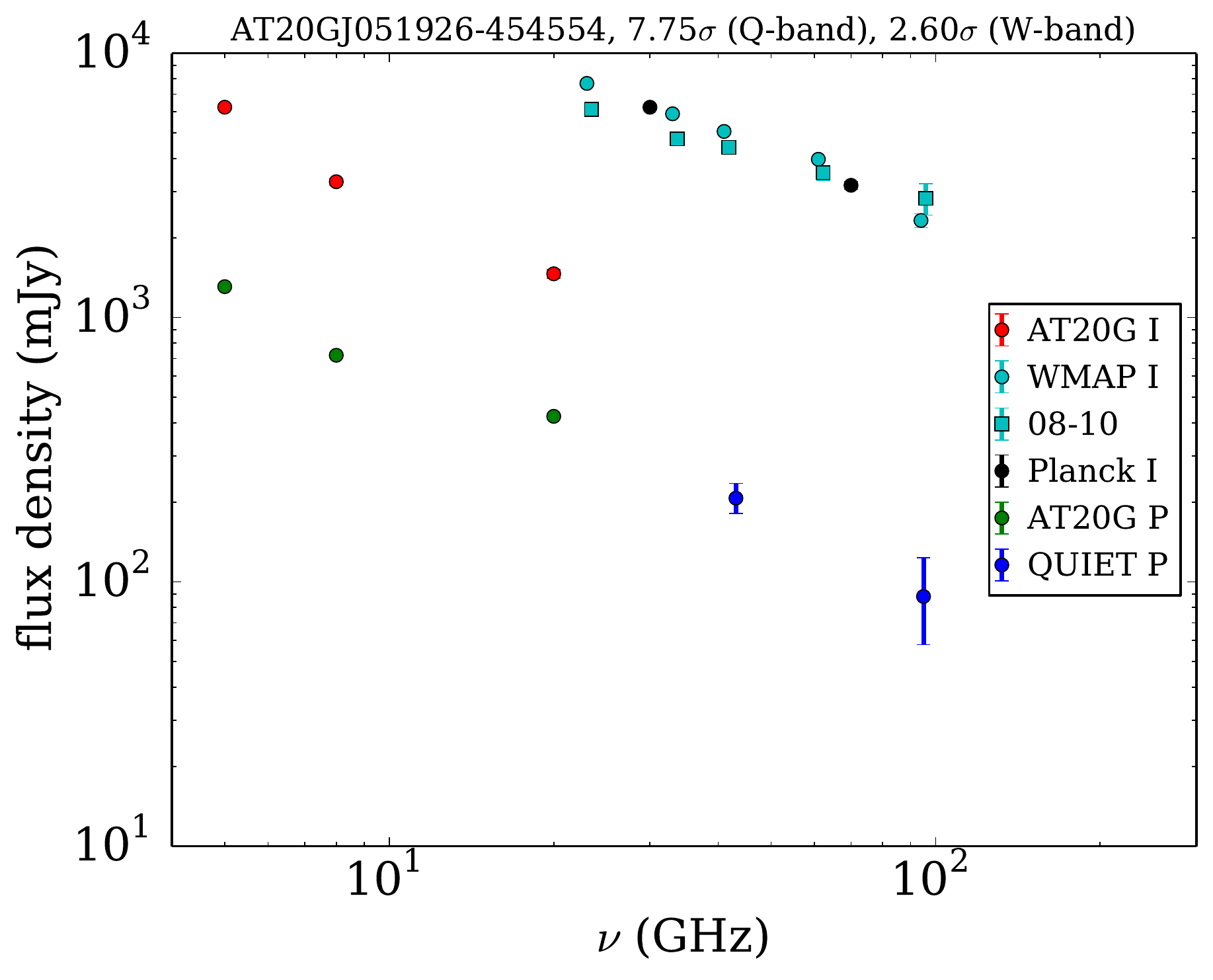}
\includegraphics[width=0.33\textwidth]{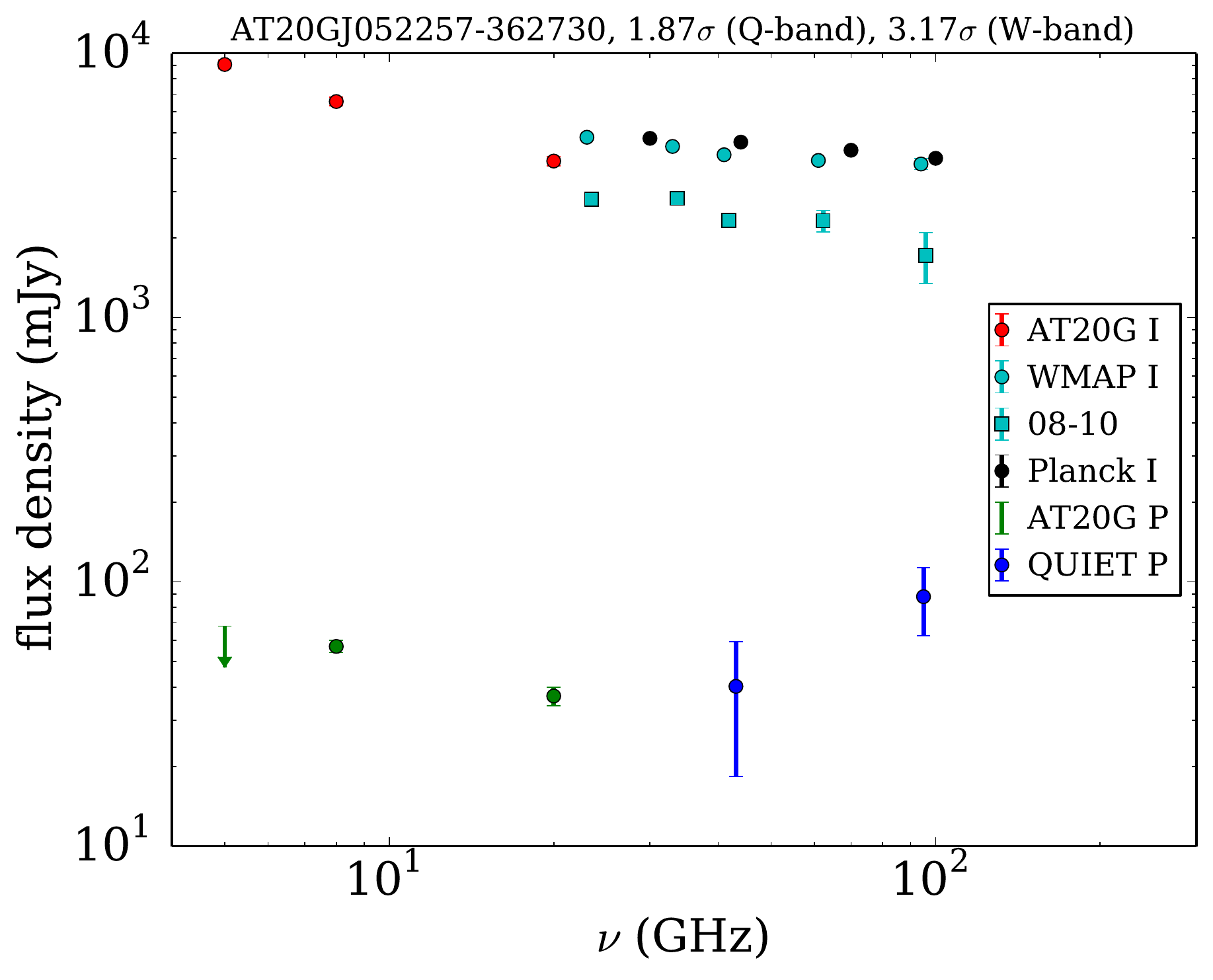}
\includegraphics[width=0.33\textwidth]{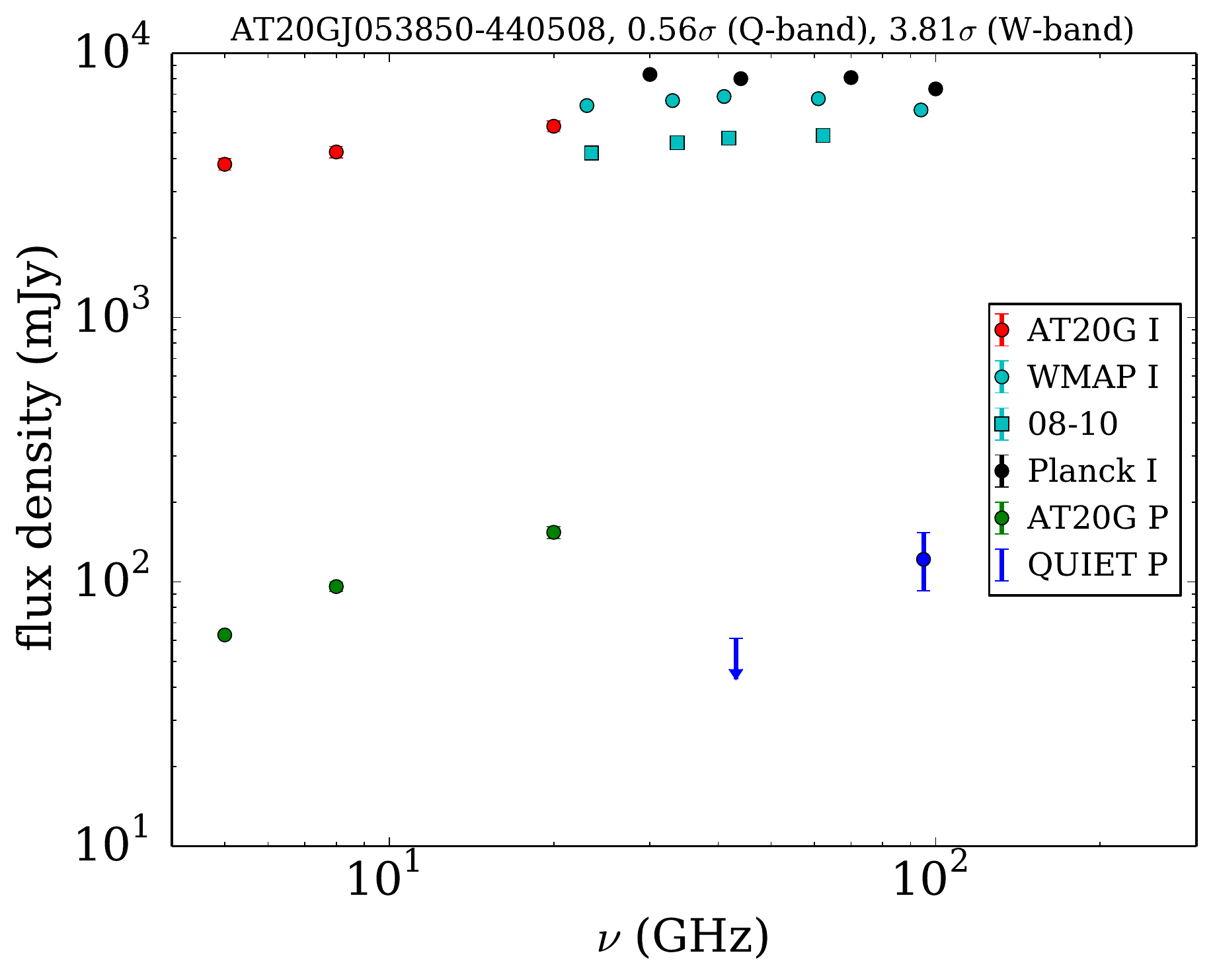}
\includegraphics[width=0.33\textwidth]{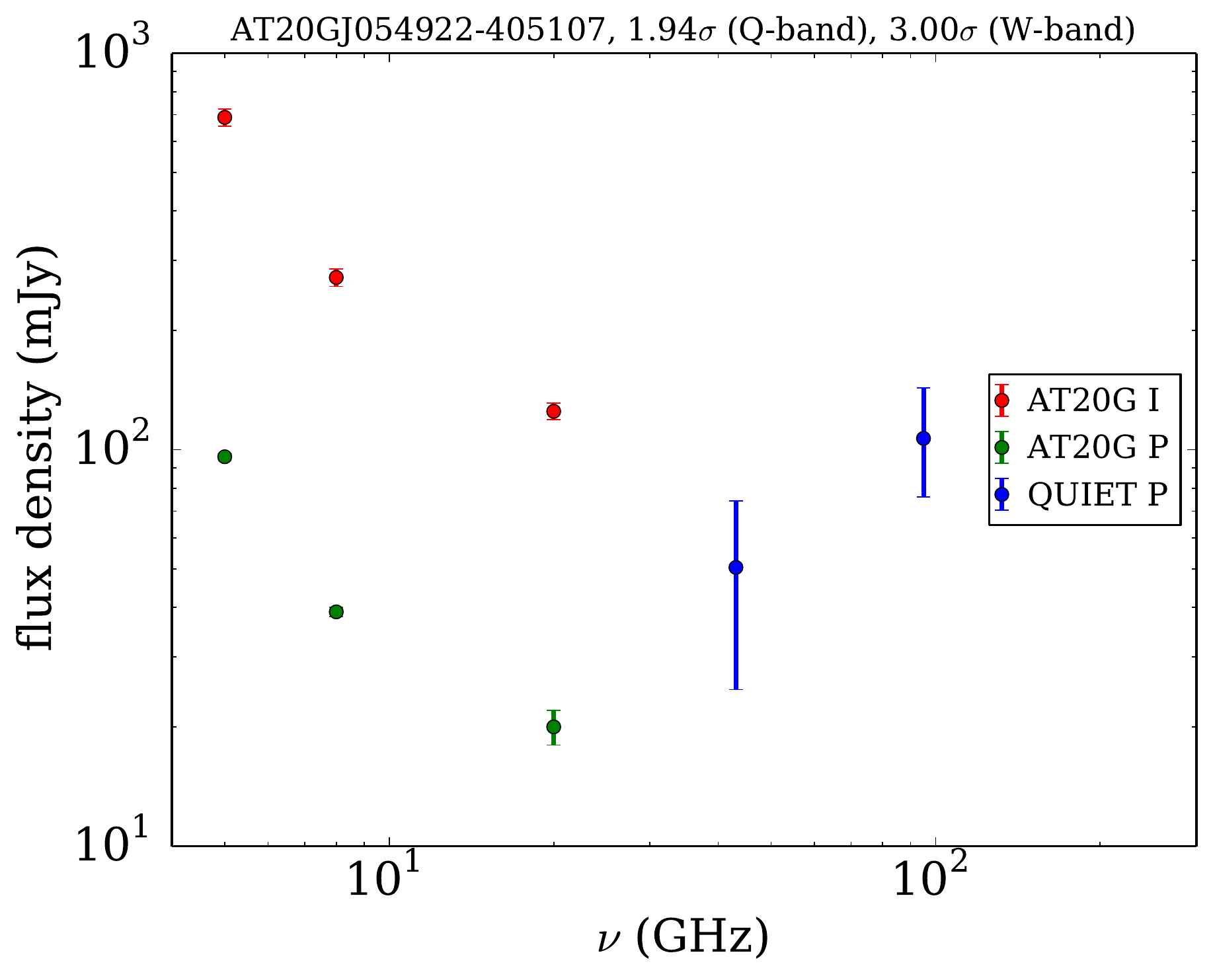}
\includegraphics[width=0.33\textwidth]{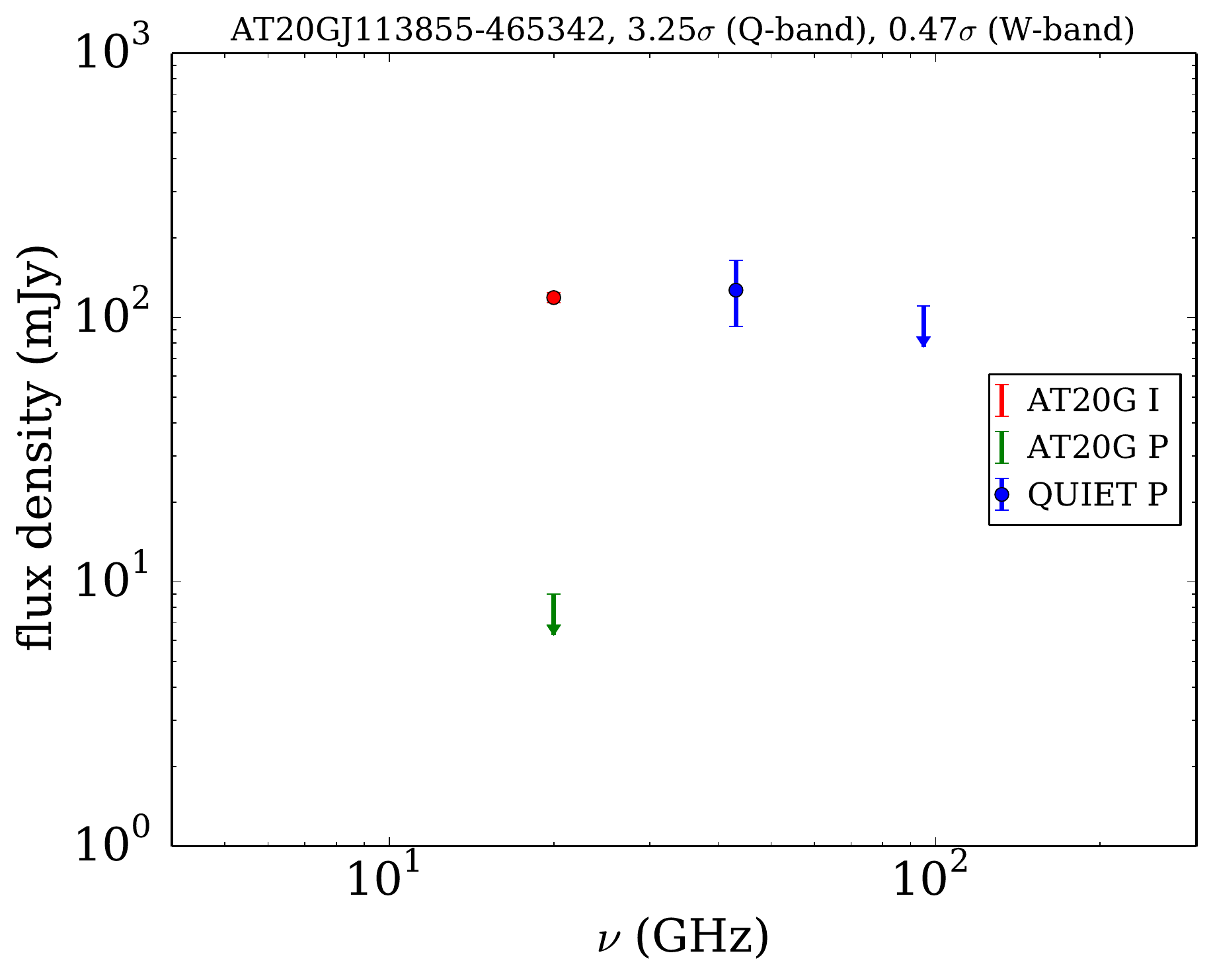}
\includegraphics[width=0.33\textwidth]{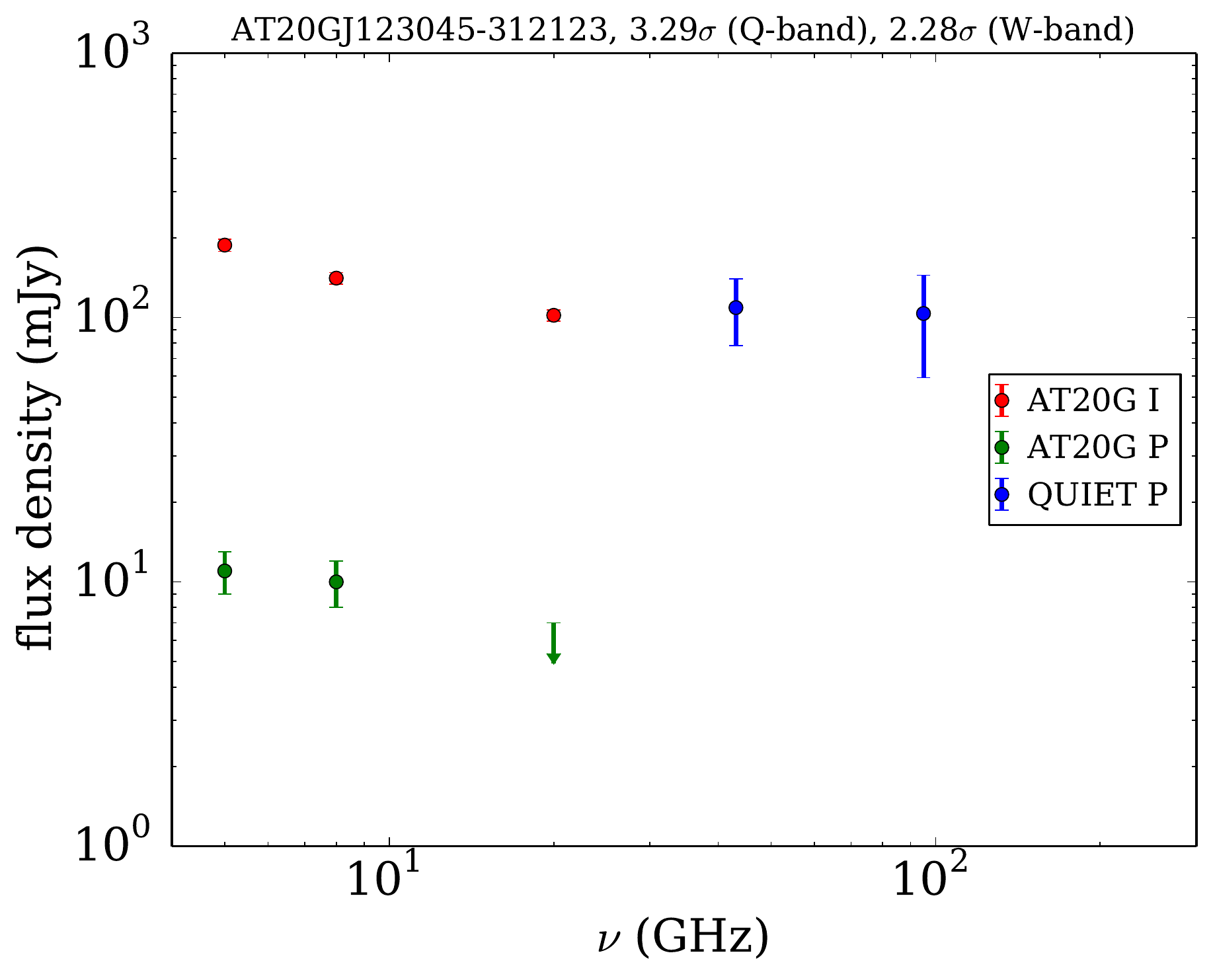}
\end{center}
\caption{Spectral energy distributions in temperature and polarization for point-like sources that are S/N~$\geq 3$ or better in either Q-band or W-band.  Left-to-right and top-to-bottom, the plots are sorted by right ascension.  Two points are shown for \textit{WMAP} when available: the nine-year catalog value (\textit{WMAP} I) and an average of \textit{WMAP} individual year catalogs (2008--2010) overlapping in time with QUIET observations. }
\label{fig:sed}
\end{figure*}
Ten of our highest S/N sources have their SEDs plotted in Figure \ref{fig:sed}, all having S/N$\geq 3$ in at least one QUIET band.  We expect about 1.3 spurious detections at S/N~$>3$ for each band, so $2.6 \pm 1.6$ of these SEDs may be selected due to a spurious fluctuation.  We plot two combinations of \textit{WMAP} data, namely values from the nine-year catalog and an average of years 2008--2010, which are contemporaneous to the QUIET observations. 

The SEDs are diverse and interesting.  Half the sources have rising SEDs, half are flat or falling.  All four sources that have a total intensity amplitude above 1$\,$Jy as observed by \textit{WMAP} and \textit{Planck} also have a polarization fraction above 1\%, and a few have very large polarization fractions.  For instance, AT20GJ042840-375619 in W-band has a polarization of $12.1^{+2.1}_{-2.2}$ percent compared to the \textit{Planck} total intensity catalog value at 100 GHz.  The polarization fraction compared to \textit{WMAP} is higher, $19.3^{+5.7}_{-4.6}$ percent, but \textit{WMAP} averages the total intensity over a longer duration. The source AT20GJ010645-403419 has a polarization fraction at Q-band of $11.1^{+4.1}_{-3.3}$ percent compared to \textit{Planck}, but we do not detect it in W-band.  The total intensity measurement in \textit{WMAP} Q-band is higher, so the comparative polarization fraction is lower, $6.3^{+1.9}_{-1.7}$ percent. For Pictor A's western lobe, AT20GJ051926-454554, the \textit{WMAP} and \textit{Planck} data pick up contributions from the nucleus and eastern lobe, which reduces the effective polarization fraction in $>10'$ resolution imaging to $4.1\pm{0.6}$ percent at Q-band.

Several of the SEDs appear peculiar. For instance, the QUIET polarized flux density is sometimes equal to or greater than the \textit{total intensity} measurement at 5--20$\,$GHz.  Variability could explain this, but the relevant sources are for the most part not listed in the \textit{WMAP} or \textit{Planck} catalogs, which implies that the total intensity flux density must be below $\sim 1$\, Jy.  This in turn would mean that the polarization fractions are very large.  Some of these fainter sources may be spurious (such as AT20GJ005645-445102, AT20GJ050838-330853, and AT20GJ113855-465342), because they are significant in only one band.  However, AT20GJ123045-312123 is detected with significances of S/N~$=3.29$ and S/N~$=2.28$ in the Q- and W-bands, and the probability is 1.1\% that we should encounter in both bands  simultaneously a spurious fluctuation this large or larger.

\section{Conclusions} \label{sec:conclusions}

We have measured polarization at 43 and 95$\,$GHz at the locations of $\sim 480$
 20-GHz-selected radio sources.
We have several S/N~$>3$ detections of polarized emission.  Since we know from the 20$\,$GHz data that sources are present, our many upper limits also provide useful constraints for the 43 and 95$\,$GHz polarized source populations.
We find no immediately obvious trend between the 20$\,$GHz intensity or polarization and the higher frequency polarization.  These results do not support simple models that assume source populations with uniform SEDs and polarization fractions.

We find similar numbers of high-significance polarization detections at 43 and 95$\,$GHz, despite a lower sensitivity in the higher frequency band.  This may suggest a flat spectrum in polarization.  Several of the bright sources show the same trend.  However the signal-to-noise ratio is too low for most individual sources to allow us to draw firm conclusions from direct band-to-band comparisons of flux densities.
The SEDs of bright sources are diverse and interesting, and  may in some cases require significant variability to make sense.

With these observations we have in hand a probability distribution for the polarization for each source. That allows us to set constraints on properties of the whole population.  In the future we will test models of polarized source counts, and to assess the implied impact on CMB polarization.

Future CMB surveys 
(e.g.~AdvACTPol\footnote{\url{http://www.princeton.edu/act/}}, 
Simons Array\footnote{\url{http://cosmology.ucsd.edu/simonsarray.html}}, 
SPT3G\footnote{\url{http://pole.uchicago.edu/}}, 
COrE+\footnote{\url{http://www.core-mission.org/}}, 
LiteBIRD\footnote{\url{http://litebird.jp/eng/}}, 
PIXIE\footnote{\citet{2011JCAP...07..025K}}) 
will make even more sensitive polarization measurements over large areas of the sky.  The methods presented here for detection and analysis of extragalactic, polarized sources can be readily applied to those data sets, revealing more about the properties of AGN emission.

\section*{Acknowledgments}

Bruce Winstein, who led the QUIET project, died in 2011, soon after observations concluded.  The project's success owes a great debt to his intellectual and scientific leadership.

Support for the QUIET instrument and operation was provided through the NSF
cooperative agreement AST-0506648. Support was also provided by NSF awards
PHY-0855887, PHY-0355328, AST-0448909, AST-1010016, and PHY-0551142; KAKENHI 20244041,
20740158, and 21111002; PRODEX C90284; a KIPAC Enterprise grant; and by the Strategic Alliance for
the Implementation of New Technologies (SAINT).  
This research used resources
of the National Energy Research Scientific Computing Center (NERSC), which is supported by the
Office of Science of the U.S. Department of Energy under Contract No. DE-AC02-05CH11231.

Some work was performed on the Joint Fermilab-KICP Supercomputing
Cluster, supported by grants from Fermilab, the Kavli Institute for
Cosmological Physics, and the University of Chicago.  Some work was
performed on the Abel Cluster, owned and maintained by the University
of Oslo and NOTUR (the Norwegian High Performance Computing
Consortium), and on the Central Computing System, owned and operated
by the Computing Research Center at KEK.
Portions of this work were
performed at the Jet Propulsion Laboratory (JPL) and California
Institute of Technology, operating under a contract with the National
Aeronautics and Space Administration. The Q-band modules
were developed using funding from the JPL R\&TD program.  We acknowledge 
the Northrop Grumman Corporation for collaboration in the development and 
fabrication of HEMT-based cryogenic temperature-compatible MMICs.

C.D.~acknowledges an STFC Advanced Fellowship, an EU Marie-Curie IRG grant under the FP7 and an ERC Starting Grant (no.~307209).
H.K.E.~acknowledges an ERC Starting Grant under FP7.
A.D.M.~acknowledges a Sloan foundation fellowship. 
J.Z.~gratefully acknowledges a South Africa National Research Foundation Square Kilometre Array Research Fellowship.

PWV measurements were provided by the Atacama Pathfinder Experiment
(APEX). We thank CONICYT for granting permission to operate within the
Chajnantor Scientific Preserve in Chile, and ALMA for providing site
infrastructure support.
Field operations were based at the Don Esteban facility run by Astro-Norte.
We are particularly indebted to the engineers
and technician who maintained and operated the telescope: Jos\'e Cort\'es,
Cristobal Jara, Freddy Mu\~noz, and Carlos Verdugo.

In addition, we would like to acknowledge the following people for
their assistance in the instrument design, construction,
commissioning, operation, and in data analysis: 
Augusto Gutierrez Aitken, 
Colin Baines, 
Phil Bannister, 
Hannah Barker, 
Matthew R. Becker, 
Alex Blein, 
Alison Brizius,
L.~Bronfman,
Ricardo Bustos,
April Campbell, 
Anushya Chandra, 
Sea Moon Cho, 
Sarah Church,
Joelle Cooperrider,
Mike Crofts, 
Emma Curry, 
Maire Daly, 
Fritz Dejongh, 
Joy Didier, 
Greg Dooley,
Robert Dumoulin,
Hans Eide, 
Will Grainger, 
Jonathon Goh, 
Peter Hamlington,
Takeo Higuchi, 
Seth Hillbrand, 
Ben Hooberman, 
Kathryn D. Huff,
K.~Ishidoshiro,
Norm Jarosik,
M.~E.~Jones,
P.~Kangaslahti,
D.~J.~Kapner,
Eiichiro Komatsu,
Jostein Kristiansen, 
Donna Kubik,
Richard Lai, 
C.~R.~Lawrence,
David Leibovitch, 
Kelly Lepo, 
Siqi Li, 
M.~Limon,
Martha Malin, 
Mark McCulloch,
 J.~J.~McMahon,
Oliver Montes, 
David Moore, 
M.~Nagai,
H.~Nguyen,
G.~Nixon,
Ian O'Dwyer, 
Gustavo Orellana, 
Stephen Osborne, 
Stephen Padin,
T.~J.~Pearson,
Felipe Pedreros,
Ashley Perko, 
L.~Piccirillo,
J.~L.~Richards,
Alan Robinson,
Dorothea~Samtleben,
Jacklyn Sanders, 
Dale Sanford, 
Yunior Savon, 
Michael Seiffert,
Martin Shepherd,
Kendrick Smith
Mary Soria, 
Alex Sugarbaker, 
David Sutton,
Keith Vanderlinde,
Matias Vidal, 
Liza Volkova, 
Ross Williamson,
Stephanie Xenos, 
Octavio Zapata, 
and 
Mark Zaskowski.

Some of the results in this paper have been derived using the HEALPix package \citep{2005ApJ...622..759G}.

\bibliography{quiet_ptsrc}{}
\bibliographystyle{hapj}
 
\end{document}